\documentclass[aps,prb,twocolumn,showpacs,superscriptaddress,final]{revtex4-1}

\usepackage{graphicx} % Fuer das Einbinden von Grafiken ueber \includegraphics
\usepackage{color}
\usepackage{amsmath} % z.B. fuer \DeclareMathOperator{}{} und \substack
\usepackage{amstext} % z.B. $\text{}$
\usepackage{amssymb}
\usepackage{cases}
\usepackage{booktabs}
\usepackage{grffile}
\usepackage{xfrac} 	% for nice fractions with \sfrac{}{}
\usepackage{dcolumn}	% for the new column types

\definecolor{violet}{rgb}{0.5803922,0,0.827451}

\newcolumntype{t}[1]{D{.}{.}{#1}}
\newcolumntype{.}{D{.}{.}{-1}}
\def\d{\mathrm{d}}
\def\ls{\left[}
\def\rs{\right]}
\def\mod{\,\mathrm{mod}\,}
\begin{document}

%\begin{frontmatter}
\title{One-dimensional infinite component vector spin glass with long-range interactions}
%%\tnotetext[t1]{}
\author{Frank Beyer}
\email{beyerf@uni-mainz.de}
\affiliation{Institut f\"ur Physik, Johannes Gutenberg-Universit\"at Mainz,
  Staudinger Weg 7, D-55099 Mainz, Germany}

\author{Martin Weigel}
\email{martin.weigel@coventry.ac.uk}
\affiliation{Institut f\"ur Physik, Johannes Gutenberg-Universit\"at Mainz,
  Staudinger Weg 7, D-55099 Mainz, Germany}
\affiliation{Applied Mathematics Research Centre, Coventry University,
  Coventry, CV1~5FB, England}

\author{M. A. Moore}
\email{m.a.moore@manchester.ac.uk}
\affiliation{School of Physics and Astronomy, University of
  Manchester, Manchester M13 9PL, UK}

\date{\today}

\begin{abstract}
  We investigate zero and finite temperature properties of the one-dimensional
  spin-glass model for vector spins in the limit of an infinite number $m$ of spin
  components where the interactions decay with a power, $\sigma$, of the distance.  A
  diluted version of this model is also studied, but found to deviate significantly
  from the fully connected model.  At zero temperature, defect energies are
  determined from the difference in ground-state energies between systems with
  periodic and antiperiodic boundary conditions to determine the dependence of the
  defect-energy exponent $\theta$ on $\sigma$. A good fit to this dependence is
  $\theta =\frac{3}{4} -\sigma$. This implies that the upper critical value of
  $\sigma$ is $3/4$, corresponding to the lower critical dimension in the
  $d$-dimensional short-range version of the model. For finite temperatures the large
  $m$ saddle-point equations are solved self-consistently which gives access to the
  correlation function, the order parameter and the spin-glass
  susceptibility. Special attention is paid to the different forms of finite-size
  scaling effects below and above the lower critical value, $\sigma =5/8$, which
  corresponds to the upper critical dimension $8$ of the hypercubic short-range
  model.
\end{abstract}

%75.50.Lk Spin glasses and other random magnets
%64.60.Fr Equilibrium properties near critical points, critical exponents
%02.60.Pn Numerical optimization

%\begin{keyword}
%vector spin glasses \sep spherical limit \sep lower critical dimension \sep 1d power-law
%\end{keyword}
%
%\end{frontmatter}

\maketitle

\section{Introduction}

The problem of understanding the physics of spin glasses in the form of simple model
systems incorporating frustration and random disorder has challenged theoretical
physicists for the last forty years \cite{binder:86a}. Although significant progress
has been made \cite{kawashima:03a}, predominantly through extensive numerical
simulations, a number of important puzzles are still unsolved, and we do not have a
clear understanding yet of the nature of the spin-glass phase and whether replica-symmetry 
breaking, the hallmark of the spin-glass state in mean-field theory
\cite{parisi:79, *parisi:83}, carries over to systems in low dimensions $d$.
\cite{moore:11}

Unlike the case of ferromagnets, we have currently no means of performing a
well-behaved perturbative expansion of the replica-field theory of spin glasses in
dimensions below their upper critical dimension $d_u = 6$. While this program
effectively starts from the $d=\infty$ Sherrington-Kirkpatrick (SK)
\cite{sherrington:75} model to understand behavior in finite dimensions, an
alternative approach is to consider vector spin glasses with an infinite number of
spin components $m=\infty$, but arbitrary spatial dimension $d$, and extend these
results in a $1/m$ expansion to the physically more relevant cases with finite
$m$. \cite{aspelmeier:04a} This approach appears particularly suitable as the
$m=\infty$ limit implies a number of simplifications as compared to the models with
finite $m$. Firstly, the model is replica symmetric even in the mean-field limit
\cite{dealmeida:78}, in contrast to the Ising, {\em XY\/} or Heisenberg spin glasses
usually considered (with $m=1,2$ and $3$ spin components, respectively).  Secondly,
it is tractable by analytical and numerical means. For calculations at zero
temperature, it turns out to be very useful that the metastability afflicting
finite-$m$ spin glasses disappears \cite{bray:81,morris:86a}, making it numerically
straightforward to determine ground states. At finite temperatures, the
$m\rightarrow\infty$ limit leads to saddle-point equations which allow for the exact
calculation of correlation functions of finite samples for both mean-field and
non-mean-field models \cite{bray:82b}.

On the other hand, the $m=\infty$ model has peculiarities. It has been shown that the
upper critical dimension, $d_u$, which is six for spin glasses with a finite number
of spin components, is elevated to eight \cite{green:82}. Likewise, the {\em lower\/}
critical dimension at which a finite-temperature transition first occurs, appears to
be also increased from that of systems with a finite number of spin components.  It
has been estimated from numerical studies that \cite{lee:05a, beyer:11} $d_l \approx
6$ . The mechanism of effective dimensional reduction that is at work in lifting
$d_u$ also leads to a violation of hyperscaling even below the upper critical
dimension \cite{green:82}.  Finally, regarding the numerical calculations considered
here, one should note that an order of limits, $m\rightarrow\infty$ before
$N\rightarrow\infty$, is used which is opposite to that used in field theoretic
calculations \cite{green:82,viana:88}. Taking the infinite-component limit first
might be considered the zeroth-order term in a $1/m$ expansion around the
field-theoretic calculation \cite{lee:05}.

From studies of ferromagnets it was realized many years ago \cite{dyson:69,*dyson:71}
that systems with long-range, power-law interactions in low dimensions could be used
to model the non-trivial critical behavior of the kind expected in short-range
systems of higher dimensions. Similar observations were later made for spin glasses
\cite{kotliar:83,bray:86}. It was subsequently realized that such models are useful
for numerical studies, as finite-size corrections, known to be strong for spin-glass
systems, depend on the {\em linear\/} extension of the lattice. Hence, studying, for
instance, a one-dimensional system with interactions which fall off with distance
with a power $\sigma$ allows one to access significantly larger (linear) system sizes
than studying similar systems on hypercubic lattices
\cite{katzgraber:03a,katzgraber:05b,katzgraber:05}. While these first works
considered Ising spin glasses, Potts \cite{andrist:11} and Heisenberg
\cite{kawamura:07a,viet:10,sharma:11,*sharma:11a} models have also recently received
some attention. For the $m=\infty$ limit considered here, this approach appears to be
well suited, as reasonable system sizes in the dimensions $d>6$ where a finite
temperature phase transition occurs are nearly inaccessible with current
computational resources. For the case of Ising spin glasses, diluted lattices have
been used to reach even larger linear dimensions
\cite{leuzzi:08,katzgraber:09,leuzzi:09}, and the usefulness of this approach for the
$m=\infty$ model will be discussed in some detail below.

The paper is organized as follows. In Sec.~\ref{sec:model} the model and
some theoretical preliminaries will be introduced. Section \ref{sec:theory} discusses
the phase diagram of the one-dimensional spin glass with power-law interactions for
finite and infinite $m$, and the scaling and finite-size scaling in the vicinity of
the critical point. In Sec.~\ref{sec:GS}, we report on the results of ground-state
calculations and a study of the defect energies. The critical behavior is examined
with finite-temperature methods in Sec.~\ref{sec:finite}. Finally,
Sec.~\ref{sec:concl} contains our conclusions.

\section{The model\label{sec:model}}

\begin{figure}
        \scalebox{0.15}{\includegraphics[trim= 4cm 1cm 3cm 1cm]{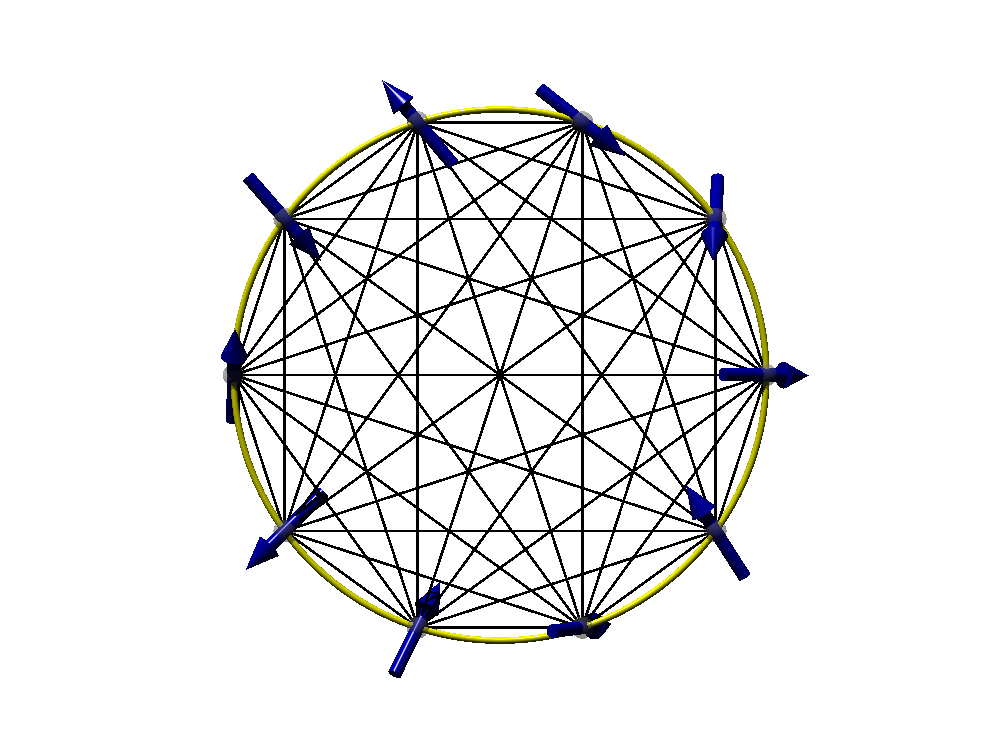}}
        \scalebox{0.15}{\includegraphics[trim= 4cm 1cm 4cm 1cm]{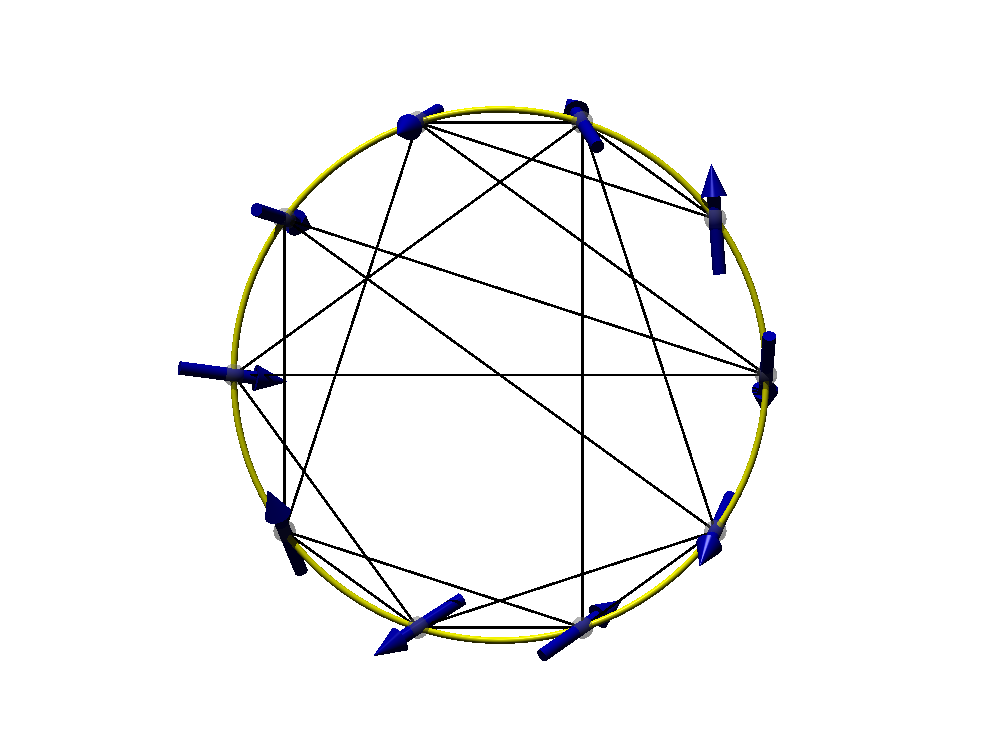}}
        \caption{(Color online). The 1d power-law spin-glass model on a ring
          geometry. The left panel shows the fully connected version where the
          magnitude of the interaction strength falls off with distance. The right
          panel shows the diluted model with the bond existence probability falling
          off with distance, whereas the bond strengths are distance independent.}
	\label{Fig:RingGeometry_full_vs_dil}
\end{figure}

In this paper we study flavors of the well-known Edwards-Anderson Hamiltonian
\begin{equation}
  \mathcal{H} = -\frac{1}{2}\sum_{i,j=1 \atop i\ne j}^L J_{ij}\,\mathbf{S}_i\cdot\mathbf{S}_j,
  \label{eq:hamiltonian}
\end{equation}
where the $\mathbf{S}_i\in \mathbb{R}^{m},i=1,\ldots,L$, are vector spins with $m$
components, normalized as $|\mathbf{S}_i| = \sqrt{m}$. The spins are organized in an
effectively one-dimensional (1d) geometry, either chosen to be a chain with periodic
boundary conditions or a ring as depicted in
Fig.~\ref{Fig:RingGeometry_full_vs_dil}. In this fully connected version the exchange
interactions are between all spin pairs, decaying as a power law with distance,
\begin{equation}
  \label{eq:distance_law}
  J_{ij} \sim \frac{\varphi_{ij}}{r_{ij}^\sigma},
\end{equation}
where $r_{ij} = |\mathbf{r}_i-\mathbf{r}_j|$ and $\varphi_{ij}$ is a standard normal
random variable.

\begin{figure}
  \includegraphics[scale=0.78, trim= 0cm 21cm 24cm 0cm]{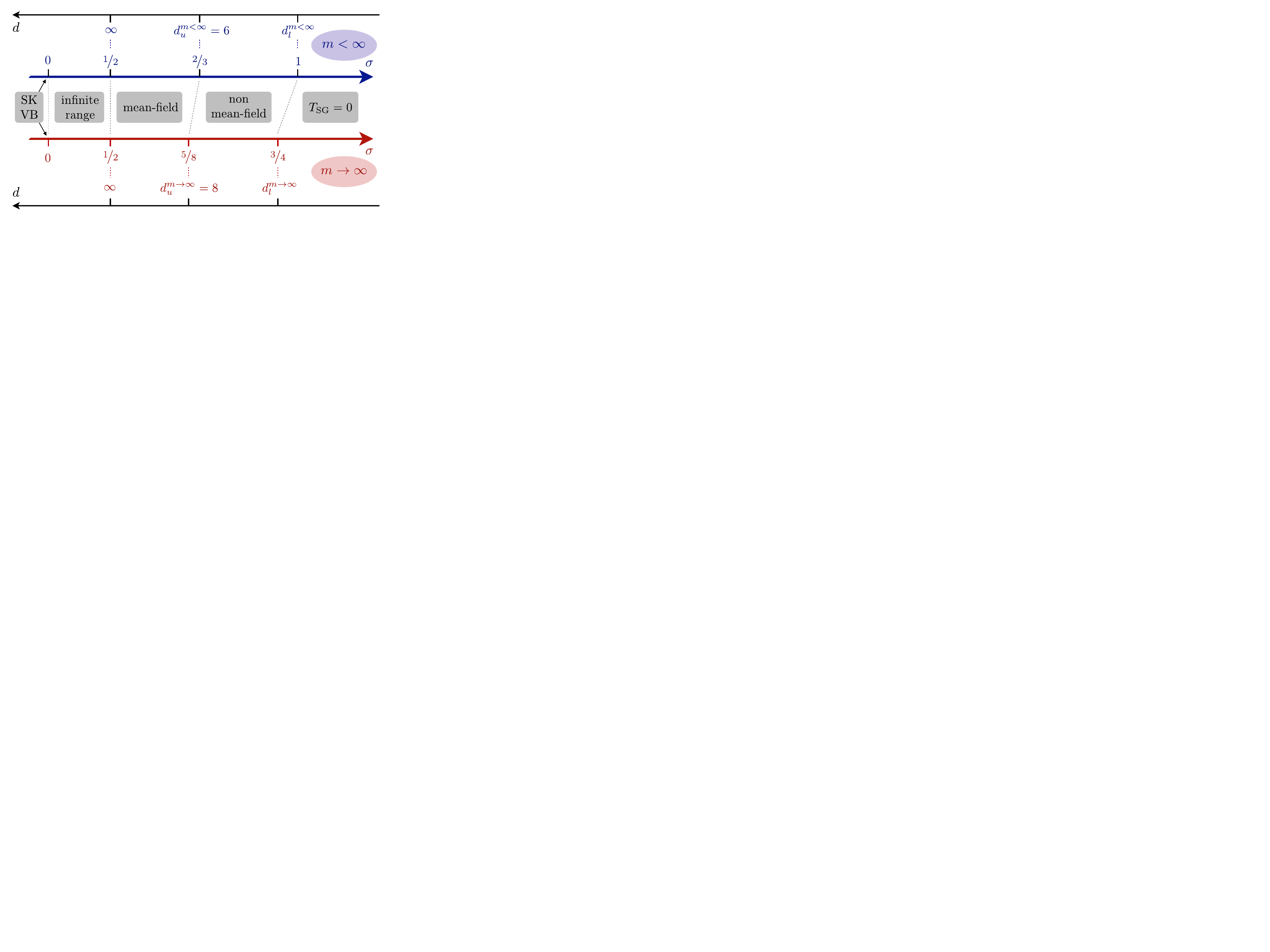}
  \caption{ (Color online). Correspondence between the 1d spin-glass model with
    power-law interactions characterized by an exponent $\sigma$ and the short-range
    model on a hypercubic lattice of dimension $d$. The upper part of the figure
    applies to finite spin dimensions $m$, whereas the lower part describes the limit
    $m=\infty$ discussed here. Increasing $\sigma$ corresponds to decreasing the
    analogous lattice dimension $d$.
  }
  \label{Fig:d_sigma_plot_1}
\end{figure}

For the case of Ising spins ($m=1$)
\cite{kotliar:83,bray:86,leuzzi:99,katzgraber:03a,katzgraber:05b,leuzzi:08,katzgraber:09}
and, more recently, Heisenberg spins ($m=3$) \cite{viet:10,sharma:11,sharma:11a},
this model has been extensively investigated. It is found that, as the range of
interactions is tuned by varying $\sigma$, the model has a behavior which mimics that
of the short-range spin glass as its dimension $d$ is tuned.  For large $\sigma$ the
spin-glass transition temperature $T_\mathrm{SG} = 0$.  This corresponds to
dimensions below the lower critical dimension $d_l$. A non-mean field regime is
adjacent at intermediate $\sigma$ continuing on to a mean-field region for small
$\sigma$, which corresponds to the dimensions above the upper critical dimension,
$d_u$. Finally the SK model \cite{sherrington:75} is obviously reached in the limit
$\sigma\rightarrow 0$. The phase diagram of the model as a function of $\sigma$ is
shown in Fig.~\ref{Fig:d_sigma_plot_1}, and will be discussed in more detail below. A
dictionary can be set up relating the behavior of the 1d power-law model at a given
$\sigma$ and a corresponding short-range model on hypercubic lattices of dimension
$d$. \cite{leuzzi:08,katzgraber:09} We shall see here that for the limit of an
infinite number $m$ of spin components, the phase diagram is modified as shown in the
lower part of Fig.~\ref{Fig:d_sigma_plot_1}.

\subsection{Choice of couplings}

We studied the model \eqref{eq:hamiltonian} in different variants and on different
geometries. 
\subsubsection*{Fully connected model}
The fully connected system implied by Eq.~\eqref{eq:hamiltonian} is
realized with interaction constants
\begin{equation}
  J_{ij} = c(\sigma,L)\frac{\varphi_{ij}}{r^{\sigma}_{ij}}, \label{Eq:Jij_1d_pow_fullConnect}
\end{equation}
where $\varphi_{ij}\in \mathcal{N}(0,1)$ are standard normal random variables. It is
the strength of interactions that falls off as $1/r^{\sigma}$ here. The mean-field
transition temperature,
\begin{equation}
  \left[ T^{\text{MF}}_{\text{SG}}(c) \right]^2 = \frac{1}{L}\sum_{i,j=1 \atop i\ne j}^L \left[J_{ij}^{2}\right]_{\text{av}}
  = \frac{c(\sigma,L)^{2}}{L}\sum_{i,j=1 \atop i\ne j}^L \frac{1}{r^{2\sigma}_{ij}}, \label{Eq:T_SG_1d_pow_fullConnect}
\end{equation}
diverges for $\sigma \le 1/2$, unless we prevent this by an appropriate $L$ dependent
choice of the normalization factor $c(\sigma,L)$, for instance by requiring that
\begin{equation}
  %\left[ T^{\text{MF}}_{\text{SG}}(c) \right]^2 \stackrel{\text{!}}=1, \label{Eq:T_SG_1d_pow_fullConnect2}
  T^{\text{MF}}_{\text{SG}}(c) \stackrel{\text{!}}=1, \label{Eq:T_SG_1d_pow_fullConnect2}
\end{equation}
which fixes $c(\sigma,L)$. While this is only strictly necessary for $\sigma\le 1/2$,
we apply the same normalization for all $\sigma$. Clearly, the limit $\sigma
\rightarrow 0$ corresponds to the SK model. In fact, it can be shown that mean-field
theory is exact (at any temperature) for all $\sigma \le 1/2$. \cite{mori:11} For
numerical simulations employing single-spin manipulations, this fully connected model
is slow as the number of bonds equals $L(L-1)/2$, so
that the cost of a lattice sweep of updates scales quadratically with the system size
$L$.

\subsubsection*{Bond-diluted model}

To improve on this costly update for the fully connected model and allow numerical
studies to get closer to the large system limit, a number of authors have considered
a diluted version of the 1d power-law spin glass \cite{leuzzi:08}. Its Hamiltonian reads
\begin{equation}
  \mathcal{H} = -\frac{1}{2\sqrt{z}}
  \sum_{i,j=1 \atop i\ne j}^L \varepsilon_{ij}J_{ij}\,\mathbf{S}_i\cdot\mathbf{S}_j,
  \label{eq:hamiltonian_diluted}
\end{equation}
where now $J_{ij}\in \mathcal{N}(0,1)$, but the probability distribution of the
dilution variables $\varepsilon_{ij}\in\{0,1\}$ falls off with the distance $r_{ij}$
as
\begin{eqnarray}
        \varepsilon_{ij} &=& 
        \begin{cases}
                1, & p<p_{ij}, \\
                0, & \text{otherwise}, 
        \end{cases} \label{Eq:epsilon_1d_pow_wo_field_dilute} \\
        p_{ij} &\sim& r_{ij}^{-2\sigma},
\end{eqnarray}
with $p\in \mathcal{U}[0,1]$ a uniform random number from the interval $[0,1]$. To
ensure that the form of $p_{ij}$ is a proper probability density function, we
normalize \cite{katzgraber:09}
\begin{equation}
  p_{ij} = 1-\exp(-A/r_{ij}^{2\sigma}), \label{Eq:pij_1d_pow_wo_field_dilute}
\end{equation}
and determine $A$ by fixing the \textit{average} coordination number
\begin{equation}
  z = \sum_{i=1}^{L-1}p_{iL}. \label{Eq:coord_1d_pow_wo_field_dilute} 
\end{equation}
Unless stated otherwise, for the data discussed here we used $z=12$, corresponding to
a hypercubic lattice at the probable lower critical dimension $d_l = 6$
\cite{beyer:11}. We apply the Newton method \cite{numrec} in $A$ in order to iterate
the probabilities $p_{iL}$ until their sum equals the desired coordination number to
a certain precision. The factor $1/\sqrt{z}$ in Eq.~\eqref{eq:hamiltonian_diluted}
ensures that $T^{\text{MF}}_{\text{SG}} = 1$, consistent with the fully connected
model. This diluted version of the model was previously studied for the Ising
\cite{katzgraber:09}, Heisenberg \cite{sharma:11,sharma:11a} and $p$-spin
\cite{larson:10} spin-glass models. The authors of Ref.~\onlinecite{leuzzi:08} and
subsequent studies claimed this model to be in the same universality class as the
fully connected system. As we will see below, however, this is not the case for
$\sigma > 1$. Also, certain properties differ for $\sigma < 1/2$. In the limit
$\sigma\rightarrow 0$, the diluted system corresponds to the Viana-Bray (VB) model
\cite{bray:85}. Numerically, the diluted system with $zL/2$ bonds reduces the sweep
time from quadratic to linear in $L$.

\subsection{Choice of geometry\label{sec:geometry}}

\begin{figure}
  \begin{center}
    \includegraphics[trim=0 15 0 0 ]{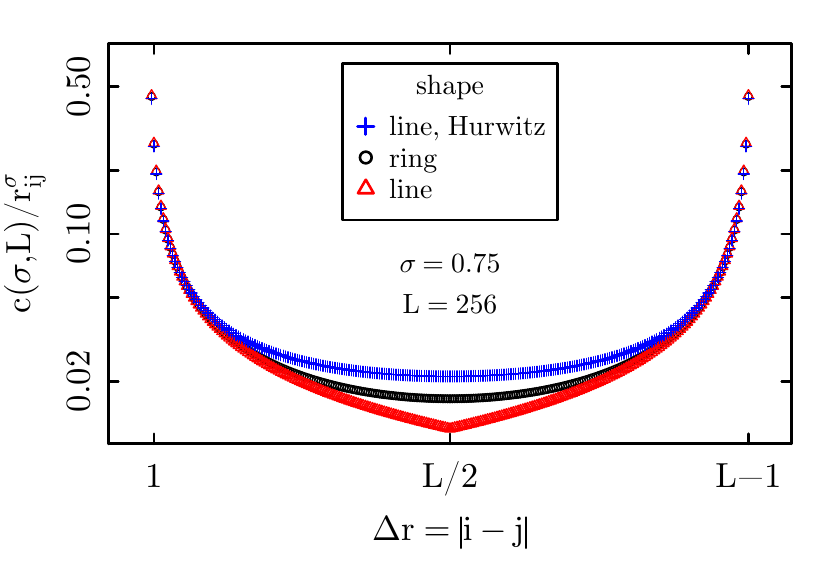}\\
  \end{center}
  \caption {(Color online). The non-random part $c(\sigma,L)/r_{ij}^{\sigma}$ of the
    interaction constants in Eq.~\eqref{Eq:Jij_1d_pow_fullConnect}, i.e.\ for the
    fully connected model, for the line and ring geometries and the resummed
    (Hurwitz) version ($\sigma = 3/4$ and $L=256$). Note the logarithmic scale of
    the ordinate.  }
  \label{Fig:geometries_Hurwitz}
\end{figure}

Two different effectively one-dimensional geometries have been previously considered
in studying power-law spin glasses: a ring of spins \cite{katzgraber:03a} as depicted
in Fig.~\ref{Fig:RingGeometry_full_vs_dil} and the possibly more natural linear chain
with periodic boundary conditions without any embedding space \cite{bhatt:86}. In the
ring model, the distances are measured according to the Euclidean metric in the
plane,
\begin{equation}
        r_{ij}^{\circ} = \frac{L}{\pi} \sin{\left(\frac{\pi |i-j|}{L}\right)},
        \label{Eq:spin_dist_1d_pow_ring}
\end{equation}
and periodic boundaries are incorporated automatically. In the chain formulation,
spins are located at integer positions $i$ on a straight line with distances
\begin{equation}
r_{ij}^{\shortmid}=\min(|i-j|,L-|i-j|), 
        \label{Eq:spin_dist_1d_pow_line}
\end{equation}
again assuming periodic boundary conditions. While one expects the specific form of
the geometry to influence the finite-size behavior, in the limit of large distances
on large chains or rings, both formulations become equivalent. Universal properties,
of course, should not depend on these details. On the other hand, one might argue
that finite-size corrections, which are notoriously important in the study of
spin-glass systems, will differ between the two formulations and might thus lead to
an effective advantage for one or the other form.

When studying long-range interactions one needs to be careful about defining a
controlled approach to the thermodynamic limit. Using the periodic boundary
conditions preferred to suppress boundary effects, each spin effectively interacts
with an infinite set of periodic images. The resulting infinite sums are usually
performed in reciprocal space (Ewald summation). For the fully connected
one-dimensional chain, i.e., Eq.~\eqref{Eq:spin_dist_1d_pow_line},
they can be performed without cut-off. Summing over images for the couplings of
Eq.~\eqref{Eq:Jij_1d_pow_fullConnect} one obtains the effective coupling
\begin{eqnarray}
        \tilde{J}_{ij}  &=& c(\sigma,L) \sqrt{ \sum_{n=-\infty}^{\infty} \frac{1}{|r_{ij}^{\shortmid}+Ln|^{2\sigma}}} \hspace{4pt} \varphi_{ij} \nonumber \\
                        &=& \frac{c(\sigma,L)}{|L|^{\sigma}} \sqrt{ \sum_{n=-\infty}^{\infty} \frac{1}{|r_{ij}^{\shortmid}/L+n|^{2\sigma}}} \hspace{4pt} \varphi_{ij} \nonumber \\
                        &=& \frac{c(\sigma,L)}{|L|^{\sigma}} \sqrt{ \zeta\left(2\sigma,\frac{r_{ij}^{\shortmid}}{L}\right) + 
                                        \zeta\left(2\sigma,1-\frac{r_{ij}^{\shortmid}}{L}\right) } \hspace{4pt} \varphi_{ij},
\end{eqnarray}
with the Hurwitz Zeta function \cite{abramowitz:book}
\begin{equation}
        \zeta\left(s,q\right) :=  \sum_{k=0}^{\infty} \frac{1}{(k+q)^{s}}.
\end{equation}
The corresponding  mean-field critical temperature is then
\begin{eqnarray}
        \left[T^{\text{MF}}_{\text{SG}}(c)\right]^2 = \frac{1}{L} \sum_{i\ne j} \left[\tilde{J}_{ij}^{2}\right]_{\text{av}} 
                        = c(\sigma,L)^{2} \frac{1}{|L|^{2\sigma+1}} \times \nonumber \\
		 \times	\sum_{i\ne j} {\left[ \zeta\left(2\sigma,\frac{r_{ij}^{\shortmid}}{L}\right) + 
                                        \zeta\left(2\sigma,1-\frac{r_{ij}^{\shortmid}}{L}\right)\right]},
 \label{Eq:T_SG_1d_pow_fullConnect_Hurwitz}
\end{eqnarray}
which, with the normalization $T^{\text{MF}}_{\text{SG}}(c) = 1$,
fixes $c(\sigma,L)$.

While for ferromagnetic systems summation over image ``charges'' is crucial, for a
spin-glass system with average magnetization $\langle m\rangle = 0$ it should not
change the asymptotic behavior \cite{alonso:10}. It might modify finite-size
corrections, however. In Fig.~\ref{Fig:geometries_Hurwitz} we compare the non-random
part of the interactions for the ring geometry as well as the bare and summed chain
interactions. The differences away from $r=0$ are small and, as we shall see below,
the alterations of the finite-size scaling (FSS) behavior are rather minor. Note that
for the special case $\sigma=1$ the constants $c(\sigma,L)$ for the ring and the
summed line geometries coincide, which is easily understood from the identity
\cite{abramowitz:book}
$$
	\zeta\left(2,r_{ij}^{\shortmid}/L\right) +
	\zeta\left(2,1-r_{ij}^{\shortmid}/L\right) = \frac{\pi^2}{\sin^{2}(\pi
	r_{ij}^{\shortmid}/L)} = \left(\frac{L}{r_{ij}^{\circ}}\right)^2.
$$
Unless stated otherwise, all of the calculations presented below have been performed
for the ring geometry.

\section{Phase diagram and critical behavior\label{sec:theory}}

To understand the expected critical behavior of the model in the $m=\infty$ limit, it
is useful to review and generalize the results for the $m<\infty$ case. The most
distinct feature of the $m=\infty$ limit on hypercubic lattices is the elevation of
the upper critical dimension (UCD) to $d_u = 8$ and the accompanying violation of
hyperscaling \cite{green:82}. 

\subsection{Mean-field critical exponents}

Recall that the mean-field exponents of (say) the Ising ferromagnet are
\begin{equation}
	\alpha = 0,\;\;\;\beta = 1/2,\;\;\;\gamma = 1,\;\;\;\nu = 1/2,\;\;\;\eta = 0.
\end{equation}
These satisfy hyperscaling,
\begin{equation}
  \label{eq:hyperscaling}
  d\nu = 2-\alpha,  
\end{equation}
(only) at the standard upper critical dimension $d_u = 4$. 

For the ($m < \infty$) spin glass, the upper critical dimension is \cite{binder:86a}
$d_u=6$ and the mean-field value of the exponents are
\begin{equation}
	\alpha = -1,\;\;\;\beta = 1,\;\;\;\gamma = 1,\;\;\;\nu = 1/2,\;\;\;\eta = 0.
\end{equation}
These again satisfy hyperscaling only at the upper critical dimension,
$d_u=6$. $\alpha$ and $\beta$ are the exponents of the SK model \cite{binder:86a}.

For the $m=\infty$ spin glass the upper critical dimension is \cite{green:82}
$d_u=8$. This model, however, violates hyperscaling also in the non-mean-field
regime, and the hyperscaling relation is replaced by a dimensionally reduced version,
\begin{equation}
	(d-2)\nu = 2-\alpha.
\end{equation}
The mean-field  exponents for the $m=\infty$ model are the same as those of
the Ising spin glass i.e.
\begin{equation}
	\alpha = -1,\;\;\;\beta = 1,\;\;\;\gamma = 1,\;\;\;\nu = 1/2,\;\;\;\eta = 0.
\end{equation}

\subsection{Finite-size scaling above the upper critical dimension}

To understand the behavior of the model in the mean-field regime, it is useful to
recall the relevant form of scaling and FSS above the UCD. Below the UCD, finite-size
corrections depend on the ratio of correlation lengths in the finite and infinite
systems, $\xi_L/\xi_\infty \sim Lt^\nu$, where $t=(T-T_c)/T_c$ is the reduced
temperature. For a singular quantity $A$, we therefore expect the FSS form
\cite{privman:privman}
\begin{equation}
	A \sim L^{\kappa/\nu} {\cal A}(L^{1/\nu}t),\;\;\;d<d_u,
\end{equation}
where $\kappa$ is the critical exponent associated to $A$. For a dimensionless quantity
such as the finite-size correlation length normalized by the system size, we expect
\begin{equation}
	\xi/L \sim {\cal X}(L^{1/\nu}t),\;\;\;d<d_u.
\end{equation}
At and above the UCD, FSS should hold with mean-field exponents with the role of the
correlation length $\xi_L \sim L$ taken on by some effective length
\cite{binder:85,jones:05} $\zeta_L \sim \ell \sim N^{1/d_u} = L^{d/d_u}$, such that
\begin{equation}
   \label{eq:fss_above_ucd1}
   A \sim N^{\kappa/d_u\nu} {\cal A}(L^{d/d_u\nu}t),\;\;d\ge d_u,
\end{equation}
where $\nu$ and $\kappa$ take on their mean-field values. Similarly, for the case of
a dimensionless quantity, we arrive at
\begin{equation}
  \xi/L^{d/d_u} \sim {\cal X}(L^{d/d_u\nu}t),\;\;d\ge d_u.
  \label{eq:fss_above_ucd2}  
\end{equation}
We can therefore extend the hyperscaling law beyond its usual range of validity $d\le
d_u$ by replacing the correlation length exponent $\nu$ with a renormalized value
\begin{equation}
  \label{eq:effective_nu}
  \nu' = \left\{
    \begin{array}{rl}
      \nu, & d<d_u,\\
      d_u\nu/d = d_u/2d, & d \ge d_u,
    \end{array}
  \right.  
\end{equation}
since then $d\nu' = 2-\alpha$ in {\em all\/} dimensions. As a consequence, at $d=d_u$
we find
\begin{equation}
  \label{eq:modified_hyperscaling}
  d\nu' = d_u/2 = 2-\alpha,  
\end{equation}
leading to $\alpha = 0$ for $d_u = 4$ and $\alpha = -1$ for $d_u = 6$.

Comparing the critical scaling of the Landau-Ginzburg-Wilson (GLW) effective
Hamiltonian for the one-dimensional long-range model and the short-range model in
general dimensions $d$, one infers that close to the UCD one has \cite{leuzzi:08}
\begin{equation}
  \label{eq:effective_dimension}
  d_\mathrm{eff} = \frac{2}{2\sigma-1}.
\end{equation}
As we will see below, $\nu = 1/(2\sigma-1) = d_\mathrm{eff}/2$ in the mean-field
region, such that $d/d_u = d_\mathrm{eff}/d_u = \nu/3$ and $d/d_u\nu = 1/3$ and it
follows from Eqs.~\eqref{eq:fss_above_ucd1} and \eqref{eq:fss_above_ucd2} that
\begin{subequations}
\begin{eqnarray}
   \chi_\mathrm{SG} & \sim &  N^{1/3}{\cal C}(tN^{1/3}),  \\
   q_\mathrm{EA} & \sim & N^{-1/3} {\cal Q} (tN^{1/3}),  \\
   \frac{\xi}{L^{\nu/3}} & \sim & {\cal X}(tN^{1/3}). 
\end{eqnarray}
\end{subequations}

\subsection{1d long-range spin glass\label{sec:ising_theoretical}}

The long-range Ising spin glass was discussed analytically in
Refs.~\onlinecite{kotliar:83,bray:86,fisher:88,moore:10}. In
Ref.~\onlinecite{vanenter:85} it was proven rigorously that there is no phase
transition for $\sigma > 1$. Studying the effective GLW Hamiltonian in replica space
\cite{kotliar:83}, it was inferred that there is a finite-temperature phase
transition for $1/2\le \sigma \le 1$, which is of mean-field type for $1/2 \le \sigma
\le 2/3$ and of non-mean-field type for $2/3<\sigma \le 1$. Therefore, the lower
critical $\sigma_l = 2/3$ corresponds to the upper critical $d_u = 6$ for systems on
hypercubic lattices and similarly for the upper critical $ \sigma_u = 1$ and the
lower critical $d_l$.  It is useful to set up a dictionary of correspondences between
the 1d long-range model and the short-range models on hypercubic lattices, cf.\
Fig.~\ref{Fig:d_sigma_plot_1}.  Arguments were given in Larson {\em et al.\/}
\cite{larson:10} that the effective dimensionality for $2/3\le \sigma \le 1$ was
approximately given by \cite{katzgraber:09}
\begin{equation}
  \label{eq:effective_dimension2}
  d_\mathrm{eff} = \frac{2-\eta(d_\mathrm{eff})}{2\sigma-1},  
\end{equation}
where $\eta(d_\mathrm{eff})$ is the exponent of the corresponding short-range
model. The upper critical $\sigma_u = 1$ can be inferred from the result
\cite{bray:86}
\begin{equation}
  \label{eq:theta_LR_ising}
  \theta_\mathrm{LR} = 1-\sigma
\end{equation}
for the long-range defect-energy exponent (see Sec.\ \ref{Sec:Defectenergies}). Since
$\theta_\mathrm{SR} = -1$ for the short-range Ising spin glass in 1d, a
finite-temperature transition ceases to exist at $\theta_\mathrm{LR} = 0$, that is,
at $\sigma_u = 1$.

In order to determine the correlation length $\xi$ for long-range models with
power-law interactions, one uses the fact that the propagator is modified from the
well-known Ornstein-Zernicke form to (in reciprocal space) \cite{stell:72,suzuki:73}
\begin{equation}
G(k) \sim \frac{1}{m^2+k^{2\sigma-1}}. \label{Eq:Ornstein-Zernicke_propagator}
\end{equation}
For the spin-glass,  the analogous form is for  the spin-glass correlator
$\chi_\mathrm{SG}(k)$. Consequently, the second-moment definition of the correlation
length is modified to
\begin{equation}
\xi_\mathrm{SG} =
\frac{1}{2\sin(k_\mathrm{min}/2)}\left[\frac{\chi_\mathrm{SG}(0)}{\chi_\mathrm{SG}({\mathbf
      k}_\mathrm{min})}-1\right]^{1/(2\sigma-1)} ,\label{Eq:xi_normal}
\end{equation}
with ${\mathbf k}_\mathrm{min}=(2\pi/L)\, \vec{e}_1\in\mathbb{R}^d$, where
$\vec{e}_1$ is a lattice basis vector.
%with ${\mathbf k}_\mathrm{min}=(2\pi/L,0,0)$.

At criticality (where $m=1/\xi = 0$), it is found that the Gaussian propagator $\sim
1/k^{2 \sigma-1}$ does not receive any corrections away from mean field
\cite{fisher:72} and hence
\begin{equation}
2-\eta = 2\sigma - 1,\;\;\;\sigma\le 1.
\end{equation}
The upper critical value $\sigma_u=1$ also follows directly from observing that at
the lower critical dimension (LCD), we expect the critical correlation function decay
$G(r) \sim 1/r^{d-2+\eta}$ to be constant, i.e., $d_l-2+\eta=0$. Since we have $d=1$
and $2-\eta=2 \sigma-1$, it follows that $\sigma_u=1$.

In the mean-field regime of the Ising model, Kotliar, Anderson and Stein (KAS) find
\cite{kotliar:83}
\begin{equation}
  \nu = 1/(2\sigma-1),\;\;\; 1/2\le \sigma \le 2/3.  
  \label{eq:mf_nu_ising}
\end{equation}
This implies that
\begin{equation}
\gamma = (2-\eta)\nu = 1,\;\;\;1/2\le \sigma \le 2/3.
\end{equation}
Using modified hyperscaling, Eq.~\eqref{eq:modified_hyperscaling}, for $\sigma<2/3$,
we expect
\begin{equation}
d\nu' = 1\cdot\frac{1}{2\sigma-1}\cdot\frac{6}{2/(2\sigma-1)} = 3 \stackrel{!}{=}
2-\alpha,
\end{equation}
i.e., $\alpha = -1$ and $\nu' = 3$ for $1/2\le \sigma \le 2/3$ and, consequently,
\begin{equation}
\beta = \frac{(2-\alpha)-\gamma}{2} = 1,\;\;\;1/2\le \sigma \le 2/3.
\end{equation}
Note that with
\begin{equation}
2-\eta = 2\sigma-1 = (2-\eta_\mathrm{MF})/d_\mathrm{eff} \label{Eq:nu_LR_version}
\end{equation}
and
\begin{equation}
\nu = \frac{1}{2\sigma-1} = \nu_\mathrm{MF} d_\mathrm{eff},
\end{equation}
where $d_\mathrm{eff} = 2/(2\sigma-1)$, all exponents take their expected mean-field
values.

%\begin{figure}
 % \begin{center}
  %  \includegraphics[trim=0 15 0 0 ]{pics/1dpl_O1_oonu_p1_embd.pdf}\\
  %\end{center}
  %\caption
 %{ (Color online). Previous estimates of $1/\nu$ for the 1d power-law Ising spin
   % glass in comparison with various theoretical predictions discussed in
  %  Sec.~\ref{sec:ising_theoretical}, in particular the exact mean-field form
    %\eqref{eq:mf_nu_ising} and the estimates \eqref{eq:kas_wrong1},
    %\eqref{eq:kas_wrong2} and \eqref{eq:moore_nu_ising} in the non-mean-field regime.
    %The numerical results are taken from Bhatt et al.\ \cite{bhatt:86}, Leuzzi A
   % \cite{leuzzi:99}, Katzgraber et al.\ \cite{katzgraber:03a}, Leuzzi et al.\ B
    %\cite{leuzzi:08} and Sharma et al. \cite{sharma:11}. Recall that for $m<\infty$
  %  the mean-field regime is $1/2<\sigma<2/3$, see Fig.~\ref{Fig:d_sigma_plot_1}.
   % \label{Fig:m_is_1_published_results}
  %} 
%\end{figure}

In the non-mean-field regime, $2/3<\sigma <1$, KAS showed that an expansion around
the lower critical value $\sigma_l =2/3$ in the variable $\epsilon = \sigma-2/3$ was
possible and yielded to first order in $\epsilon$
\begin{equation}
  \label{eq:kas_wrong1}
  \frac{1}{\nu} = \frac{1}{3}-4\epsilon.
\end{equation}
Expansions around the upper critical $\sigma_u = 1$ have also been proposed
\cite{kotliar:83,moore:10}.

%\begin{equation}
 % \label{eq:kas_wrong2}
%  \frac{1}{\nu} = 1.1\sqrt{2\epsilon} = 1.1\sqrt{2(1-\sigma)}.
%\end{equation}
%Comparing these to the previously found numerical results, both estimates appear to
%be rather poorly converged, cf.\ the data collected in
%Fig.~\ref{Fig:m_is_1_published_results} (see also the discussion in
%Ref.~\cite{sharma:11}). A suggestion by Moore in Ref.~\cite{moore:10},
%\begin{equation}
  %\label{eq:moore_nu_ising}
 % \nu = 1/(1-\sigma),\;\;\;2/3\le\sigma\le 1
%\end{equation}
%seems to work significantly better, cf.\ the data collected in
%Fig.~\ref{Fig:m_is_1_published_results}. This would imply
%$$
%\gamma = (2-\eta)\nu = \frac{2\sigma-1}{1-\sigma},\;\;\;2/3\le\sigma\le 1. 
%$$
%Since hyperscaling should work in this regime, we also expect (assuming $1/\nu =
%1-\sigma$)
%$$
%2-\alpha = d\nu = \nu = 1/(1-\sigma),\;\;\;2/3\le\sigma\le 1,
%$$
%and
%$$
%\beta = \frac{(2-\alpha)-\gamma}{2} = 1,\;\;\;2/3\le\sigma\le 1.
%$$

\subsection{1d long-range $m=\infty$ spin glass \label{Sec:1dpl_mInf_SG}}

For the $m=\infty$ model we know \cite{green:82} that the UCD is elevated from the
usual (spin-glass) $d_u = 6$ to $d_u = 8$ and that, additionally, there is a failure
of hyperscaling, even below the UCD. For hypercubic lattices at the UCD, FSS should
work in $N$ (see above), e.g.,
\begin{equation}
\chi_\mathrm{SG} \sim L^{\gamma/\nu} = L^{2-\eta} = N^{(2-\eta)/d} =
N^{1/4},\;\;\;d\ge 8.
\end{equation}
For the 1d long-range model, we expect the long-range form of the exponent of the
correlation function, Eq.~\eqref{Eq:nu_LR_version},
%$$
%2-\eta = (2-\eta_\mathrm{MF})/d_\mathrm{eff} = 2\sigma-1
%$$
to carry over to the $m=\infty$ model. At the lower critical $\sigma_l$,
where mean-field behavior first becomes modified, we should find
\begin{equation}
(2-\eta)/d = 2-\eta = 2\sigma_l - 1 \stackrel{!}{=} 1/4
\end{equation}
or $\sigma_l = 5/8$. Therefore, the mean-field regime is here defined as
\begin{equation}
1/2\le \sigma \le 5/8.
\end{equation}
Of course, this range can be also obtained directly via the calculational methods in
Green {\em et al.\/} \cite{green:82}.  The effective correlation length exponent thus
becomes $\nu' = 4$ and, due to dimensional reduction, we expect a modified
hyperscaling relation to hold,
\begin{equation}
(d-\Theta)\nu' = 2-\alpha,
\end{equation}
with some violation-of-hyperscaling exponent $\Theta$ for the long-range case. Since
we should have $\alpha = -1$, we infer $\Theta = 1/4$ for $1/2\le\sigma\le 5/8$. The
``bare'' correlation length exponent should  be unaltered,
\begin{equation}
\nu = \frac{1}{2\sigma-1} = \nu_\mathrm{MF} d_\mathrm{eff},
\end{equation}
so that then $\gamma = 1$ and $\beta = 1$ as expected from mean-field theory.

The FSS forms of the critical quantities become modified by $d_u = 8$ according to
the discussion outlined above to read
\begin{subequations}
\begin{eqnarray}
  \chi_\mathrm{SG} & \sim &  N^{1/4}{\cal C}(tN^{1/4}), \label{Eq:chiSG_mInf_infRange_MF} \\
  q_\mathrm{EA} & \sim & N^{-1/4} {\cal Q} (tN^{1/4}),  \\
  \frac{\xi}{L^{\nu/4}} & \sim & {\cal X}(tN^{1/4}).
\end{eqnarray}
\end{subequations}

A consistent definition of the violation-of-hyperscaling exponent $\Theta$  is
given by
\begin{equation}
\Theta = \left\{
  \begin{array}{rl}
    2\sigma-1 = 2/d_\mathrm{eff}, & 5/8 \le \sigma, \\
    1/4,                          & 1/2\le \sigma < 5/8.
  \end{array}
  \right.
\end{equation}
When $\sigma >5/8$, this follows from the form of the propagators at $T_c$, which go
as $~1/k^{2 \sigma-1}$, and the results in Ref.~\onlinecite{green:82}.  $\Theta$ is
used in scaling relations which involve the dimensionality $d$ when one replaces $d$
by $d-\Theta$.  Thus the scaling relation $\beta/\nu =(d-2+\eta)/2$, with the
replacements $d \rightarrow d-\Theta$, $2-\eta=2 \sigma-1$, and $d =1 $ becomes
\begin{equation}
\beta/\nu= (3-4\sigma)/2. 
\label{Eq:beta_over_nu_prediction_nonMF}
\end{equation}
This is consistent with our numerical results shown in
Fig.~\ref{Fig:results_from_qEA_FSS_1}.  It is possible to determine exactly the value
of the upper critical $\sigma_u$ from generalizing the argument that at the lower
critical dimension $d-2+\eta = 0$. Replacing once again $d \rightarrow d-\Theta$ and
$2-\eta =2\sigma-1$ , one has at $\sigma = \sigma_u$,
$1-(2\sigma_u-1)-(2\sigma_u-1)=3-4\sigma_u=0$, so $\sigma_u =3/4$.

As will be discussed below, the defect-energy calculations for $m=\infty$ presented
here, cf.\ Fig.~\ref{Fig:defE_theta_vs_sigma}, can be summarized as
$\theta(\sigma=3/4)=0$, i.e., $\sigma_u = 3/4$, and $\theta(\sigma=1/2)=1/4$, which
lead us to conjecture that
\begin{equation}
  \label{eq:minfty_theta_conjecture}
  \theta_{\mathrm{LR}} = \frac{3}{4}-\sigma.  
\end{equation}
In the following, we refer to $\theta_{\mathrm{LR}}$ simply as $\theta$. The form
\eqref{eq:minfty_theta_conjecture} works over a rather wide range of $\sigma$, even
for $\sigma > 3/4$ in the fully connected model. When $\sigma >3/4$, there is no
finite temperature transition. We have been unable to give a formal derivation of
this result, but suspect that this might be possible by generalizing the formalism of
Aspelmeier {\em et al.\/} \cite{aspelmeier:03} to spatially varying solutions. One
can however understand Eq.~\eqref{eq:minfty_theta_conjecture} from the following
considerations. From the scaling arguments in Ref.~\onlinecite{bray:86} for
long-range Ising spin glasses $ 2\theta=2d -2\sigma$. This would also follow from the
formalism of Ref.~\onlinecite{aspelmeier:03} which would result in an expression for
the variance of the defect energy, (which scales as $L^{ 2\theta}$) proportional to a
double sum over $i$ and $j$ of $[J_{ij}^2]_{\mathrm{av}}$ if a spatially non-uniform
solution for the defect energies is studied. We have to consider how the failure of
hyperscaling for the large $m$ limit might affect this relation.

$\theta$ is not a critical point exponent, but an exponent associated with the fixed
point at zero temperature. For it, we suspect that the mean-field form of
$\Theta=1/4$ is relevant for both $\sigma$ greater than and less than $5/8$, since
zero-temperature exponents like $\theta$ can usually be obtained by a simple
minimization of the defect energy, just as one determines mean-field behavior by
minimizing the total energy of the system. Thus allowing for the failure of
hyperscaling, the equation $2 \theta=2d-2\sigma$ becomes $2 \theta
=2(d-\Theta)-2\sigma$. With $\Theta =1/4$, Eq.~\eqref{eq:minfty_theta_conjecture} for
$\theta$ is obtained on setting $d=1$. We would expect this argument to still be
valid in the fully connected model even for $\sigma >3/4$ when there is no finite
temperature transition.

McMillan \cite{mcmillan:84a,fisher:88} has argued that the relevant renormalization
group equation for the flow of the temperature $T$ near the lower critical dimension
is
\begin{equation}
\frac{\d T}{\d \ln L} = -\theta T + cT^3+\ldots.
\end{equation}
For $\theta$ small and positive, (i.e., for $\sigma$ below, but close to $3/4$), one
finds a fixed point at $T_\mathrm{SG} \propto \sqrt{\theta} \propto
\sqrt{3-4\sigma}$. If we choose the proportionality constant so that
$T_\mathrm{SG}(\sigma \le 1/2) = 1$, then
\begin{equation}
T_\mathrm{SG} = \sqrt{3-4\sigma}, \label{Eq:Tc_prediction_MF_nonMF}
\end{equation}
which  fits the critical temperature quite well in the whole regime $1/2\le \sigma \le
3/4$ (cf.\ Fig.~\ref{Fig:results_from_qEA_FSS_1}).  The eigenvalue at this fixed point is
\begin{equation}
   \nu=\frac{1}{2\theta} = \frac{2}{3-4\sigma}.  
    \label{eq:nu_minfty_nonmf}
\end{equation}
This appears to be consistent with the data shown in
Fig.~\ref{Fig:xi_data_collapse_results_1} for the regime $5/8\le \sigma\le 3/4$, but
it cannot be regarded as anything but an interpolation formula, exact only at the end
points $\sigma =5/8$ and $\sigma =3/4$. For $\sigma > 3/4$ $T_\mathrm{SG} = 0$. Then
one expects \cite{bray:87a} $\nu = -1/\theta$.

From the scaling relation $\gamma=\nu(2-\eta)$ with $\nu$ given by
Eq.~\eqref{eq:nu_minfty_nonmf} one has the approximate result that
\begin{equation}
\gamma = \frac{4\sigma-2}{3-4\sigma},
\end{equation}
 in the regime $5/8\le \sigma\le 3/4$ . 
 $\gamma = 1$ is thus expected at  $\sigma = 5/8$ and $\gamma\rightarrow\infty$ as
$\sigma\rightarrow3/4$ from below.

By combining Eq.~\eqref{Eq:beta_over_nu_prediction_nonMF} with
Eq.~\eqref{eq:nu_minfty_nonmf} one finds $\beta = 1$ throughout the interval $5/8\le
\sigma \le 3/4$. Thus the expectation is that $\beta$ remains close to its mean-field
value even in the non-mean-field region.

\section{Zero-temperature calculations\label{sec:GS}}

We start our numerical investigations of the $m=\infty$ spin glass by studying its
ground-state properties as a function of $\sigma$. Due to the possibility of studying
large system sizes, we first concentrated on the diluted model of
Eq.~\eqref{eq:hamiltonian_diluted}. For most calculations, the ring arrangement was
used. Compared to the ground-state problem for generic spin glasses which is found to
be {\em NP} hard \cite{barahona:82}, ground states for the $m=\infty$ limit are much
easier to determine. Starting out with the Ising model with $m=1$, with increasing
spin dimension the energy landscape simplifies gradually until, for
$m\rightarrow\infty$, all metastability has vanished and the ground state becomes
unique. This fact was already exploited for hypercubic systems in
Ref.~\onlinecite{beyer:11}, where the lower critical dimension was determined with
the defect-energy approach and for different boundary conditions.

The possibility to realize the limit $m\rightarrow\infty$ in numerical calculations
rests on the fact that for a finite system of $N$ spins, the ground state occupies a
finite dimensional sub-manifold in spin space \cite{hastings:00,chandra:08}, the
dimension of which is limited by the rigorous upper bound
\begin{equation}
  m_{\operatorname{max}}(N)=\left\lfloor\left(\sqrt{8N+1}-1\right)/2\right\rfloor 
  \sim N^{\mu},\; \mu=1/2,
  \label{Eq:defE_mu}
\end{equation}
where $\lfloor x\rfloor$ stands for the largest integer smaller than or equal to
$x$. Hence, for each system size a {\em finite\/} number $m^{\ast}(N)\le
m_{\operatorname{max}}(N)$ of spin components is sufficient to describe the
$m=\infty$ model. For commonly used spin-glass models, the scaling is in fact weaker
than $m^\ast(N)\sim N^{1/2}$. For the SK model realized, e.g., in the limit
$\sigma\rightarrow 0$ of our fully connected 1d spin glass, one finds
\cite{hastings:00,aspelmeier:04a} $\mu = 2/5$. As the degree of connectivity is
lowered, $\mu$ is reduced. We determined the required number of spin components
$m^{\ast}(N)$ for each single realization of the bonds $J_{ij}$ and computed the
disorder average $m_{0}=[m^{\ast}(N)]_{\mathrm{av}}$. The values of $m^{\ast}(N)$ are
found to vary only slightly between disorder realizations, such that using
$m_\mathrm{act} \approx 1.1 [m^{\ast}(N)]_{\mathrm{av}}$ was sufficient to ensure
that $m=\infty$ ground states are found for all realizations. The procedure of
determining the number of necessary spin dimensions $m^{\ast}$ will be described at
the end of Sec.~\ref{Sec:GS_props}.

Due to the lack of metastability for $m\rightarrow\infty$, it is quite
straightforward to determine true ground states numerically. Here, we employ a local
spin-quench procedure, for which the spins are iteratively aligned with their
respective local molecular fields $\mathbf{H}_i$, so that the new value of the
spin $\mathbf{S}_i$ is given by
\begin{equation}
  \mathbf{S}_{i}^{\prime} \parallel \mathbf{H}_i = \sum_{j\in\mathcal{N}(i)}
  J_{ij}\mathbf{S}_j,
  \label{Eq:the_local_field}
\end{equation}
where the sum runs over the set $\mathcal{N}(i)$ of connected neighbors of the spin at
site $i$.  It is easily seen that alignment of each spin with its molecular field is
a necessary condition for the system to be in its ground state. For the present case
of a system without metastable states \cite{hastings:00}, it is also
sufficient. These updates are interspersed with sweeps of over-relaxation moves to
speed up convergence, which have also been found to improve the decorrelation of
systems with finite spin dimension $m$ in Monte Carlo simulations
\cite{campos:06}. These moves, again being local, preserve the energy of the whole
spin configuration since the updated spin is merely rotated around its local field
and therefore moves at constant energy. The simplest way of implementing such a
procedure, in particular for the case of arbitrary spin dimensions $m$, is to reflect
the spin along $\mathbf{H}_i$, such that
\begin{equation}
    \mathbf{S}_{i}^{\prime} = -\mathbf{S}_i+2\frac{\mathbf{S}_i\cdot\mathbf{H}_i}{|\mathbf{H}_i|^2}\mathbf{H}_i.
\end{equation}
This maximal movement can also be argued to lead to a maximal decorrelation effect
within the constant-energy manifold of single-spin movements. The whole procedure of
spin-quench and over-relaxation moves can be implemented very efficiently, since only
a few elementary operations are required for each step, and no random numbers are
involved.

\subsection{Ground-state properties \label{Sec:GS_props}}

\begin{figure}
  \begin{center}
    \includegraphics[scale=1, trim=0 4 0 8]{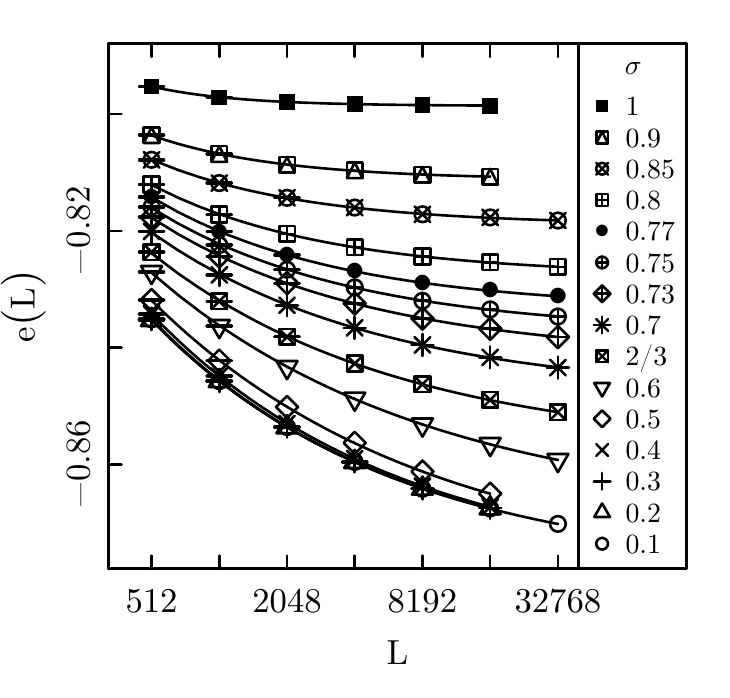}\\
    \includegraphics[scale=1, trim=0 4 0 8]{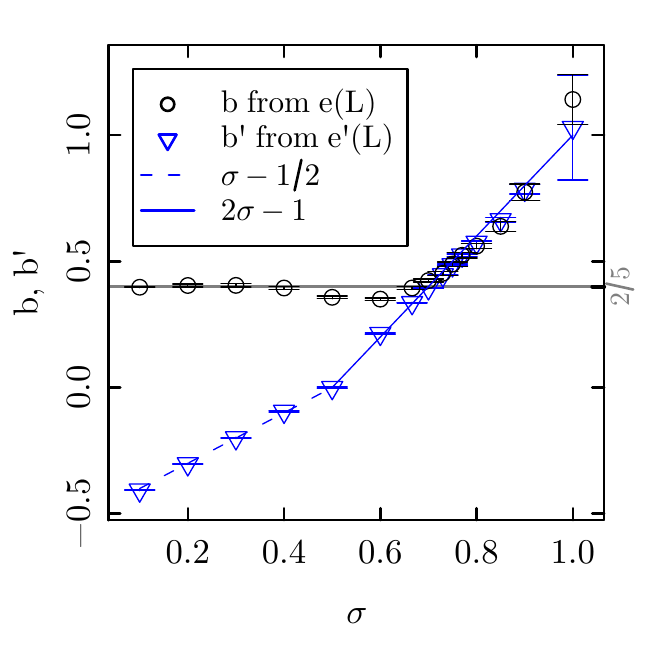}\\
    \includegraphics[scale=1, trim=0 10 0 8]{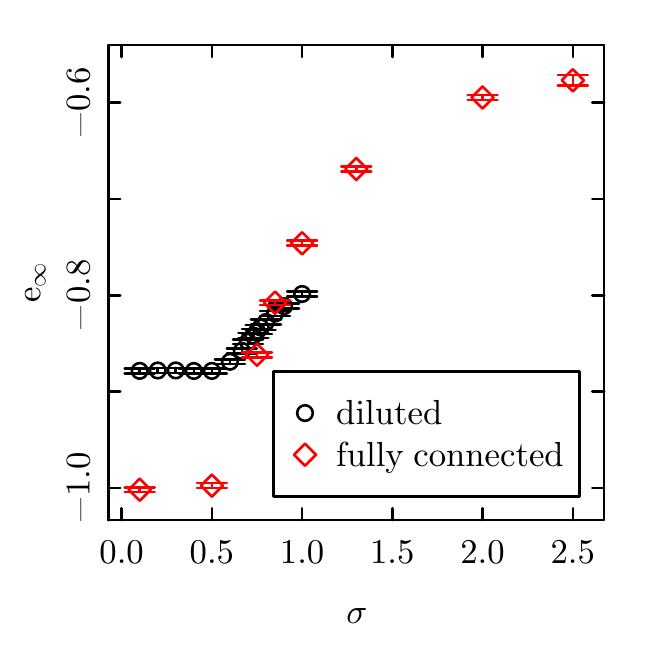}\\
  \end{center}
  \caption
  { (Color online). Average ground-state energies of the 1d power-law model in the
    $m\rightarrow\infty$ limit. Top panel: ground-state energies $e$ as a function of
    system size $L$ and interaction range $\sigma$. The lines show fits of the form
    \eqref{Eq:defE_Egs_a} to the data. The corresponding correction exponents $b$ are
    shown in the middle panel. The blue lines indicate the expectations for the
    unrenormalized energies $e'(L)$ which are $b'=\sigma-1/2$ for $\sigma < 1/2$ and
    $b'=2\sigma-1$ for $\sigma \ge 1/2$. The bottom panel shows the resulting
    asymptotic ground-state energies $e_{\infty}$, for the diluted and the fully
    connected model, respectively.
    \label{Fig:defE_Egs}
  }
\end{figure}

Finite-size corrections to the ground-state energy of spin glasses have been
extensively discussed recently for the case of the short-range Edwards-Anderson
system \cite{bouchaud:03,hartmann:04a,boettcher:11}, spin glasses on the Bethe
lattice and random graphs \cite{boettcher:10}, and the SK model
\cite{aspelmeier:08}. The dominant contribution for short-range systems is due to the
presence of domain-wall defects, leading to corrections proportional to
\cite{hartmann:04a} $L^{d-\theta}$. For the system studied here, however, these
effects, although presumably present, are masked by corrections stemming from the
power-law nature of the interactions. As indicated in
Eq.~\eqref{Eq:T_SG_1d_pow_fullConnect}, the relevant energy scale for the case of
unrenormalized coupling strengths, i.~e., $c(\sigma,L) = 1$, is set by the integral
over the couplings,
$$
\sum_{i\ne j} \left[J_{ij}^{2}\right]_{\text{av}}
= \sum_{i=1}^{L-1}\frac{1}{r^{2\sigma}_{ij}} \sim \mathrm{const.} + cL^{2\sigma-1},\;\;\;L\gg 1.
$$
Hence, the ground-state energy per spin and spin-component, i.e., $e(L)=E/Lm$,
is expected to scale as
\begin{equation}
  \label{eq:gs_scaling1}
  e'(L) = -\sqrt{{e'_\infty}^2+c'L^{1-2\sigma}} + \cdots,
\end{equation}
where further finite-size corrections stemming from the presence of domain-wall
excitations etc.\ have been neglected. We will see below that these are sub-leading
and cannot be resolved by the numerics. The primed quantities in
Eq.~\eqref{eq:gs_scaling1} are meant to indicate the unrenormalized case with
$c(\sigma,L) = 1$. For large systems, we therefore expect different limiting
behaviors depending on whether $\sigma \gtrless 1/2$, viz.
\begin{equation}
  \label{eq:gs_scaling2}
  e'(L) \sim \left\{
    \begin{array}{rl}
      \displaystyle
      e_\infty'\left(1+\frac{c'}{2 {e'_\infty}^2}L^{1-2\sigma}\right), & \sigma >
      1/2,\;\;L\gg 1,\\
      \sqrt{c'} L^{1/2-\sigma}, & \sigma < 1/2,\;\;L\gg 1,
    \end{array}
  \right.
\end{equation}
with logarithmic scaling right at $\sigma = 1/2$. 
If we choose to make the energy
scale convergent for $\sigma\le 1/2$ by setting
\begin{equation}
c(\sigma,L)^2 =  \sum_{i=1}^{L-1}\frac{1}{r^{2\sigma}_{ij}},
\end{equation}
we instead consider $e(L) = e'(L)\cdot c(\sigma,L)$ with limiting behavior
\begin{equation}
  \label{eq:gs_scaling3}
  e(L) \sim \left\{
    \begin{array}{rl}
      \displaystyle
      e_\infty+cL^{1-2\sigma}, & \sigma > 1/2,\;\;L\gg 1,\\
      \mathrm{const}, & \sigma < 1/2,\;\;L\gg 1.
    \end{array}
  \right.
\end{equation}
For the case of the diluted model, similar considerations lead to the same results,
where now $A$ of Eq.~\eqref{Eq:pij_1d_pow_wo_field_dilute} takes on the role of
$c(\sigma,L)^2$. In Fig.~\ref{Fig:defE_Egs} we present the results of the scaling
of ground-state energies. The correction exponents result from fits of the general
form
\begin{equation}
  \begin{split}
    e(L) &= e_{\infty}+cL^{-b},\\
    e'(L) &= e'_{\infty}+c'L^{-b'},
  \end{split}
    \label{Eq:defE_Egs_a}
\end{equation}
to the data. The number of disorder realizations used for the ground-state
calculations are summarized in Table \ref{Tab:defE_results}. As is seen from the
middle panel of Fig.~\ref{Fig:defE_Egs}, the predictions $b'=2\sigma-1$ for $\sigma >
1/2$ and $b'=\sigma-1/2$ for $\sigma < 1/2$ for the $e'(L)$ and $b=2\sigma-1$ for
$\sigma > 1/2$ in the renormalized case are borne out well in the data. For $\sigma <
1/2$, where we predict $b=0$, sub-leading corrections become visible. The resulting
correction exponent $b=2/5$ is consistent with the expectations for the SK model,
cf.\ Ref.~\onlinecite{braun:06} and the discussion in Sec.~\ref{sec:EV_distribution}
below. For the renormalized energies $e(L)$, we see a dip of the correction exponent
for $0.5 \le \sigma \lesssim 0.6$, which is possibly due to additional finite-size
effects resulting from the crossover between the forms for $\sigma \gtrless 1/2$. As
shown in the bottom panel of Fig.~\ref{Fig:defE_Egs}, the asymptotic ground-state
energy $e_\infty$ smoothly increases for interaction ranges $\sigma>1/2$. For
$\sigma\le 1/2$ it is independent of $\sigma$ and takes the value $-1$ in the fully
connected version of the model \cite{bray:81}. The independence of this non-universal
quantity on $\sigma$ in this regime is a clear sign of the exactness of mean-field
theory for $\sigma \le 1/2$ as proposed in Ref.~\onlinecite{mori:11}. For models of
lower connectivity, however, this energy is increased. Calculations on a Bethe
lattice \cite{braun:06} are consistent with our results, for instance
$e_\infty(\sigma=0.1)=-0.8784(1)$, if the average coordination number is taken into
account.

\begin{figure}
        \includegraphics[trim=0 2 0 2]{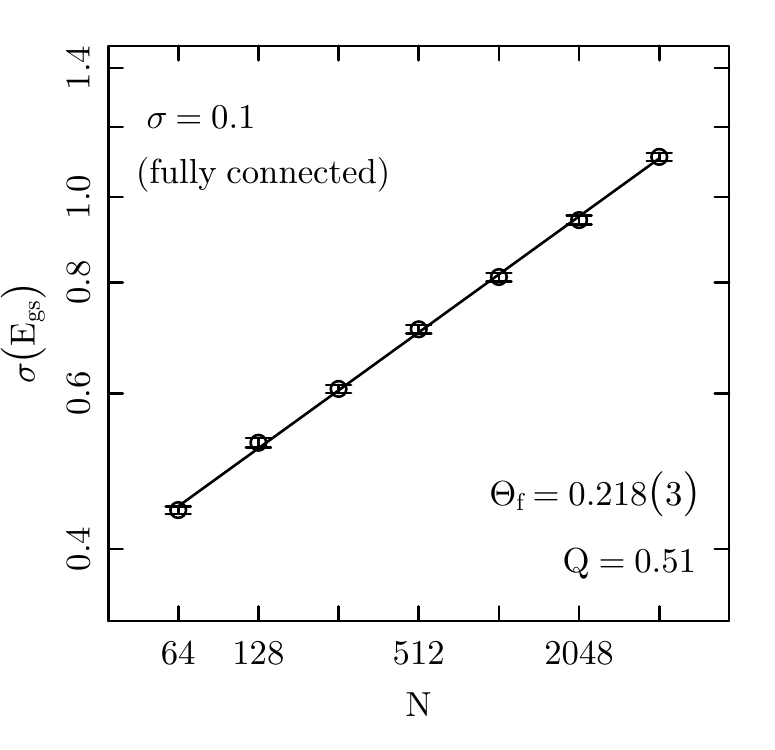}
        \caption{The scaling behavior of the sample-to-sample
			fluctuations characterized by the exponent $\Theta_f$ for
                        $\sigma = 0.1$. This showcase example
			for the fully connected model indicates the correctness of the
			prediction $\Theta_f=1/5$ in the SK limit ($\sigma\rightarrow 0$).
                       		}
	\label{Fig:sample2sample_exponent_showcase1_fully}
\end{figure}

\begin{figure}
	\includegraphics[trim=0 2 0 2]{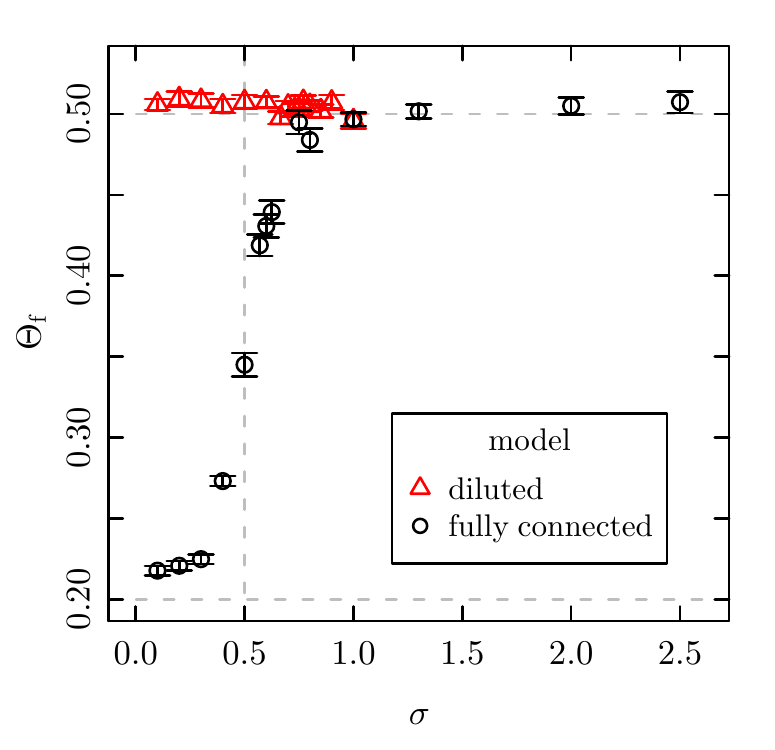}
  	\caption{ (Color online). Sample-to-sample fluctuations differ for the
		diluted and the fully connected model. Whereas the characteristic
		exponent is fixed to $1/2$ for the diluted model, it has a jump
		from $1/5$ to $1/2$ when changing from $\sigma<1/2$ to $\sigma>1/2$
                for the fully connected model.
              }

  \label{Fig:sample2sample_compare_dil_ful}
\end{figure}

At $T=0$ the free energy $F$ reduces to the internal energy. We can use this fact to
consider its sample-to-sample fluctuations,
\begin{equation}
  \sigma_N \sim N^{\Theta_f}
\end{equation}
in a FSS analysis.  The scaling of this quantity has been the subject of a number of
recent analytical and numerical studies see, e.g.,
Refs.~\onlinecite{bouchaud:03,andreanov:04,boettcher:05a,parisi:09,parisi:10,aspelmeier:10}. For
the $(m=1)$ Ising SK model, there has been some debate as to whether \cite{kondor:83}
$\Theta_f = 1/6$ or \cite{aspelmeier:03} $\Theta_f = 1/4$, but now there is growing
consensus that \cite{aspelmeier:08} $\Theta_f = 1/6$.  For the $m=\infty$ model
discussed here, the situation is less well studied.  For the replica symmetric
spherical SK model it was shown \cite{andreanov:04} that $\Theta_f = 1/3$.  As the
$m=\infty$ limit is also replica symmetric \cite{dealmeida:78}, one might suspect
this value to carry over to the present case. In Ref.~\onlinecite{aspelmeier:10},
however, it was argued on the basis of connections of the problem to bond chaos that,
instead, $\Theta_f = 1/5$.  For the long-range model studied here, we therefore
expect $\Theta_f = 1/5$ in the infinite-range regime $\sigma \le 1/2$ and a trivial
$\Theta_f = 1/2$ for $\sigma > 1/2$.  Using our data for the ground-state energies to
study this problem, we find clear power-law scaling of the distribution widths
$\sigma_N$ irrespective of the chosen value of $\sigma$. An example is presented in
Fig.~\ref{Fig:sample2sample_exponent_showcase1_fully} for $\sigma = 0.1$. As shown in
Fig.~\ref{Fig:sample2sample_compare_dil_ful}, the value of $\Theta_f$ for the fully
connected model is found to approach $\Theta_f = 1/5$ in the SK limit as predicted,
hence confirming that the $m=\infty$ model is not in the same universality class as
the spherical spin-glass. In the short-range limit $\sigma \to\infty$, we arrive at a
trivial $\Theta_f = 1/2$. The transition between these two extremes, expected to be
sharp at $\sigma = 1/2$ in the thermodynamic limit, is found to be rather smeared
out, however, indicating the presence of strong corrections to scaling. In contrast,
the situation for the diluted model is found to be rather different with the
fluctuation exponent being compatible with $\Theta_f = 1/2$ irrespective of $\sigma$,
cf.\ Fig.~\ref{Fig:sample2sample_compare_dil_ful}.  This is some first evidence of a
lack of universality between the fully connected and diluted versions of the 1d
power-law model. In fact, a similar behavior was predicted in
Ref.~\onlinecite{parisi:10} for the case of diluted Ising systems, where it was
attributed to the local heterogeneities caused by the fluctuating coordination
number.

\begin{figure}
  \begin{center}
    \includegraphics[scale=1, trim=0 0 0 0]{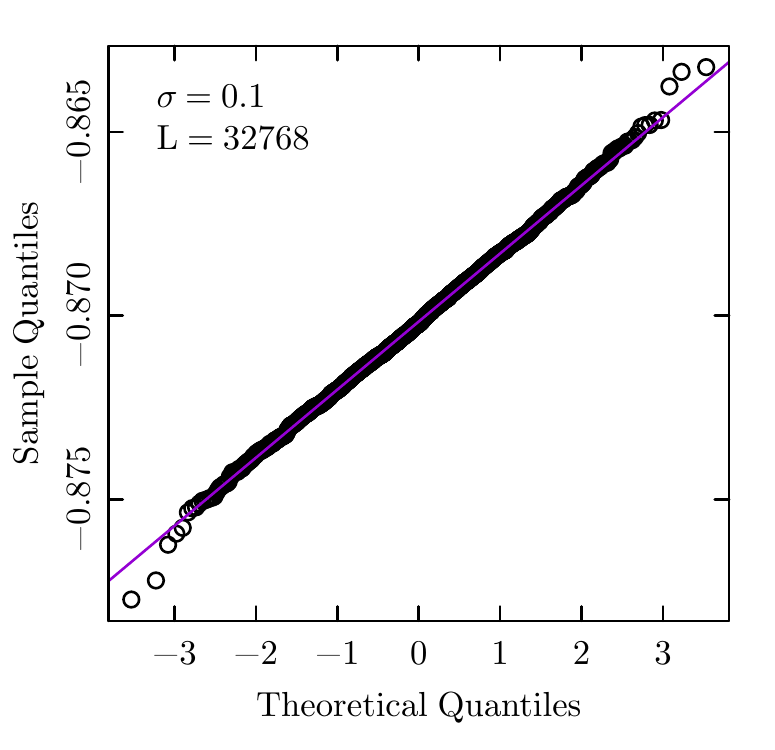}\\
    \includegraphics[scale=1, trim=0 10 0 0]{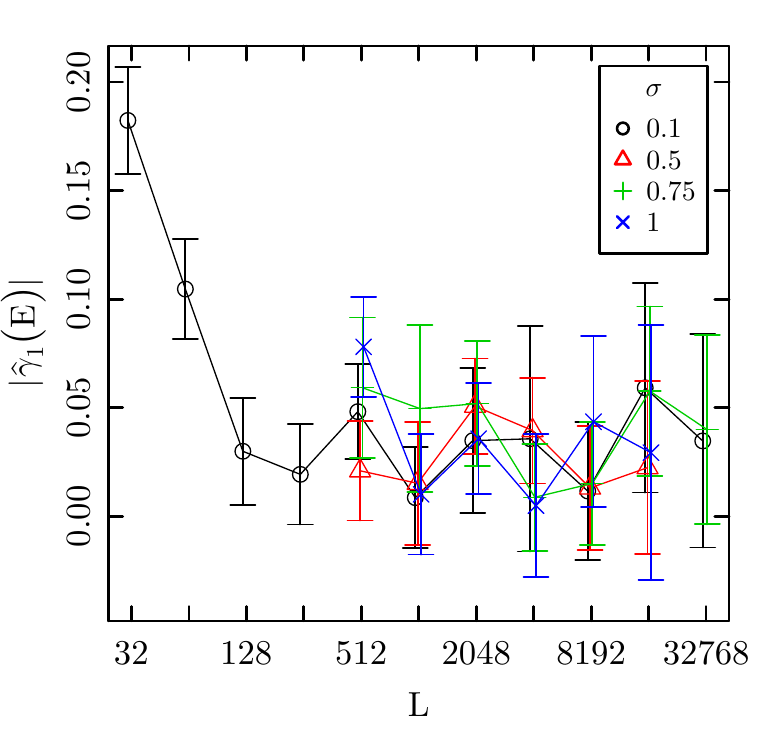}
  \end{center}
  \caption
  { (Color online).  Evidence of the Gaussian nature of the ground-state energy
    distribution for a range of different values of $\sigma$.  The top panel shows a
    normal quantile-quantile-plot in a showcase example for $\sigma=0.1$ and the
    largest available lattice size $N=32\,768$.  Sample quantiles do not deviate from
    the theoretical straight line assuming a Gaussian distribution function.  The
    lower panel shows the scaling of the (modulus of the) skewness according to
    Eq.~\eqref{Eq:skewness} for $\sigma=0.1$, $0.5$, $0.75$, and $1$. It vanishes for
    all $\sigma$ verifying the Gaussian.  }
  \label{Fig:defE_Egs_qqplot}
\end{figure}

We also checked the distribution functions of the ground-state energies for the diluted as well
as the fully connected model
and each power-law exponent $\sigma$.
%and found them to be Gaussian throughout.\\
%Another feature of interest is the distribution of ground-state energies. 
While these show a non-trivial form for the (Ising) SK model \cite{boettcher:05a},
Gaussian distributions have been reported for short-range models
\cite{bouchaud:03}. For the 1d Ising power-law chain, a crossover from Gaussian to
non-trivial has been found on moving into the mean-field regime
\cite{katzgraber:05b}. The distribution of ground-state energies for the 1d
$m=\infty$ model is analyzed in Fig.~\ref{Fig:defE_Egs_qqplot}. For all values of the
power-law exponent $\sigma$ considered here ($0.1\le\sigma\le 1$)
%as well as for the actual $m=\infty$ SK model,
the distributions seem to be compatible with a Gaussian. This is indicated by a
showcase example for $\sigma=0.1$ and $N=32\,678$ with a quantile-quantile plot as
well as an analysis of the skewness of the distribution, estimated by
\begin{equation}
    \hat{\gamma}_{1}(E) = \frac{1}{N_{\mathrm{s}}} \sum_{i=1}^{N_{\mathrm{s}}}
    \left( \frac{E_{i}-\hat{E}}{\hat{\sigma}_{E}} \right)^{3} \label{Eq:skewness}
\end{equation}
in dependence of the system size for several values of $\sigma$ spanning the
mean-field as well as non-mean-field regimes. Here, $N_s$ denotes the number of
disorder realizations. $\hat{E}=(1/N_{\mathrm{s}})\sum_{j=1}^{N_{\mathrm{s}}} E_{j}$
is the usual estimator for the expectation value $\langle E\rangle$ and
$\hat{\sigma}_{E}^{2}=(1/(N_{\mathrm{s}}-1))\sum_{j=1}^{N_{\mathrm{s}}}
(E_{j}-\hat{E})^2$ estimates the variance $\sigma_{E}^{2}$. Comparing to the results
of Ref.~\onlinecite{katzgraber:05b}, it is worthwhile to note that lattice sizes used
there were considerably smaller ($L\le192$) due to the fact that a Monte Carlo
simulation was employed on the \textit{fully connected} 1d power-law model. As shown
in the lower panel of Fig.~\ref{Fig:defE_Egs_qqplot} for such small systems we see a
significant decay for increasing lattice size similar to the one reported in
Ref.~\onlinecite{katzgraber:05b}.  The data shown stem from calculations using the
diluted version of the model.  We also checked the fully connected version, however,
and found identical results (not shown).

\begin{figure}
  \begin{center}
    \includegraphics[scale=0.96, trim=0 5 0 5]{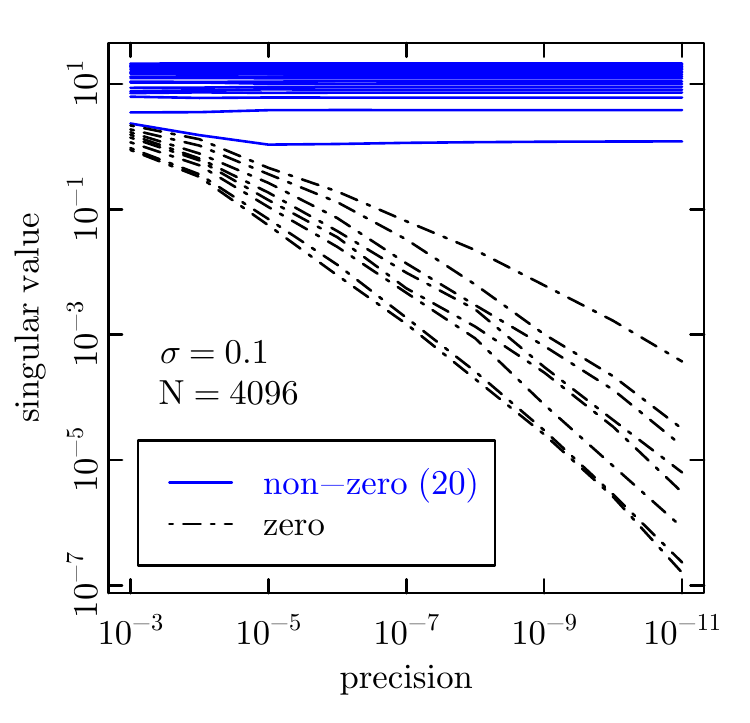}\\[-2ex]
    \includegraphics[scale=0.96, trim=0 5 0 0]{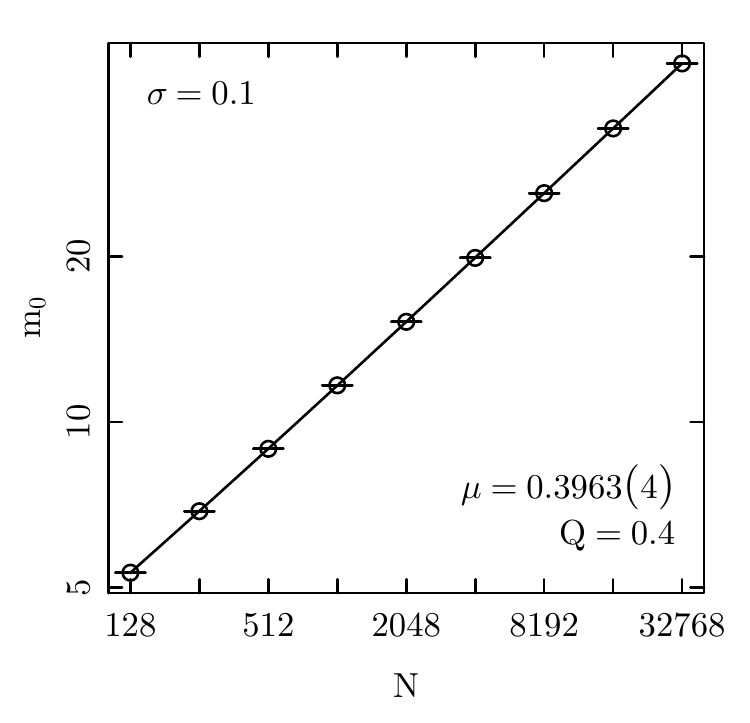}\\[-1ex]
    \includegraphics[scale=0.96, trim=0 0 0 0]{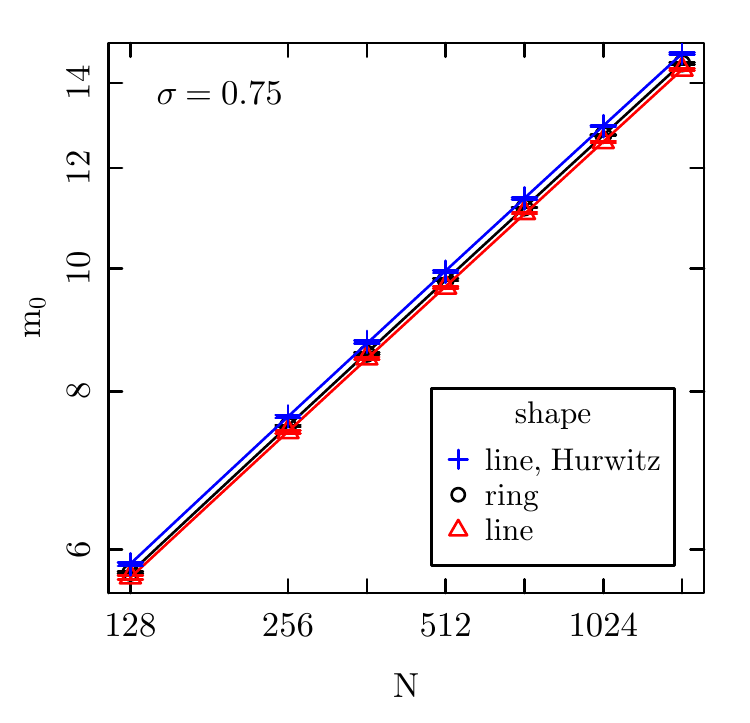}
  \end{center}
  \vspace*{-0.55cm}
  \caption{
    (Color online). Top panel: singular values of the spin matrix $M$ for a
    sample with $N=4096$, $\sigma = 0.1$ for a quench into the ground
    state. The ``precision'' refers to the relative change of the configurational
    energy after a fixed number of sweeps. Middle panel: average number of
    ground-state spin dimensions $m_{0}$ for
    $\sigma=0.1$. The line shows a fit of the form
    \eqref{Eq:defE_mu_fit} to the data. Bottom panel: comparison of $m_0$
    for the ring geometry and the line geometry with bare and resummed
    interactions (Hurwitz) for $\sigma = 3/4$.
    \label{Fig:defE_mu_sig100}
  }
\end{figure}

For the case of the $m=\infty$ model considered here another zero-temperature
property concerns the average number $m_{0}=[m^{\ast}]_\mathrm{av}$ of spin
components required to form the ground state.  We determine $m^{\ast}$ for each
realization by ordering all spin vectors of the ground state configuration into an
$m\times N$-matrix $M=\{\mathbf{S}_1,\ldots,\mathbf{S}_N\}$ and performing a singular
value decomposition (SVD) in order to calculate the number of non-zero singular
values, which is equal to the rank of the matrix. Since the rank of a matrix
determines the number of linearly independent columns, i.e., spin-vectors, it is the
desired number $m^{\ast}$. In practice, we monitor the size of all singular values of $M$ online
as the quench into the ground state proceeds. As the precision of the ground-state
determination is increased, those singular values that vanish in the exact ground state
will scale to zero, whereas all the other singular values reach finite limiting
values. This is illustrated for a sample of size $N=4096$ and $\sigma = 0.1$ in the
top panel of Fig.~\ref{Fig:defE_mu_sig100}.

\begin{figure}
  \begin{center}
    \includegraphics[scale=1, trim=0 10 0 0]{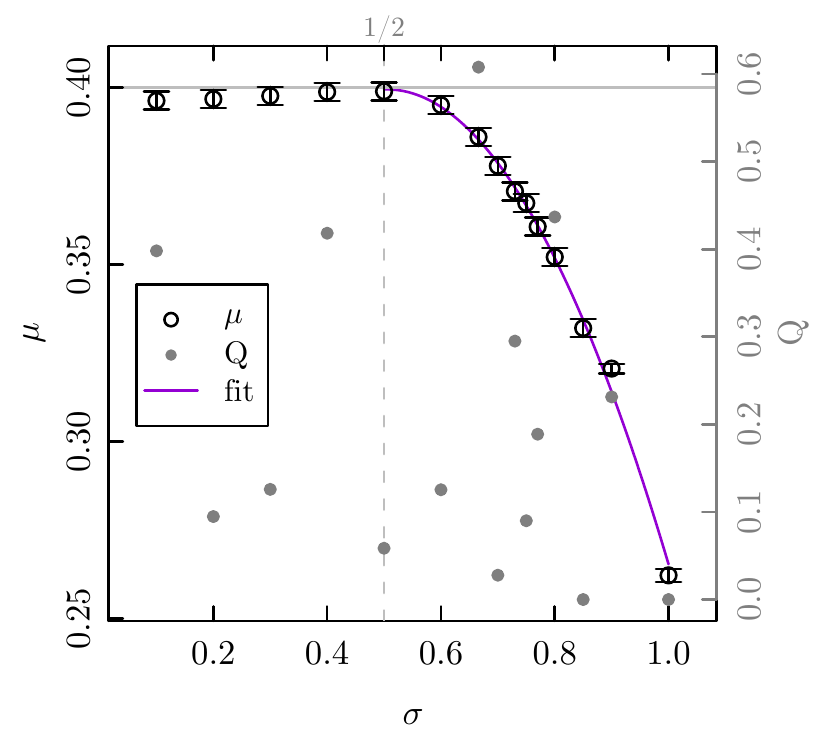}\\
  \end{center}
  \caption
  { (Color online). Variation of the exponent $\mu$ with the interaction range
    $\sigma$.  The points are results of fits of the form Eq. \eqref{Eq:defE_mu_fit}
    to the data for the average number of occupied spin components in the ground
    state for the diluted version of the model.  Gray full points show the
    quality-of-fit parameter $Q$ (right scale). The purple line corresponds to the
    functional form $\mu(\sigma) = 0.3995-0.55(\sigma-0.504)^2$.  }
  \label{Fig:defE_mu_fit}
\end{figure}

For the average number of spin components in the ground state, we assume the scaling
form
\begin{equation}
    m_{0} = [ m^{\ast} ]_{\mathrm{av}} = \mathrm{const.}\times N^{\mu} + c,
    \label{Eq:defE_mu_fit}
\end{equation}
where the additive correction $c$ can account for the fact that for small systems
$m_{0}$ will not scale to zero, but will be $m_0 = 1$ for tiny systems with $N=1$ and
$N=2$.  We present the results of this analysis for the infinite-range value $\sigma
= 0.1$ and the diluted model in the middle panel of
Fig.~\ref{Fig:defE_mu_sig100}. Since mean-field theory is exact there, we expect to
see the value $\mu = 2/5$ found for the SK model \cite{hastings:00,aspelmeier:04a},
which is indeed borne out rather well. For $\sigma > 1/2$, $\mu$ continuously
decreases below $\mu = 2/5$, cf.\ the summary of our data for the diluted model in
Fig.~\ref{Fig:defE_mu_fit}. A fit to a parabola yields $\mu(\sigma) =
0.3995-0.55(\sigma-0.504)^2$. Note that since $m_{0}\sim N^{\mu}$ and due to the
boundedness of $\mu\le \sfrac{2}{5}<1$, it follows that $m_{0}/N\sim
N^{\mu-1}\rightarrow0$ in the thermodynamic limit. Hence, as $T\rightarrow 0$ the
spins condense into a subspace of vanishing relative size, just as in the more
familiar Bose-Einstein condensation \cite{aspelmeier:04a}.

To check for the influence of the different geometries introduced above in
Sec.~\ref{sec:geometry} on the scaling results, we performed some calculations for
the bare and resummed line geometries. The effect of these changes on $m_0$ is
illustrated for $\sigma = 3/4$, where we expect the largest deviations, in the bottom
panel of Fig.~\ref{Fig:defE_mu_sig100}. We find a small overall shift in $m_0$ but,
as expected, no change in $\mu$. Moreover, there appear to be no significant
alterations of corrections to the leading FSS behavior.

\subsection{Defect energies \label{Sec:Defectenergies}}

The defect-energy approach \cite{banavar:82a} is widely used in studying systems with
spin-glass phases. It is based on the assumption that the cost $E_{\mathrm{def}}$ of
the insertion of a system-size defect into a state of the ordered phase scales as
\cite{bray:84}
\begin{equation}
    E_{\mathrm{def}} \propto L^{\theta}, \label{eq:defE}
\end{equation}
where $\theta$ is known as the spin-stiffness exponent. Generalizing Peierls'
argument for the stability of the ordered phase of a ferromagnet, one predicts
$T_\mathrm{SG} = 0$ whenever $\theta < 0$, whereas the ordered phase is stable at
finite temperatures for $\theta > 0$. The limiting case $\theta = 0$ corresponds to
the LCD of the system. Additionally, for the case $\theta < 0$ with a
zero-temperature transition, $\theta$ is related to the correlation length exponent
as \cite{bray:84} $\nu = -1/\theta$.

Numerically, defect energies are conventionally determined by comparing ground states
of systems with a pair of different boundary conditions (BCs) chosen such that the
respective ground states must differ by a relative domain-wall type excitation. Then 
the defect energy corresponds to the energy difference. The most commonly used
set of such BCs are periodic and antiperiodic boundaries. The defect energy of a
given realization is then
\begin{equation}
    \Delta E = |E_\mathrm{AP}-E_\mathrm{P}|,
\end{equation}
where the modulus is required since, for symmetric coupling distributions, the two
boundary conditions are statistically equivalent. For the case of the long-range ring
geometry considered here, a ground-state search is performed for the original
coupling configuration, yielding $E_\mathrm{P}$. In the second step, the boundary
exchange couplings are flipped to the antiperiodic state by choosing \textit{one
  arbitrary} nearest-neighbor pair $(\mathbf{S}_{a}$, $\mathbf{S}_{a+1})$,
$a\in{1,\ldots,N}$ (without having them necessarily interact in the diluted version
of the model) and changing the sign of all interaction constants $J_{ij}$ between
spin $\mathbf{S}_{i}$ and $\mathbf{S}_{j}$ for all $i\ne j$ if the shorter path
between those two spins falls on top of the path between $\mathbf{S}_{a}$ and
$\mathbf{S}_{a+1}$.  \cite{katzgraber:03a} A second ground-state search for this
altered configuration then yields $E_\mathrm{AP}$. There has been some discussion in
the past about whether this setup is suitable for the case of continuous spins, since
both periodic and antiperiodic boundaries induce some defects, such that the energy
difference $E_\mathrm{P}-E_\mathrm{AP}$ does not directly correspond to a defect
energy \cite{kosterlitz:99a,weigel:05f,weigel:06c,beyer:11}. Therefore, boundary
conditions that allow directly to measure the energy of a single defect seem
preferable. For the case of the $m=\infty$ spin glass on hypercubic lattices we
indeed found different stiffness exponents from such ``domain wall'' BCs
\cite{beyer:11}. For the 1d long-range system, however, it is not obvious how to
implement such alternative prescriptions.

\begin{figure}[t]
  \begin{center}
    \includegraphics[scale=1, trim=0 10 0 0]{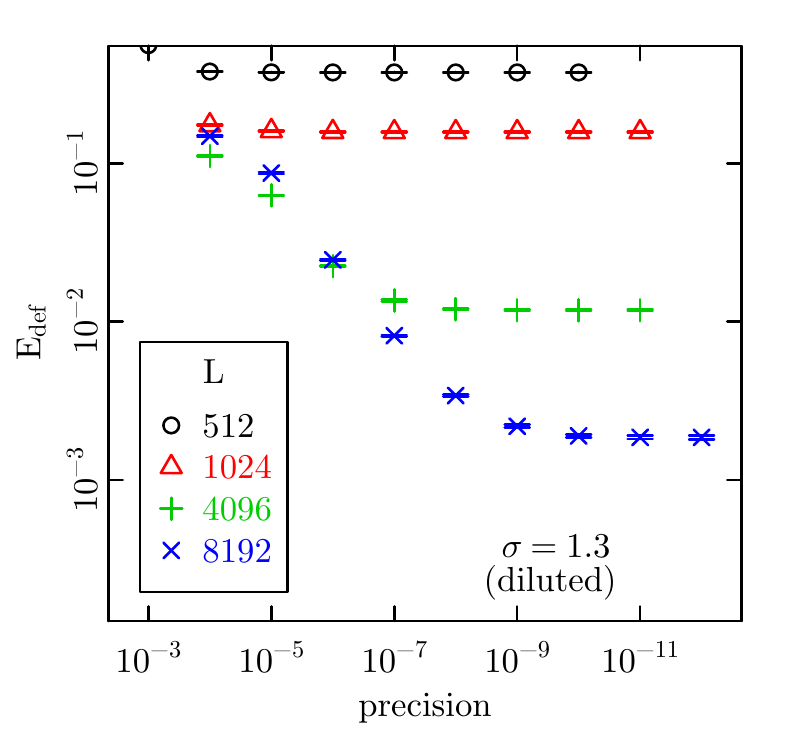}\\
  \end{center}
  \caption
  { (Color online). Convergence of the defect energy for the diluted model as a
    function of the precision of the ground-state energy calculation. For large $\sigma$
    this should be monitored accurately.
    \label{Fig:defE_evolution}
  }
\end{figure}

The resulting energy differences are averaged,
\begin{equation}
    E_{\mathrm{def}} = \left[\Delta E\right]_{\mathrm{av}} =
    \left[|E_\mathrm{AP}-E_\mathrm{P}|\right]_{\mathrm{av}}
\end{equation}
to yield an estimate $E_\mathrm{def}$ of the defect energy. Ground states were
computed for $\sigma$ in the interval $0.1\le\sigma\le 1$ for the diluted version of
the model. For most of the values of $\sigma$ a range of system sizes $512\le L\le
32\,768$ has been considered with about $3\,000$ disorder realizations for larger and
up to $10\,000$ samples for smaller system sizes, see the parameters collected in
Table \ref{Tab:defE_results}. We ensured convergence by monitoring $E_{\mathrm{def}}$
as the ground-state quench proceeds, cf.\ Fig.~\ref{Fig:defE_evolution}. This is of
particular importance for large $\sigma$, where choosing a fixed precision fails to
produce converged results for sufficiently large systems.

\begin{figure}[t]
  \begin{center}
    \includegraphics[scale=1, trim=0 10 0 0]{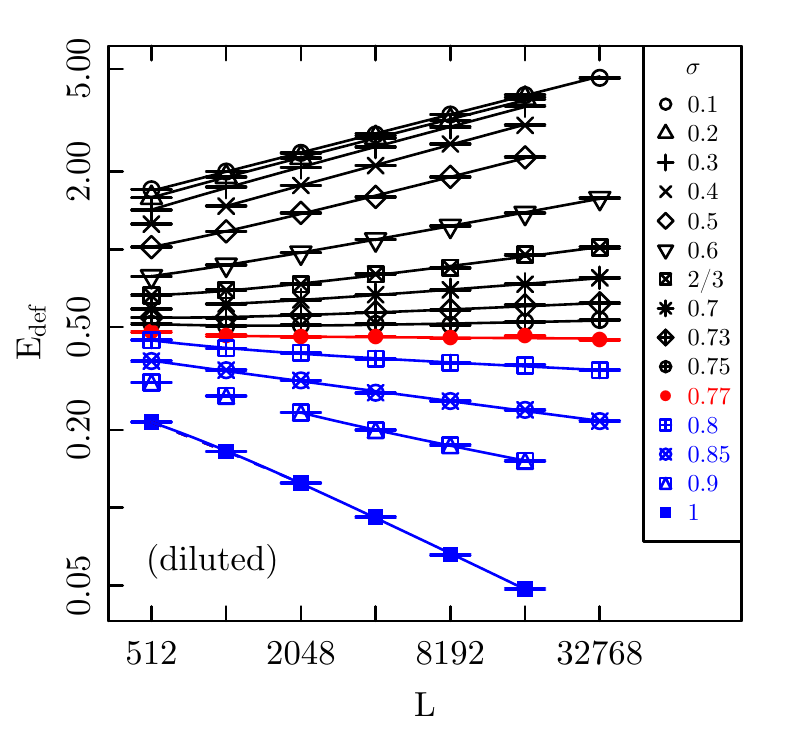}\\
  \end{center}
  \caption
  { (Color online). Defect energies for the diluted 1d power-law $m=\infty$ spin
    glass as a function of lattice size $L$ for a number of different interaction
    ranges $\sigma$. The solid lines show fits of the functional form
    \eqref{eq:defE_corr} to the data. The corresponding fit parameters are summarized
    in Table \ref{Tab:defE_results}. The defect-energy exponent $\theta$ changes
    sign in the vicinity of $\sigma=0.77$, cf.\ Fig.~\ref{Fig:defE_theta_vs_sigma}.
    \label{Fig:defE_theta}
  }
\end{figure}

The averaged defect energies are shown for the available interaction ranges $\sigma$
in Fig.~\ref{Fig:defE_theta}. To extract the stiffness exponents, we performed fits
of the functional form
\begin{equation}
  E_{\mathrm{def}} = aL^{\theta}(1+b/L) \label{eq:defE_corr}
\end{equation}
to the data. Similarly to our experience from the hypercubic systems, we found this
form to describe the corrections rather well \cite{beyer:11}. While this form
parametrizes the leading analytical correction for $\theta < 0$, a constant would be
asymptotically dominant over $b/L$ for $\theta > 0$. Fits including a constant but no
$1/L$ correction, however, are not found to describe the data well for $\theta > 0$,
such that we stick with the form \eqref{eq:defE_corr} for all $\sigma$. Any
non-analytic corrections, if present, appear to be sub-leading. These fits are shown
in Fig.~\ref{Fig:defE_theta} and the corresponding fit parameters are collected in
Table \ref{Tab:defE_results}. Fit qualities are found to be high throughout,
indicating the suitability of the form chosen in Eq.~\eqref{eq:defE_corr}.

\begin{table}[b]
  \caption
  {
    Estimates of the spin stiffness exponent $\theta$
    in the \textit{diluted} model resulting from fits of the functional form \eqref{eq:defE_corr} to the
    data, including system sizes $N_\mathrm{min} \le N \le N_\mathrm{max}$. $Q$
    denotes the quality-of-fit.
    \label{Tab:defE_results}
  }
  \begin{ruledtabular}
    \begin{tabular}{lccrrt{7}l}
      \toprule
      & \phantom{a} &                   &  \multicolumn{4}{c}{$aL^{\theta}(1+b/L)$} \\
      \cmidrule{6-7}
      $\sigma$  &       & \multicolumn{1}{c}{samples$/10^{3}$}& $N_{\mathrm{min}}$      & $N_{\mathrm{max}}$    & \multicolumn{1}{c}{$\theta$}  & \multicolumn{1}{c}{$Q$} \\
      \hline
      0.1           &       & $2.4-10$              & 32        & 32768 & 0.245(02)             & 0.12  \\
      0.2           &       & $2.9-10$              & 512       & 16384 & 0.243(09)             & 0.39  \\
      0.3           &       & $3.1-9.1$             & 512       & 16384 & 0.255(08)             & 0.70  \\
      0.4           &       & $3.3-9.5$             & 1024      & 16384 & 0.260(12)             & 0.54  \\
      0.5           &       & $3.6-10$              & 512       & 16384 & 0.245(08)             & 0.31  \\
      0.6           &       & $3.1-9.5$             & 512       & 32768 & 0.177(06)             & 0.98  \\
      \sfrac{2}{3}  &       & $3.2-9.2$             & 512       & 32768 & 0.126(07)             & 0.13  \\
      0.7           &       & $3.2-10$              & 1024      & 32768 & 0.076(08)             & 0.96  \\
      0.73          &       & $3.2-9.1$             & 512       & 32768 & 0.046(07)             & 0.89  \\
      0.75          &       & $3.2-9.1$             & 512       & 32768 & 0.021(07)             & 0.56  \\
      0.77          &       & $3.2-9.3$             & 1024      & 32768 & -0.006(09)            & 0.28  \\
      0.8           &       & $3.2-9.0$             & 512       & 32768 & -0.046(06)            & 0.75  \\
      0.85          &       & $3.1-6.4$             & 512       & 32768 & -0.127(07)            & 0.44  \\
      0.9           &       & $2.8-10$              & 2048      & 16384 & -0.183(26)            & 0.65  \\
      1.0           &       & $1.8-9.7$             & 512       & 16384 & -0.467(09)            & 0.67  \\
    \end{tabular}
  \end{ruledtabular}
\end{table}

\begin{table}[b]
  \caption
  {
    Estimates of the spin stiffness exponent $\theta$
    in the \textit{fully connected} model resulting from fits of the
    functional form \eqref{eq:defE_corr2} (for $\sigma\le 0.75$) and
    \eqref{eq:defE_corr} (for $\sigma > 0.75$), respectively, to the
    data.
    \label{Tab:defE_results_fully}
  }
  \begin{ruledtabular}
    \begin{tabular}{lccrrt{7}l}
      \toprule
      %& \phantom{a} &                   &  \multicolumn{3}{c}{$aL^{\theta}(1+b/L)$} \\
      %\cmidrule{5-6}
      $\sigma$  &       & \multicolumn{1}{c}{samples$/10^{3}$}  & $N_{\mathrm{min}}$& $N_{\mathrm{max}}$    & \multicolumn{1}{c}{$\theta$}  & \multicolumn{1}{c}{$Q$} \\
      \hline
      0.1       &       & $3.9-4$     & 64    & 4096  & 0.238(19)   & 0.98  \\
      0.2       &       & $3.9-4$     & 64    & 4096  & 0.238(19)   & 0.06  \\
      0.3       &       & $3.9-4$     & 64    & 4096  & 0.275(20)   & 0.12  \\
      0.4       &       & $3.9-4$     & 64    & 4096  & 0.247(21)   & 0.31  \\
      0.5       &       & $3.9-4$     & 64    & 4096  & 0.262(24)   & 0.76  \\
      0.57      &       & $3.9-4$     & 128    & 4096  & 0.180(41)   & 0.86  \\
      0.6       &       & $3.8-4$     & 128    & 4096  & 0.200(46)   & 0.52  \\
      0.625     &       & $3.9-4$     & 64    & 4096  & 0.168(36)   & 0.52  \\
      0.75      &       & $3.8-4$     & 64    & 4096  & 0.052(95)   & 0.87  \\
	  \hline
      0.8       &       & $3.8-4$     & 256    & 4096  & 0.014(15)  & 0.73  \\
      1.0       &       & $2$    	  & 256    & 4096  & -0.209(21)  & 0.28  \\
      1.3       &       & $1.9-2$     & 256    & 4096  & -0.436(33)  & 0.99  \\
      2.0       &       & $1-2$    	  & 256    & 2048  & -1.132(37) & 0.96  \\
      2.5       &       & $1-2$    	  & 128    & 1024  & -1.518(47) & 0.96  \\
    \end{tabular}
  \end{ruledtabular}
\end{table}

\begin{figure}[t]
  \begin{center}
    \includegraphics[scale=1, trim=0 10 0 0]{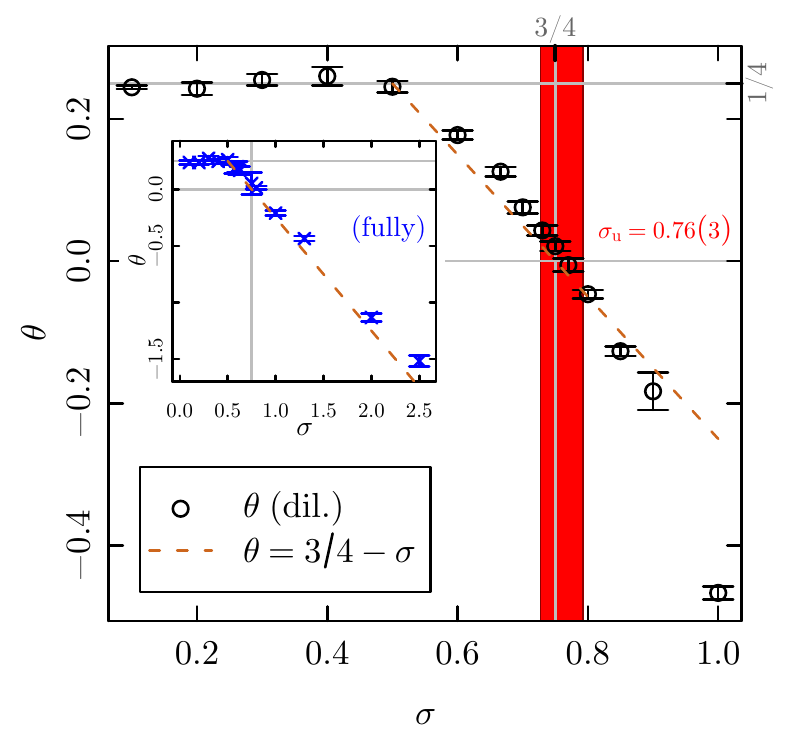}\\
  \end{center}
  \caption
  { (Color online).  The stiffness exponent $\theta$ extracted from the diluted model
    as a function of the interaction range exponent $\sigma$. See Table
    \ref{Tab:defE_results} for the corresponding fit parameters. $\theta$ changes
    sign for the critical value $\sigma_{u}=0.76(3)$ marked by the shaded area.
    Below $\sigma = 1/2$, the spin stiffness exponent levels off at the 
    infinite-range value $\theta = 1/4$. The dashed line denotes the conjectured form
    $\theta = 3/4-\sigma$, cf.\ Eq.~\eqref{eq:minfty_theta_conjecture}. The inset
    shows analogous results for the fully connected model and a wider range of
    $\sigma$.
    \label{Fig:defE_theta_vs_sigma}
  }
\end{figure}

Figure \ref{Fig:defE_theta_vs_sigma} summarizes our results for $\theta$ as a
function of $\sigma$. The stiffness exponent $\theta$ clearly becomes constant at a
value compatible with $\theta = 1/4$ in the infinite-range regime $\sigma\le 1/2$. To
determine the upper critical $\sigma_u$ where $\theta(\sigma_u) = 0$, we performed a
linear fit to the results in the range $0.5\le\sigma\le0.9$, resulting in an
intercept of $\sigma_{u}=0.76(3)$. In connection with the observation of a linear
behavior in the regime $1/2\le \sigma \le 3/4$, this is compatible with the
conjectured form $\theta_\mathrm{LR} = 3/4-\sigma$, cf.\
Eq.~\eqref{eq:minfty_theta_conjecture}. This behavior is clearly different from the
corresponding Ising spin-glass model with $\theta_\mathrm{LR} = 1-\sigma$. Notice
that the change to the value $1/4$ at $\sigma =1/2$ is not due to a failure of
Eq.~\eqref{eq:minfty_theta_conjecture} for $\sigma <1/2$, but because for
$\sigma<1/2$ we have rescaled the bonds down by a factor of $L^{1/2-\sigma}$. If we
had not done that $\theta$ would have continued to be fitted by
Eq.~\eqref{eq:minfty_theta_conjecture}.

\begin{figure}
       \includegraphics[trim= 0.25cm 0.4cm 0.25cm 0cm]{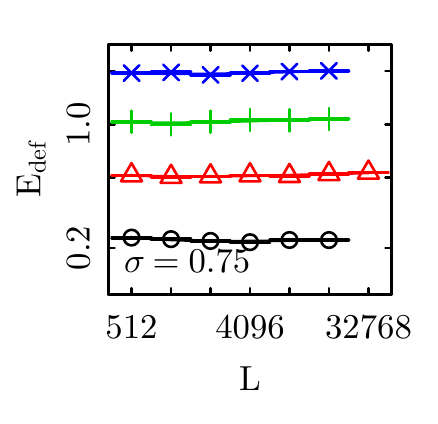}
       \includegraphics[trim= 0.25cm 0.4cm 0.25cm 0cm]{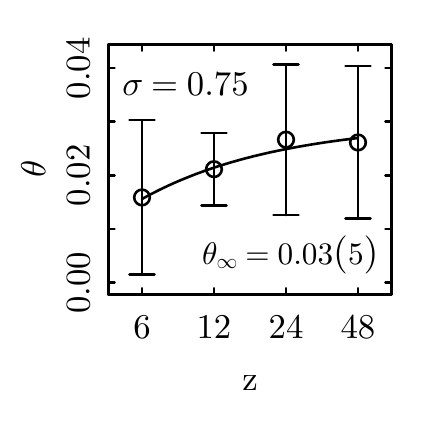} \\
       \includegraphics[trim= 0.25cm 0.4cm 0.25cm 0cm]{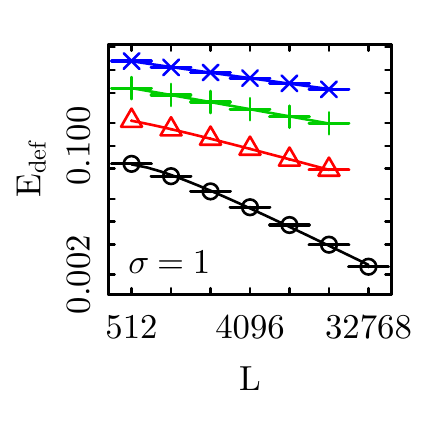}
       \includegraphics[trim= 0.25cm 0.4cm 0.25cm 0cm]{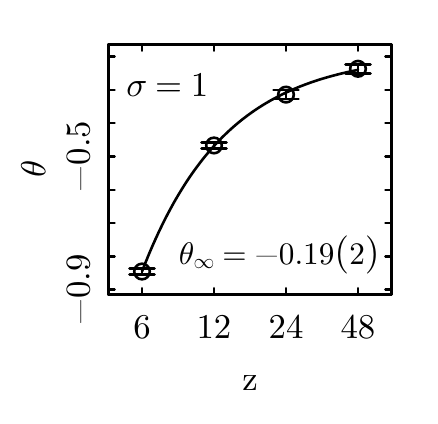} \\
       \includegraphics[trim= 0.25cm 0.4cm 0.25cm 0cm]{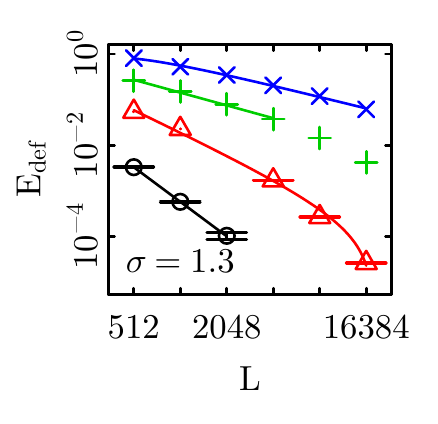}
       \includegraphics[trim= 0.25cm 0.4cm 0.25cm 0cm]{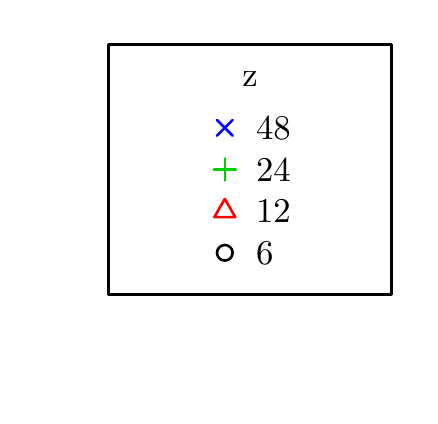}
       \caption{(Color online).
         Dependence of defect energies and stiffness exponents in the diluted model
         on the average coordination number $z$. While for $\sigma < 1$, this is
         merely a finite-size correction, we observe a non-universal $\theta$ for
         $\sigma = 1$, and non-power-law scaling implying an effective
         $\theta=-\infty$ for $\sigma > 1$.
       }
   \label{Fig:dil_z-dependence}
\end{figure}

For larger $\sigma$, however, we observe clear deviations of the data for the diluted
model from $\theta = 3/4-\sigma$. In fact, the data for $E_\mathrm{def}$ at $\sigma =
1.3$, not shown in Fig.~\ref{Fig:defE_theta} (but see the lower left panel of
Fig.~\ref{Fig:dil_z-dependence}), show a strong downward curvature, more resembling
an exponential decay.  A closer look reveals that there is no universality between
the diluted and fully connected model for $\sigma > 1$, where the properties of the
diluted graphs change significantly. As has been shown in
Refs.~\onlinecite{schulman:83,newman:86}, 1d graphs defined by
Eq.~\eqref{Eq:pij_1d_pow_wo_field_dilute} always percolate for $\sigma \le 1/2$ and
they percolate for sufficiently large $A$ (namely, for any $z > 1/2$) in the regime
$1/2 < \sigma \le 1$. In contrast, percolation is (asymptotically) absent for $\sigma
> 1$. For the defect-energy calculations considered here, this means that such
non-percolating samples contribute $\Delta E = 0$ to $E_\mathrm{def}$, leading to
much smaller averages than expected from the scaling Eq.~\eqref{eq:defE}. For the
average coordination numbers considered here, such breakdown of percolation is only
observed for very large systems, mostly beyond the reach of our numerical
calculations. If we remove the links up to a finite range, however, for instance all
nearest-neighbor links, the remaining graph does not percolate for $\sigma > 1$
already for moderate sizes such that the long-range nature is lost. An alternative
way of understanding this phenomenon is to note that the diameters of the graphs
considered here grow proportional to $(\log L)^\delta$ for $1/2\le \sigma < 1$,
corresponding to an infinite-dimensional or small-world graph, whereas they grow
proportional to $L$ for $\sigma > 1$, corresponding to a truly one-dimensional graph
\cite{benjamini:01}. This explains the strong downwards deviations of $\theta$ from
the form $\theta = 3/4-\sigma$ seen in Fig.~\ref{Fig:defE_theta_vs_sigma} for $\sigma
\ge 1$. Right at $\sigma = 1$, we expect non-universality with $\theta$ depending on
the average coordination number $z$ which, in turn, is a function of the parameter
$A$ in Eq.~\eqref{Eq:pij_1d_pow_wo_field_dilute}.

This is illustrated in Fig.~\ref{Fig:dil_z-dependence} showing our estimates for
$\theta$ from diluted lattices with different average coordination numbers
$z$. While, for $\sigma < 1$, the measured $\theta$ is essentially independent of
$z$, as expected, there is a dramatic $z$ dependence right at $\sigma = 1$, with the
non-universal $\theta(z)$ approaching the value expected for the fully connected
model in the limit $z\to\infty$. Similar $z$ dependent critical behavior at $\sigma =
1$ was also recently found for random walks on such graphs \cite{juhasz:12}. For
$\sigma > 1$, on the other hand, the curves $E_\mathrm{def}$ cease to follow power
laws and, instead, cross over to the $\theta = -\infty$ or exponential decay expected
for short-range one-dimensional systems \cite{morris:86a}.

These deviations are specific to the diluted model: independent ground-state
calculations for the fully connected model, summarized in the inset of
Fig.~\ref{Fig:defE_theta_vs_sigma}, are consistent with $\theta = 3/4-\sigma$ also
for $\sigma > 1$.  The relevant parameters and results for this model are summarized
in Table \ref{Tab:defE_results_fully}. In contrast to the diluted model, a constant
was found to be a good description of the leading scaling corrections for $\theta >
0$, such that we used the form
\begin{equation}
  E_{\mathrm{def}} = aL^{\theta}+c \label{eq:defE_corr2}
\end{equation}
for $\sigma \le 3/4$ and the form \eqref{eq:defE_corr} for $\sigma > 3/4$. Apart
from the fact that only the fully connected model does represent the long-range
universality class for $\sigma \ge 1$, we also find scaling corrections for $\sigma <
1$ to be less pronounced there than for the diluted system, such that it appears
questionable whether considering the diluted model offers a significant advantage in
terms of the precision and accuracy of the final results.

Note that the results for $\theta$ as $\sigma$ is increased are in contrast to those
for the Ising case, where $\theta_\mathrm{SR} = -1$, so that the (fully connected)
long-range Ising system is governed by short-range behavior for $\theta_\mathrm{SR} >
\theta_\mathrm{LR}$ viz.\ $\sigma > 2$ (whereas the diluted Ising system would be
truly short ranged already for $\sigma > 1$). The $m=\infty$ model, instead, is truly
long-range everywhere, and crossover to the $\theta_\mathrm{SR} = -1$ of the
nearest-neighbor 1d chain system is not seen. Instead, the limit $\sigma\to\infty$ of
our fully connected model corresponds to the 1d ladder system with \cite{morris:86a}
$\theta=-\infty$.

\section{Finite-temperature calculations\label{sec:finite}}

The critical behavior which arises when $\sigma < 3/4$ can only be studied with
techniques appropriate to finite temperatures. While the ground-state calculations
were greatly facilitated by the disappearance of metastability in the
$m\rightarrow\infty$ limit, the use of a saddle-point procedure permits the exact
solution at finite temperatures. As will be discussed in this section, this approach
leads to an iterative set of matrix equations that allow one to determine the
thermally averaged spin-spin correlation function which, in turn, gives access to the
Edwards-Anderson order parameter, the spin-glass susceptibility and the spin-glass
correlation length. Due to the nature of the matrix equations, use of a diluted model
does not have any computational advantages, so that all calculations have been
performed on the fully connected model of Eq.~\eqref{Eq:Jij_1d_pow_fullConnect}.

\subsection{Saddle-point equations}

The saddle-point equations for the $m=\infty$ model were first derived in
Ref.~\onlinecite{bray:82b} and later discussed again in
Refs.~\onlinecite{morris:86a,hastings:00,aspelmeier:04a}. Starting point is the partition
function corresponding to the Hamiltonian in Eq.~\eqref{eq:hamiltonian},
\begin{equation}
  \label{eq:partition_integral}
  Z = \int_{-\infty}^\infty \prod_{i,\mu}\d S_{i}^{\mu}
  e^{\frac{\beta}{2}\sum_{i,j,\mu} J_{ij} S_{i}^{\mu}S_{j}^{\mu}}
  \prod_i \delta(m-\sum_\mu (S_{i}^{\mu})^2).
\end{equation}
The spin integrations can be performed using integral representations of the $\delta$
functions to yield
\begin{equation}
  Z = \int_{-i\infty}^{i\infty} \prod_i\frac{\beta\d H_i}{4\pi}\exp\left\{\frac{m}{2}
    \left[\sum_i \beta H_i+\ln\det \frac{\chi}{\beta}\right]\right\},
  \label{eq:partition_integral2}
\end{equation}
where the susceptibility matrix $\chi$ is defined by
\begin{eqnarray}
  \chi_{ij} &=& (A^{-1})_{ij} \label{Eq:chi_as_fct_of_A} \\
  A_{ij} &=& H_{i}\delta_{ij}-J_{ij}. \label{Eq:the_holy_matrix_A} 
\end{eqnarray}
The $H_i$ in Eq.~\eqref{eq:partition_integral2} are initially just new integration
variables introduced by the Fourier representation of the $\delta$ constraints, but
it will turn out that they have a profound physical meaning. The matrix $A$ is
obviously symmetric, since $J_{ij}=J_{ji}$. The correlation matrix,
\begin{equation}
  C_{ij}=\frac{1}{m}\langle\mathbf{S}_{i}\cdot\mathbf{S}_{j}\rangle,
  \label{Eq:corr_mat_def_as_spin_avg}
\end{equation}
is identical to \cite{bray:82b}
%as can be seen from the generalization of Eq.~\eqref{eq:partition_integral} to
%include magnetic fields \cite{bray:82b}, 
\begin{equation}
  C_{ij}= T(A^{-1})_{ij},
  \label{Eq:C_as_fct_of_A}
\end{equation}
which can be seen via saddle-point approximation.
Taking the normalization of the spins $|\mathbf{S}_{i}|=\sqrt{m}$ into account and using 
Eq.~(\ref{Eq:corr_mat_def_as_spin_avg}), the main diagonal
of $C$ needs to be such that
\begin{equation}
        C_{ii}= 1.\label{Eq:Cii_is_one}
\end{equation}
Then Eqs.~\eqref{Eq:the_holy_matrix_A}, \eqref{Eq:C_as_fct_of_A}, and
\eqref{Eq:Cii_is_one} can be solved self-consistently for the values of the $N$
variables $H_{i}$. Eq.~\eqref{Eq:Cii_is_one} is the saddle-point
equation for the integrals in  \eqref{eq:partition_integral2} and is valid as $m \rightarrow \infty$ at fixed $N$.

\begin{figure*}
  \includegraphics[trim= 0cm 0cm 0cm 0cm]{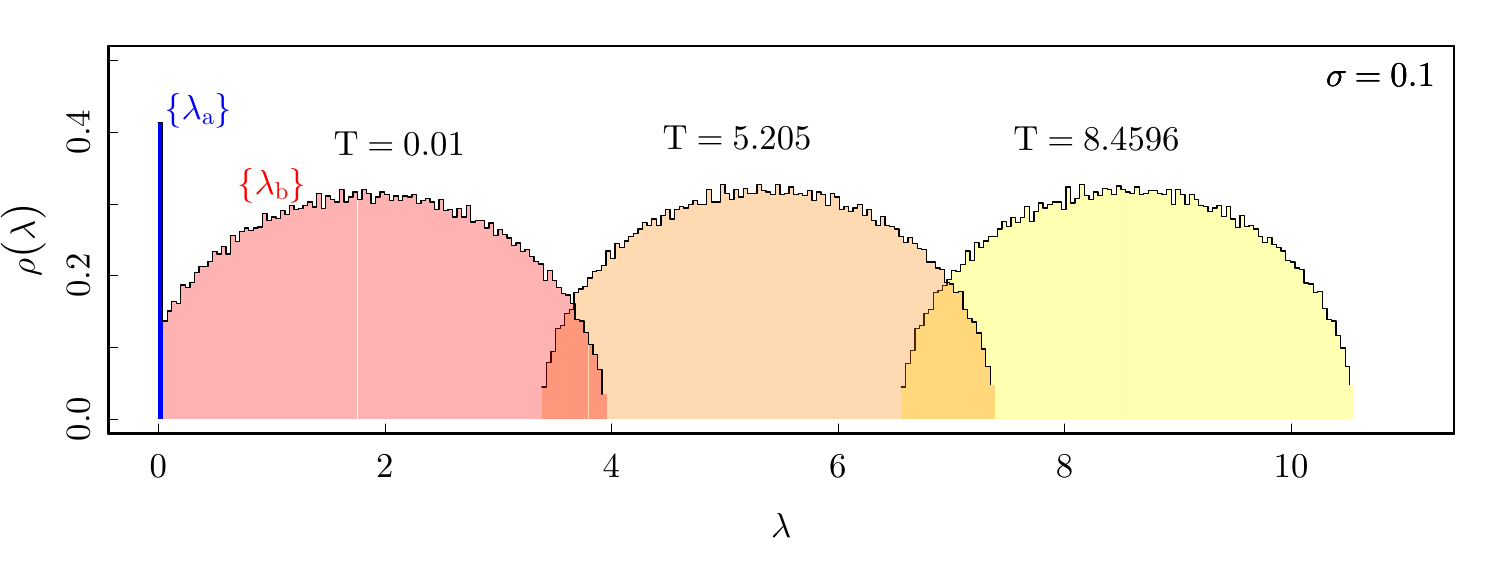}
  \caption{
    (Color online).
    The eigenvalue density $\rho(\lambda)$ of the matrix $A$, see Eq.~\eqref{Eq:the_holy_matrix_A},
    for the 1d $m=\infty$ model at $\sigma = 0.1$. At high temperatures, the distribution follows
    Wigner's semicircle law \cite{aspelmeier:04a}. Lowering the temperature shifts
    the whole distribution to the left, until for $T\rightarrow0$
    $m_{0}$ eigenvalues $\{\lambda_{a}\}$ vanish, while the majority ($N-m_{0}$) of
    eigenvalues $\{\lambda_{b}\}$ stay finite. 
    In the thermodynamic limit the Wigner semicircle law is restored again also there,
    since $m_{0}/N\rightarrow0$.
    \label{Fig:evs_lambda_histogram}
  }
\end{figure*}

At $T=0$ these equations are no longer well defined. Instead, as described above, it
is a necessary condition for the system to be in a ground state for each spin to be
aligned with its local molecular field $\mathbf{H}_{i}$, cf.\
Eq.~\eqref{Eq:the_local_field}. Due to the lack of metastability, for
$m\rightarrow\infty$ this condition is also sufficient. As was first noted in
Ref.~\onlinecite{hastings:00}, for $T\rightarrow 0$ the variables $H_{i}$ of
Eq.~\eqref{Eq:the_holy_matrix_A} are \textit{equivalent} to the rescaled amplitudes
$|\mathbf{H}_{i}|/\sqrt{m}$ of the local fields.

It was suggested in Ref.~\onlinecite{bray:82b} that the density $\rho(\lambda)$ of
(real) eigenvalues of $\chi^{-1}$ or $A$ was useful in discussions of the physics of
random spin systems. For example, it was shown that the smallest eigenvalue would
vanish at the critical temperature for $N\rightarrow\infty$. For the case of the
$m=\infty$ SK model in the thermodynamic limit, $\rho(\lambda)$ follows a Wigner
semicircle. Decreasing the temperature from $T_\mathrm{SG}$, where the first
eigenvalue vanishes, a fraction $m_0 \sim N^{2/5}$ of eigenvalues becomes zero as
$T\rightarrow 0$. \cite{aspelmeier:04a} This corresponds to the contraction of the
spin orientations into an $m_0$-dimensional subspace in the limit of zero
temperature, an effect reminiscent of the Bose-Einstein condensation in atomic
systems. The behavior of the eigenvalue density on cooling the system from high
temperatures is illustrated for our 1d system in the infinite-range regime $\sigma <
1/2$ in Fig.~\ref{Fig:evs_lambda_histogram}.

\subsection{Order parameter and spin-glass susceptibility\label{sec:AM_formalism}}

The saddle-point equations for systems on hypercubic lattices in two and three
dimensions (as well as in the SK limit) have been previously examined by Lee, Dhar
and Young in Ref.~\onlinecite{lee:05}. They considered the correlation function
$C_{ij}$ of Eq.~\eqref{Eq:C_as_fct_of_A} and determined the Edwards-Anderson order
parameter as
\begin{equation}
  \label{eq:EA_wrong}
  q_\mathrm{EA}^2 = \lim_{r_{ij}\rightarrow\infty}[C_{ij}^2]_\mathrm{av}
\end{equation}
or as $q_\mathrm{EA}^2 = [C_{ij}^2]_\mathrm{av}$, $i\ne j$ for the SK model. Taking
into account the scaling $m_0\sim N^\mu$ of the number of non-zero spin components,
they concluded that $q_\mathrm{EA}^2 \sim N^{-\mu}$ in the ground state, i.e., that
the order parameter vanishes in the thermodynamic limit. Similarly, defining
\begin{equation}
  \label{eq:chi_wrong}
  \chi^0_\mathrm{SG} = \frac{1}{N}\sum_{i,j}[C_{ij}^2]_\mathrm{av}
\end{equation}
they inferred algebraically decaying correlations $[C_{ij}^2]_\mathrm{av}\sim
r_{ij}^{-d\mu}$ in the $m=\infty$ model, i.e., merely quasi long-range order. This is
in contrast to the findings of Ref.~\onlinecite{dealmeida:78} for the case of the
$N\rightarrow\infty$ limit being taken {\em before\/} the $m\rightarrow\infty$ limit.

We believe, however, that one needs to consider the {\em connected\/} correlation
function and {\em on-site\/} correlations to determine $\chi_{\mathrm{SG}}$ and
$q_\mathrm{EA}$, and we will see that this leads to different conclusions. The basic
idea is to separate contributions from the zero and non-zero modes. Consider
Eq.~\eqref{Eq:corr_mat_def_as_spin_avg} and factor out the cumulant part
$\tilde{C}_{ij}$ of the correlation function,
\begin{equation}
  C_{ij}=\frac{1}{m}\langle\mathbf{S}_{i}\rangle\cdot\langle\mathbf{S}_{j}\rangle +
  \underbrace{\frac{1}{m}(\langle\mathbf{S}_{i}\cdot\mathbf{S}_{j}\rangle -
    \langle\mathbf{S}_{i}\rangle\cdot\langle\mathbf{S}_{j}\rangle )}_{\tilde{C}_{ij}}.
  \label{Eq:corr_mat_def_as_spin_avg_split_up}
\end{equation}
Due to the symmetry of $A$ and Eq.~\eqref{Eq:C_as_fct_of_A} the matrix $C$ is
also real and symmetric, and hence invertible.  For the eigenvalues $\lambda$ of $A$
and $\omega$ of $C$, we have the relation
\begin{equation}
        \omega = \frac{T}{\lambda}. \label{Eq:ev_lambda_mu_connection}
\end{equation}
Using the spectral theorem, $C$ has an orthonormal basis of real independent
eigenvectors $\{\vec{v}^{n}|n=1,\ldots,N\}$ for the set of eigenvalues
$\{\omega_{n}\in\mathbb{R}\}$, which do not need to be distinct.

The spectral decomposition reads
\begin{equation}
  C = \sum_{n=1}^{N} \omega_{n}\vec{v}^{n} \cdot (\vec{v}^{n})^{\mathsf T},
  \label{Eq:spectral_decomp_for_C}
\end{equation}
where we use the outer product and the transpose denoted by $(\cdot)^{\mathsf T}$.
%Then
%\begin{equation}
%  C \cdot\vec{v}^{m} = \sum_{n=1}^{N} \omega_{n}\vec{v}^{n} \cdot (\vec{v}^{n})^{\mathsf T} \cdot \vec{v}^{m} = \omega_{m}\vec{v}^{m}
%  \label{Eq:spectral_decomp_for_C}
%\end{equation}
Thus, with $(\vec{v}^{n})^{\mathsf T} \cdot \vec{v}^{m}=\delta_{mn}$ and for
$T<T_\mathrm{SG}$, the correlation matrix $C$ has the eigenvalue decomposition
\begin{equation}
  C_{ij}% = \vec{e}_{i}^{\mathsf T}\cdot(C\cdot\vec{e}_{j}) 
    = \sum_{n=1}^{N} \omega_{n} v_{i}^{n}v_{j}^{n} 
    = T \sum_{a} \frac{v_{i}^{a}v_{j}^{a}} {\lambda_{a}} +
  T \sum_{b} \frac{v_{i}^{b}v_{j}^{b}} {\lambda_{b}},
  \label{Eq:ev_decomp_for_C}
\end{equation}
where $a=1,\ldots,m_{0}$ labels the $m_0$ eigenvalues $\lambda_a$ that vanish as
$T\rightarrow0$, and $\lambda_b$, $b=N-m_{0}+1,\ldots,N$ refers to the remaining
$N-m_0$ eigenvalues which stay finite.  Here, $\lambda_{x}$ denotes the
$x^{\mathrm{th}}$ eigenvalue and $v_{k}^{x}$ is the $k^{\mathrm{th}}$ component of
the corresponding normalized eigenvector $\vec{v}^{x}$ of $A$.

\begin{figure*}
  \begin{center}
    \includegraphics[scale=1, trim=0 13 0 0]{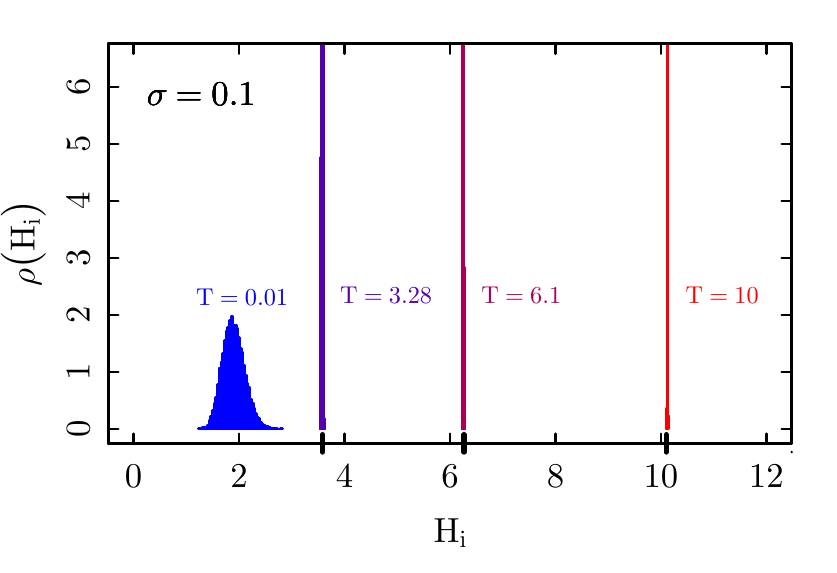}
    \includegraphics[scale=1, trim=0 13 0 0]{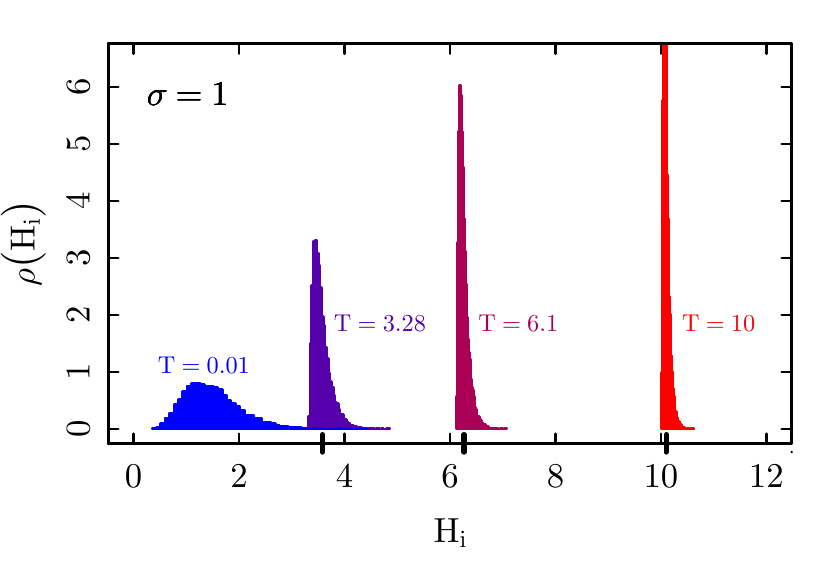}
  \end{center}
  \caption
  {
    (Color online).  Density plots for the parameters $H_{i}$ in
    Eq.~\eqref{Eq:the_holy_matrix_A} for $\sigma=0.1$ (left) and $\sigma=1.0$
    (right). In the limit $T\rightarrow0$ they correspond to the (rescaled)
    local fields $|\mathbf{H}_{i}|/\sqrt{m}$.  As is seen here,
    Eq.~\eqref{Eq:corr_mat_H_i_highT} yields suitable starting values for the
    iteration at high temperatures, cf.\ the bold ticks on the $H_{i}$ axis.  For
    increasing values of $\sigma$ the distribution broadens. The normalization of the
    histograms is arbitrary. The system size used for this plot was $N=1448$.
    \label{Fig:corr_mat_H_i_vs_T}
  }
\end{figure*}

Working at $m=m_0$ and using the reasonable assumption that the $a^{\mathrm{th}}$
component of the spin $\mathbf{S}_{i}\in\mathbb{R}^{m_{0}}$ has the form
\begin{equation}
  \langle S_{i}^{a} \rangle = \pm \sqrt{\frac{m_{0}T}{\lambda_{a}}} v_{i}^{a},
  \label{Eq:avg_expct_val_of_spin_component}
\end{equation}
the cumulant part of the correlation function can be identified with
\begin{equation}
  \tilde{C}_{ij} = T \sum_{b} \frac{v_{i}^{b}v_{j}^{b}} {\lambda_{b}},
  \label{Eq:cumulant_part_of_corr_fct}
\end{equation}
being a function of the non-vanishing eigenvalues. It is natural to define the
spin-glass susceptibility in terms of this connected correlation function,
\begin{equation}
  \begin{split}
    \chi_{\mathrm{SG}} &= \frac{1}{Nm^2}\sum_{i,j}\left[\langle\mathbf{S}_{i}\cdot\mathbf{S}_{j}\rangle
      - \langle\mathbf{S}_{i}\rangle\cdot\langle\mathbf{S}_{j}\rangle\right]^2 \\
    &= \frac{1}{N} \sum_{ij} \tilde{C}_{ij}^{2} = \frac{T^2}{N} \sum_{b} \frac{1}{\lambda_{b}^2},
  \end{split}
  \label{Eq:chi_SG_cumul_corr_fct}
\end{equation}
where the $\lambda_{b}$ are the non-vanishing eigenvalues in the limit $T\rightarrow
0$. Furthermore, the Edwards-Anderson order parameter is then given by
%\begin{eqnarray}
\begin{equation}
  \begin{split}
  q_{\mathrm{EA}} &= \frac{1}{N} \sum_{i} \frac{\langle\mathbf{S}_{i}\rangle\cdot\langle\mathbf{S}_{i}\rangle}{m_0} %\label{Eq:qEA_part_of_corr_fct_1} 
  =  \frac{T}{N} \sum_{i} \sum_{a} \frac{v_{i}^{a}v_{i}^{a}} {\lambda_{a}} \\ %\label{Eq:qEA_part_of_corr_fct_2} \\
  &=  \frac{T}{N} \sum_{a} \frac{1} {\lambda_{a}},
  \end{split}
  \label{Eq:qEA_part_of_corr_fct_3}
\end{equation}
%\end{eqnarray}
where the $\lambda_{a}$ are the vanishing eigenvalues as $T\rightarrow 0$.

The chosen formalism leads to a non-vanishing order parameter at $T=0$. To see this, consider
\begin{equation}
  \frac{1}{N}\sum_{i=1}^{N} C_{ii} = \frac{T}{N}\sum_{a} \frac{1}{\lambda_{a}} + \frac{T}{N}\sum_{b}
  \frac{1}{\lambda_{b}} = 1
  \label{Eq:sum_Cii}
\end{equation}
resulting from the normalization condition $C_{ii} = 1$,
Eq.~\eqref{Eq:Cii_is_one}. The finite eigenvalues $\lambda_b$ scale to a constant as
$N\rightarrow\infty$ and $T\rightarrow 0$, hence the second sum in
Eq.~\eqref{Eq:sum_Cii} is roughly proportional to $N^{-1}N^{1-\mu} = N^{-\mu}$ and
thus vanishes in the thermodynamic limit. Hence, $q_\mathrm{EA} = 1$ in this limit.
The ``zero'' eigenvalues $\lambda_a$ vanish as $T\rightarrow 0$ and as
$N\rightarrow\infty$. Assuming them to be proportional to $T/N^p$, we conclude from
Eq.~\eqref{Eq:sum_Cii} that $p = 1-\mu$.

\subsection{Numerical approach}

\begin{table}[b]
  \caption
  {
    Realizations used for the $T>0$ calculations.
    \label{Tab:corrmat_info}
  }
  \begin{ruledtabular}
    \begin{tabular}{lcc}
      \toprule
      $\sigma$ 	&		& \multicolumn{1}{c}{samples$/10^{3}$}	 \\
      \hline
      0.1		&		& 0.9-2.4	\\	
      0.2		&		& 0.7-1.5 	\\
      0.3		&		& 0.9-2.0 	\\
      0.4		&		& 0.9-2.0 	\\
      0.5		&		& 0.9-1.4 	\\
      0.51		&		& 0.8-1.3 	\\
      0.54		&		& 0.7-2.3 	\\
      0.57		&		& 0.7-2.4 	\\
      0.6		&		& 1.0-3.2 	\\
      \sfrac{2}{3} 	&	& 0.9-1.3 	\\
      0.7		&		& 0.8-1.4 	\\
      0.73		&		& 1.0-2.9 	\\
      0.75		&		& 1.0-1.9 	\\
      0.77		&		& 1.0-2.8 	\\
      0.8		&		& 0.9-2.8 	\\
      0.85		&		& 0.9-2.8 	\\
      0.9		&		& 0.7-2.0 	\\
      1.0		&		& 0.2-2.0 	\\
    \end{tabular}
  \end{ruledtabular}
\end{table}

We solve Eqs.~\eqref{Eq:the_holy_matrix_A}, \eqref{Eq:C_as_fct_of_A} and
\eqref{Eq:Cii_is_one} iteratively using the Newton-Raphson method for systems with
$\sigma<0.85$, see the discussion in Ref.~\onlinecite{lee:05}.  For larger $\sigma$, this
approach has some numerical instabilities leading to singular matrices in the course
of the $LU$ decomposition. We switched to a method using a $QR$ decomposition
similarly to the way Broyden's method \cite{numrec} is usually implemented. However,
in both cases we utilized the exact Jacobian to speed up calculations. Next we will
show that this extra speed-up comes for free.  According to
Eqs.~\eqref{Eq:the_holy_matrix_A}, \eqref{Eq:C_as_fct_of_A} and
\eqref{Eq:Cii_is_one} we can proceed by introducing  the $N$ functions
\begin{equation}
f_{i}(\{H_{k}\})=T(A^{-1})_{ii}-1 
\end{equation}
and solving for their zeroes. Taking into account $A^{-1}A=\mathbf{1}$ we have
\begin{equation}
  \frac{\partial \mathbf{1}}{\partial H_{j}} = \frac{\partial A^{-1}}{\partial H_{j}}A + A^{-1}\frac{\partial A}{\partial H_{j}}=\mathbf{0},
  \label{Eq:Jacob_deriv_of_const}
\end{equation}
so that after multiplying with $A^{-1}$ from the right and solving for 
the needed derivative we arrive at
\begin{equation}
  \frac{\partial A^{-1}}{\partial H_{j}} = -A^{-1}\frac{\partial A}{\partial H_{j}}A^{-1}.
  \label{Eq:Jacob_deriv_of_Ainv_A}
\end{equation} 
By virtue of Eq.~\eqref{Eq:the_holy_matrix_A} it is
\begin{equation}
  \left(\frac{\partial A}{\partial H_{j}}\right)_{nm} =\delta_{jn}\delta_{mn},
  \label{Eq:deriv_of_A}
\end{equation}
such that we finally find
\begin{eqnarray}
  \left(\frac{\partial A^{-1}}{\partial H_{j}}\right)_{ii} &=& -\sum_{n}\sum_{m}\left(A^{-1}\right)_{in}\delta_{jn}\delta_{mn}\left(A^{-1}\right)_{mi} \nonumber \\
	&=& -\left(A^{-1}\right)_{ij}\left(A^{-1}\right)_{ji} = -\left(A^{-1}\right)_{ij}^{2}.
  \label{Eq:Jacob_deriv_of_Ainv_B}
\end{eqnarray}
Hence the desired Jacobian reads
\begin{equation}
  (J_{f})_{ij} = \frac{\partial f_{i}}{\partial H_{j}} = -T\left(A^{-1}\right)_{ij}^{2}.
  \label{Eq:Jacobian}
\end{equation}
\\
We start the iterations at high temperature where we expect \cite{aspelmeier:04a}
\begin{equation}
  H_{i} = T + 1/T \label{Eq:corr_mat_H_i_highT}
\end{equation}
for the SK model, resulting in reasonable initial values also for the 1d long-range
system with $\sigma > 0$ considered here. This is illustrated in
Fig.~\ref{Fig:corr_mat_H_i_vs_T} for $\sigma = 0.1$ and $\sigma = 1$. For all systems
considered here, we found a starting temperature of $T_{1}=10$ to be a suitable and
sufficient choice.  The set of temperature points may be chosen in a geometrical
fashion, such that the inverse temperature $\beta=1/T$ is distributed
equidistantly. In all our simulations the lowest temperature was chosen to be
$T_{f}=0.01$. On decreasing the temperature, one may use the converged result of the
previous calculation as a starting point. Alternatively, decreasing the temperature
from $T_{k}$ to $T_{k+1}$ ($k>0$, $T_{k+1}<T_{k}$) a new guess for the values of the
$H_{i}$ can be obtained by using the differential equation \cite{aspelmeier:04a}
\begin{eqnarray}
  \dfrac{\operatorname{d}H_{i}}{\operatorname{d}\beta} &=& -\sum_{j}(B^{-1})_{ij},
  \label{Eq:corr_mat_diffEq_Tguess} \\
  B_{ij} &=& (\beta C_{ij})^2.
\end{eqnarray}

\begin{figure}
  \begin{center}
    \includegraphics[scale=1, trim=0 13 0 0]{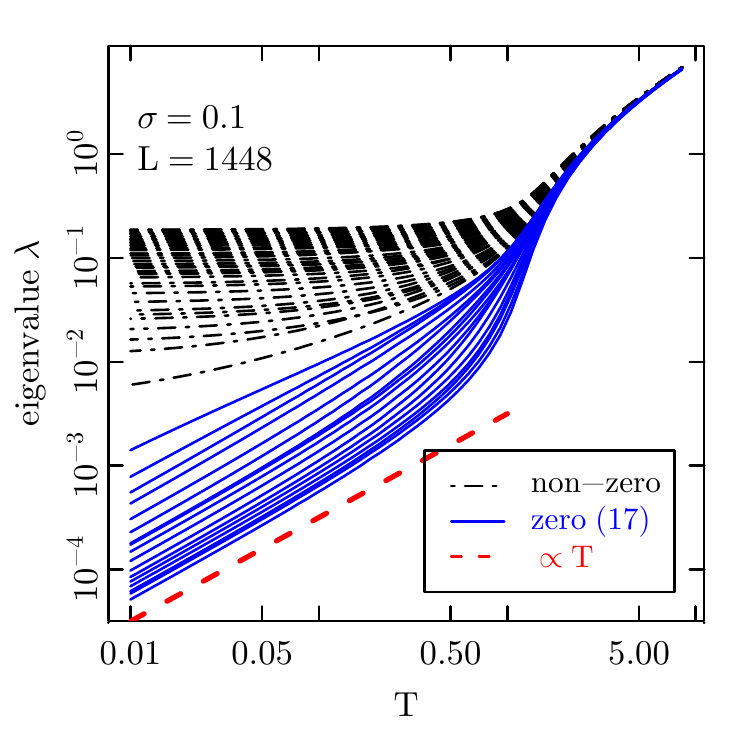}\\
  \end{center}
  \caption
  { (Color online). The evolution of eigenvalues $\lambda$ of the matrix $A$ of
    Eq.~\eqref{Eq:the_holy_matrix_A} with temperature for a $L=1448$ system with
    $\sigma = 0.1$. The non-zero eigenvalues (black, dot-dashed) stay finite in the
    limit $T\rightarrow0$ (biggest are left away).  The others (blue, solid) scale to zero as $\sim
    T$. Note that both axes have logarithmic scales.
    \label{Fig:eigenval_scaling}
  }
\end{figure}

\begin{figure}
  \begin{center}
    \includegraphics[scale=1, trim=0 10 0 0]{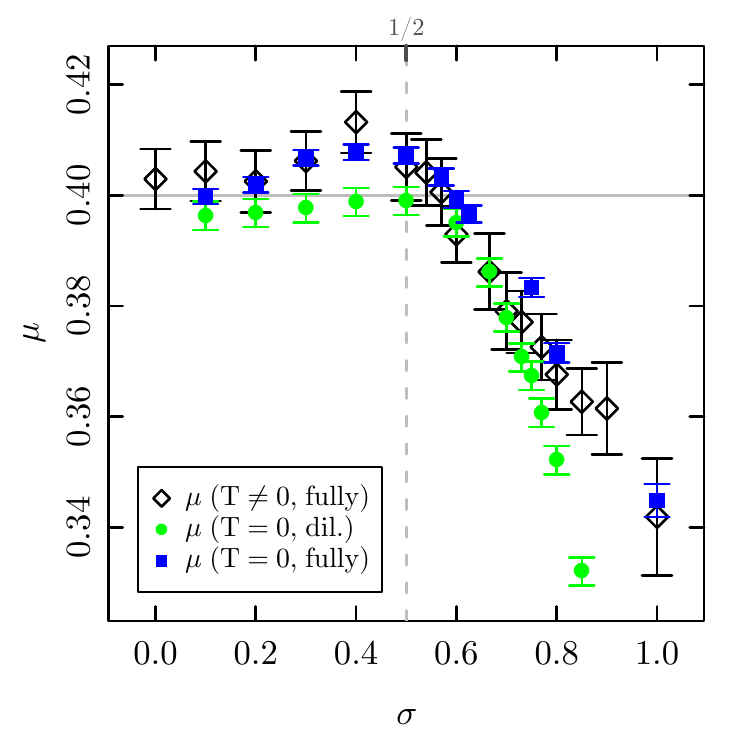}\\
  \end{center}
  \caption
  { (Color online). Spin-dimension exponent $\mu$ calculated from the eigenvalue
    density at finite temperatures ($T>0$) as compared to the result reported in
    Sec.~\ref{Sec:GS_props} from the ground-state computations ($T=0$). For $\sigma=0$ the SK
    model was considered directly. For the fully connected model, the results of the
    calculations at $T=0$ and for $T>0$ are well compatible. 
    \label{Fig:NxN_mu_fit}
  } 
\end{figure}

To perform the analysis outlined above in Sec.~\ref{sec:AM_formalism}, we need to
separate zero from non-zero eigenvalues. For finite temperatures and finite systems,
however, no eigenvalues are exactly zero. Instead, there is a difference in scaling
behavior between the ``zero'' eigenvalues that vanish proportional to $T/N^{1-\mu}$
and the other eigenvalues that scale to a constant. This is illustrated in
Fig.~\ref{Fig:eigenval_scaling}. To automatically distinguish between the two types
of eigenvalues, one might count all those as zero that fall below a chosen threshold
at the lowest considered temperature. It turned out to be more reliable, however, to
base the distinction criterion directly on the temperature scaling $\propto T$ of the
``zero'' eigenvalues. Determining the slope of $\log\lambda(T)$ at the lowest
considered temperatures, we counted those eigenvalues as scaling to zero whose slope
was above $0.5$. Even with this rather reliable criterion, however, there will
always be a certain ambiguity as the slopes change quite continuously over the
different eigenvalues, and a number of borderline cases always exists, cf.\ the
example in Fig.~\ref{Fig:eigenval_scaling}. We do not find any signs of the number of
zero eigenvalues changing with temperature. Instead, our results are compatible with
all relevant eigenvalues starting to scale to zero as soon as $T<T_\mathrm{SG}$.

\begin{figure*}
  \begin{center}
    \includegraphics[scale=1, trim=0 13 0 0]{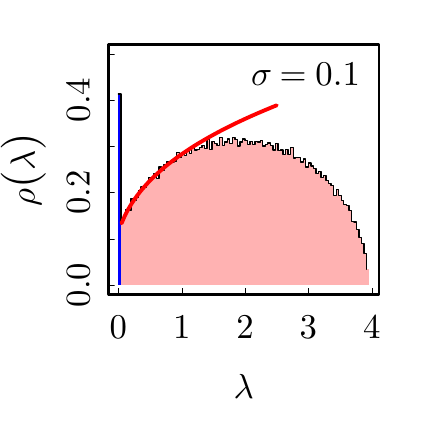}
    \includegraphics[scale=1, trim=0 13 0 0]{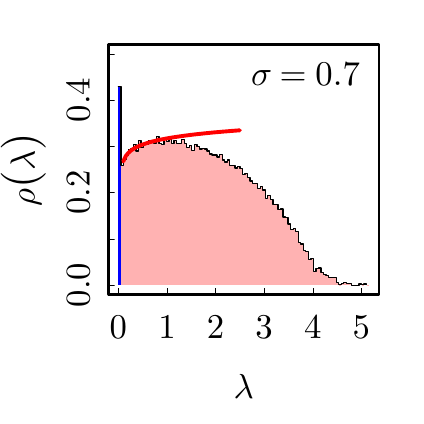}
    \includegraphics[scale=1, trim=0 13 0 0]{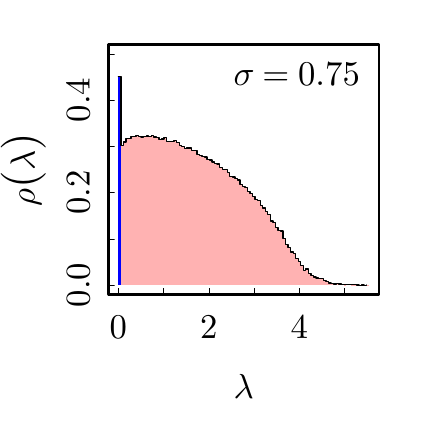}
    \includegraphics[scale=1, trim=0 13 0 0]{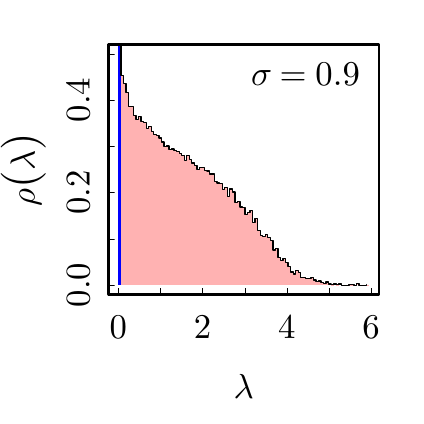}
  \end{center}
  \caption
  { (Color online).
    The eigenvalue distribution of the matrix $A$, cf.\
    Eq.~\eqref{Eq:the_holy_matrix_A}, at temperature $T=0.01$ for different values of
    the power-law exponent $\sigma$ ($N=1448$).  Wigner's semicircle form holds for the
    infinite-range region $0<\sigma\le1/2$ in the thermodynamic limit. The red, bold
    line shows fits of the functional form \eqref{Eq:eigenval_density_fit} to the
    data. A positive exponent $x$ results for $\sigma<3/4$.
    \label{Fig:semicircle_various_sigmas}
  } 
\end{figure*}

The analysis of the eigenvalue density allows for an alternative method of
calculating the spin dimension scaling exponent $\mu$ already discussed in
Sec.~\ref{Sec:GS_props}. To implement it, we determined the rank of the matrix $M$
composed of the ground-state spin vectors. Instead, we could have extracted the local
field values $|\mathbf{H}_{i}|$, fed them into Eq.~\eqref{Eq:the_holy_matrix_A} and
determined the number of zero eigenvalues. These approaches are equivalent since the
rows of $M$ correspond to the null eigenvectors of $A$ and the row and column ranks
of a matrix are identical \cite{aspelmeier:04a}. This would not have allowed us to
consider sufficiently large systems, however, since there an $N\times N$ matrix must
be inverted, which is in contrast to the $T=0$ ground-state calculations where it was
sufficient to determine the rank of the auxiliary, but smaller, $m\times N$
ground-state spin matrix $M$. The results for $m_0$ extracted from the
finite-temperature calculation are summarized in Fig.~\ref{Fig:NxN_mu_fit}. The
outcomes are mostly compatible with those of the zero-temperature approach for the
case of the fully connected model. In contrast, the $T=0$ results for the diluted
model systematically deviate from those of the fully connected model for $\sigma > 1$
as discussed above in Sec.~\ref{Sec:Defectenergies}, and some signs of this
non-universality are already seen for $\sigma \gtrsim 0.8$.

\subsection{Distribution of eigenvalues\label{sec:EV_distribution}}

Braun and Aspelmeier \cite{braun:06} suggested using the eigenvalue spectrum for a
more general understanding of scaling corrections for the case of the two competing
limits $N\rightarrow\infty$ and $m\rightarrow\infty$. Their analysis is valid for the
system on a Bethe lattice, but some results might generalize to the model considered
here. They discuss the ground-state energy $e(m,N)=E/Nm$ per spin and spin component,
which is argued to have two contributions: the ground-state energy in the limit
$N\rightarrow\infty$ with $m=m_0$ large and fixed,
$e_\infty+\frac{1}{4}m_0^{-y}+O(m_0^2)$, and the additional energy required for
forcing the $N$ spins into an $m_0$ dimensional subspace. This second contribution is
proportional to the required shift of the eigenvalue spectrum $\rho(\lambda)$ to push
$m_0$ eigenvalues to zero. Assuming the density at small $\lambda$ to scale as
$\rho(\lambda) \sim \lambda^{x}$, \cite{bray:82b} the first $m_0$ eigenvalues occupy
the interval $[0, (m_0/N)^{1/(1+x)}]$, such that the total energy is
\begin{equation}
  \label{eq:aspelmeier_energy}
  e(m,N) = e_\infty+c_1m_0^{-y}+c_2\left(\frac{m_0}{N}\right)^{1/(1+x)}.
\end{equation}
Minimizing with respect to the number $m_0$ of spin components yields the scaling
relation
\begin{equation}
  \mu=\frac{1}{y(x+1)+1}. \label{Eq:scaling_law_1}
\end{equation}
For the SK model with $x=1/2$ (the Wigner semicircle law) and $y=1$, \cite{bray:81}
we therefore arrive at the observed $\mu = 2/5$ as desired. This scaling should hold
independent of lattice structure.

\begin{figure}[tb]
  \begin{center}
    \includegraphics[scale=1, trim=0 10 0 0]{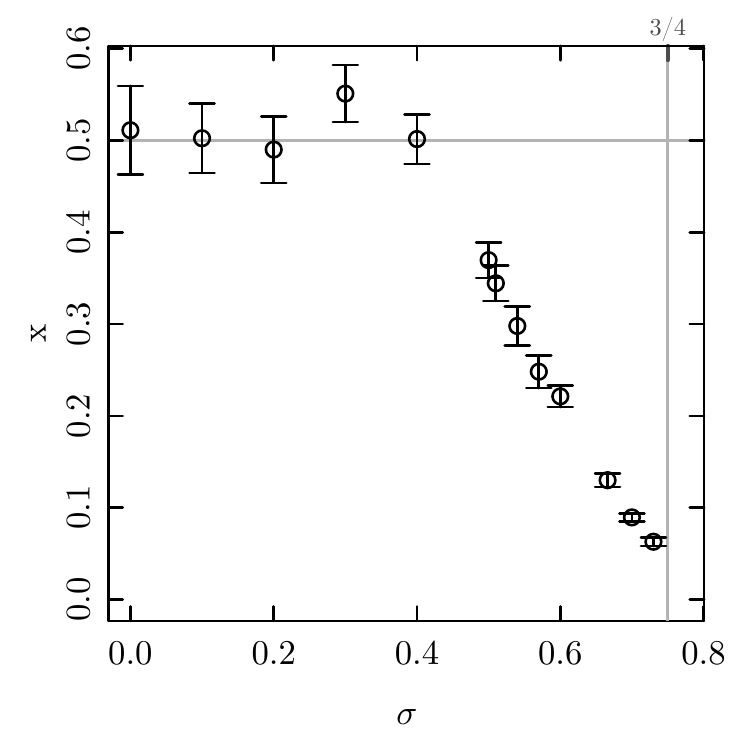}
  \end{center}
  \caption
  {(Color online). The density exponent $x$ as determined by fitting a power-law,
    Eq.~\eqref{Eq:eigenval_density_fit}, to the eigenvalue densities shown in
    Fig.~\ref{Fig:semicircle_various_sigmas}.
    \label{Fig:semicircle_nu1_for_various_sigmas}
  } 
\end{figure}

We determined the exponent $x$ of the density of eigenvalues as $\lambda \rightarrow
0$ from fits of the functional form
\begin{equation}
  \rho(\lambda) = a(\lambda+\Delta\lambda)^{x}
  \label{Eq:eigenval_density_fit}
\end{equation}
to the data. Here, the shift $\Delta\lambda$ is required to take the zero eigenvalues
into account. To perform the fits, eigenvalues from a large number of disorder
samples were accumulated, and the limit $\lambda\rightarrow 0$ was modeled by
successively omitting more of the larger eigenvalues while monitoring the resulting
estimate of $x$ as well as the goodness-of-fit. Statistical errors on the fit results
were determined using a sophisticated jackknifing analysis \cite{efron:book}. Some
example results are collected in Fig.~\ref{Fig:semicircle_various_sigmas}.  Fits of
this functional form are possible for $\sigma < 3/4$, where the vanishing of the
phase transition is signaled by $x=0$. This is expected since at the upper critical
value $\sigma_u = 3/4$ the vanishing of eigenvalues at any finite temperature ceases
to exist \cite{bray:82b}. Collecting the results for all values of $\sigma$
considered, we arrive at the data shown in
Fig.~\ref{Fig:semicircle_nu1_for_various_sigmas} which confirms our expectations of
$x=1/2$ for $\sigma\le 1/2$ and $x=0$ for $\sigma=3/4$. In view of the above
expectations which are in-line with our numerical results, it is tempting to
speculate that $x(\sigma) = 3/2-2\sigma$, but we have not been able to substantiate
this claim with a theoretical argument.

\begin{figure}[tb]
  \begin{center}
    \includegraphics[scale=1, trim=0 10 0 0]{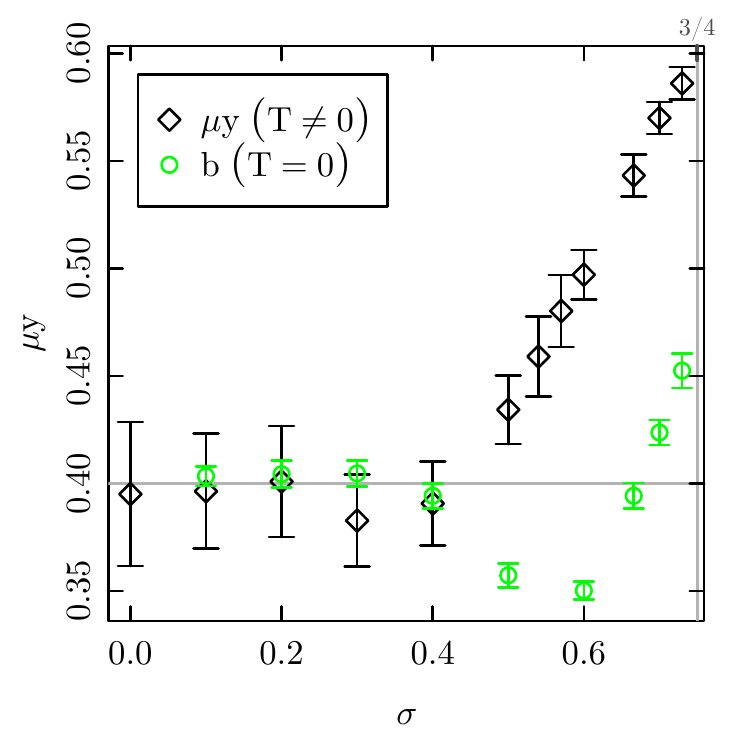}
  \end{center}
  \caption
  { (Color online). A comparison of the exponent $z=(1+\mu)/(1+x)$ determined from
    the $T> 0$ calculations (black diamonds) with the ground-state energy correction
    exponent $b$ as determined in Sec.~\ref{Sec:GS_props} (green circles).
    \label{Fig:semicircle_compare_nu_to_b}
  } 
\end{figure}

\begin{figure*}
  \begin{center}
    \includegraphics[scale=1, trim=5 13 10 0]{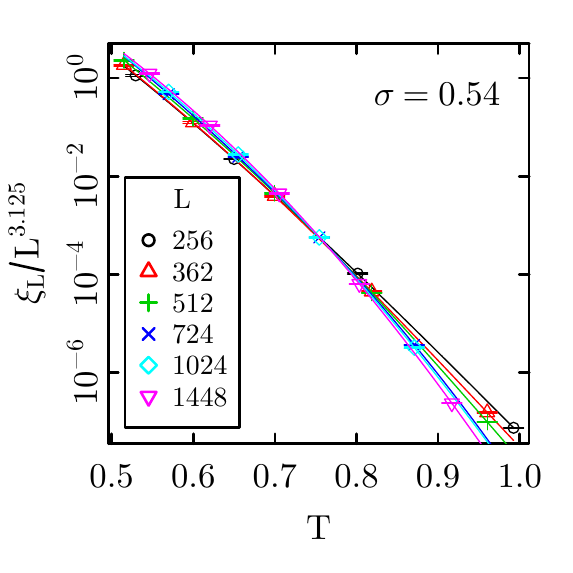} \hspace{10pt}
    \includegraphics[scale=1, trim=5 13 10 0]{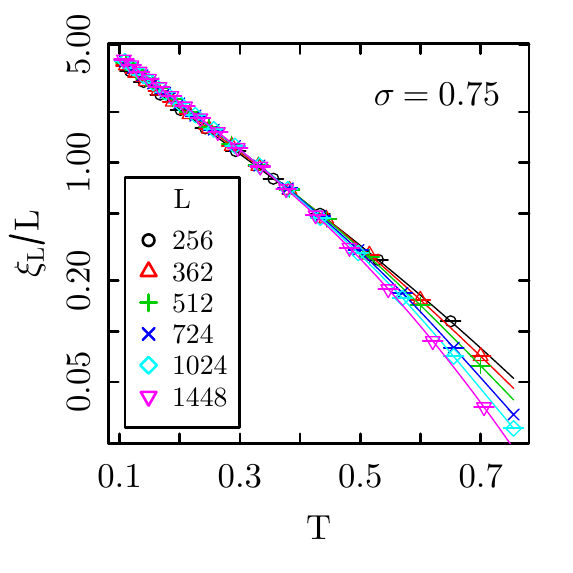} \hspace{10pt}
    \includegraphics[scale=1, trim=5 13 10 0]{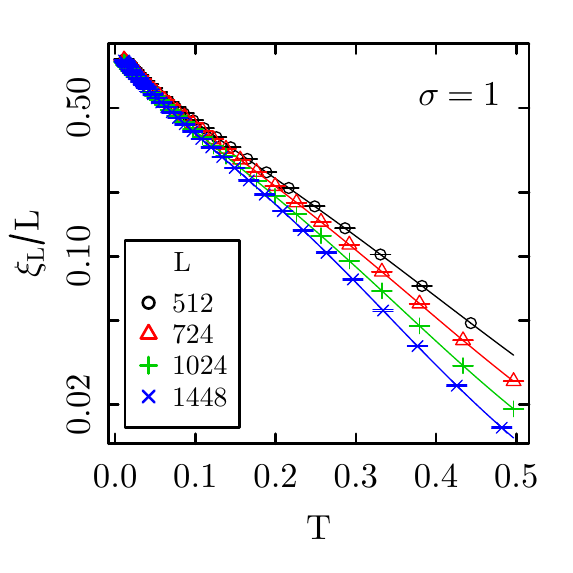}
  \end{center}
  \caption
  { (Color online). The spin-glass correlation length $\xi_L/L$ (for $\sigma \ge
    5/8$) and $\xi/L^{\nu/4}$ (for $1/2 < \sigma < 5/8$), respectively, for different
    lattice sizes $L$. For $\sigma<3/4$ lines show a clear crossing point. The
    crossing vanishes at $\sigma=3/4$, where the lines for lattice sizes $L=N>362$
    lie approximately on top of each other below a certain temperature. }
  \label{Fig:xi_crossing_wo_scaling_1}
\end{figure*}

Additionally, the authors of Ref.~\onlinecite{braun:06} suggest the scaling ansatz
\begin{equation}
  e(m,N)-e_\infty = m^{-y}F(mN^{-\mu})
\end{equation}
where $F(x)$ is a scaling function. In the relevant limit of $m\rightarrow\infty$ before
$N\rightarrow\infty$, this implies a scaling of the ground-state energy according to
\begin{equation}
  e(m=\infty,N) - e_{\infty} \sim N^{-z}
\end{equation}
with $z=\mu y = (1-\mu)/(1+x)$, where we used Eq.~\eqref{Eq:scaling_law_1}. In
Fig.~\ref{Fig:semicircle_compare_nu_to_b} we show the correction-to-scaling exponent
$b$ for the ground-state energies as determined in Sec.~\ref{Sec:GS_props} in
comparison to $z = (1-\mu)/(1+x)$ as determined from the results of $\mu$ and $x$ for
our system. Both exponents agree for $\sigma \le 1/2$ where $b = z = 2/5$. For larger
values of $\sigma$, however, $b$ is consistently smaller than $z$. Therefore, if the
corrections predicted here are present, they are sub-leading and cannot be resolved
by our numerical analysis.

\subsection{Critical behavior}

\begin{figure*}[t]
  \begin{center}
    \includegraphics[scale=1, trim=5 13 10 0]{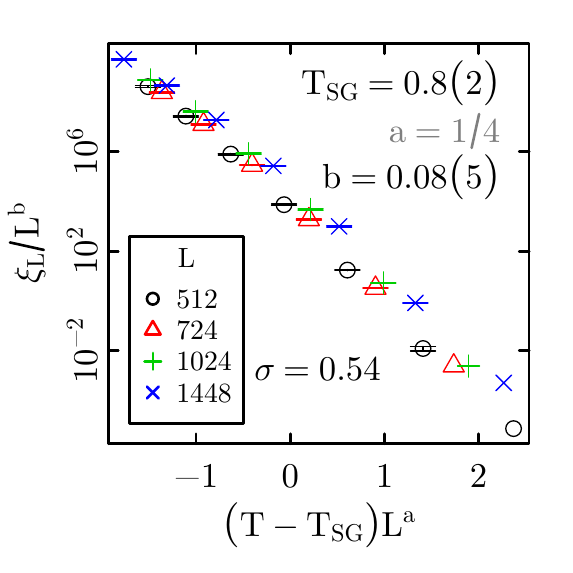} \hspace{10pt}
    \includegraphics[scale=1, trim=5 13 10 0]{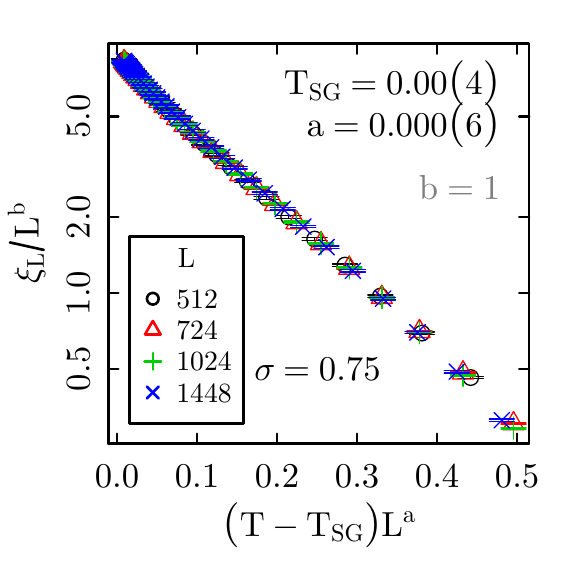} \hspace{10pt}
    \includegraphics[scale=1, trim=5 13 10 0]{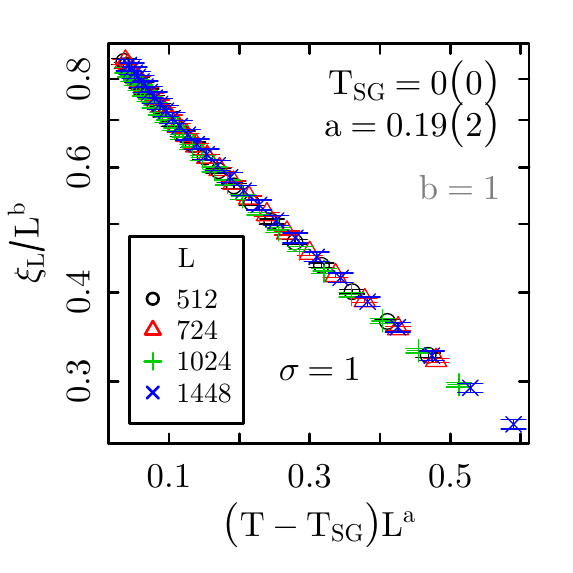}
  \end{center}
  \caption
  { (Color online). Scaling collapse of the correlation length ratios $\xi_L/L$ and
    $\xi_L/L^{\nu/4}$, respectively. For optimizing parameters, a general form $\xi_L
    = L^b{\cal X}(tL^a)$ was assumed.  For $\sigma > 3/4$ the transition temperature
    was fixed at $T_\mathrm{SG} = 0$, since the according collapses yielded values
    around zero.  The resulting parameters are summarized in
    Fig.~\ref{Fig:xi_data_collapse_results_1}.  
}
  \label{Fig:xi_data_collapse_1}
\end{figure*}

We now turn to studying the behavior of the physical observables extracted from the
solution to the saddle-point equations in the vicinity of the critical point. The
analysis of the spin-glass correlation length, the Edwards-Anderson order parameter
and the spin-glass susceptibility allows us to compare our simulations to the
theoretical predictions outlined in Sec.~\ref{Sec:1dpl_mInf_SG}.

\subsubsection{Correlation length}

\begin{figure*}
  \begin{center}
    \includegraphics[scale=1, trim=5 13 10 0]{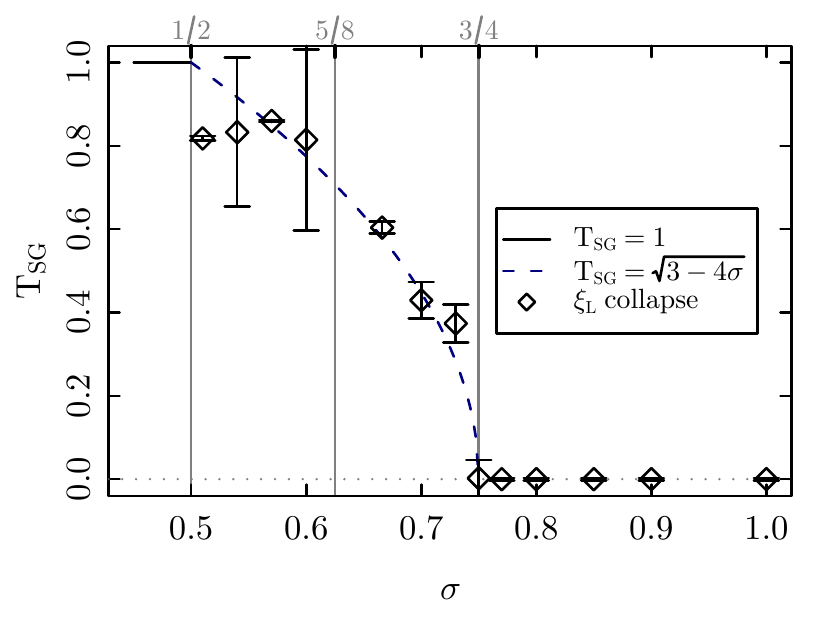} \hspace{10pt}
    \includegraphics[scale=1, trim=5 13 10 0]{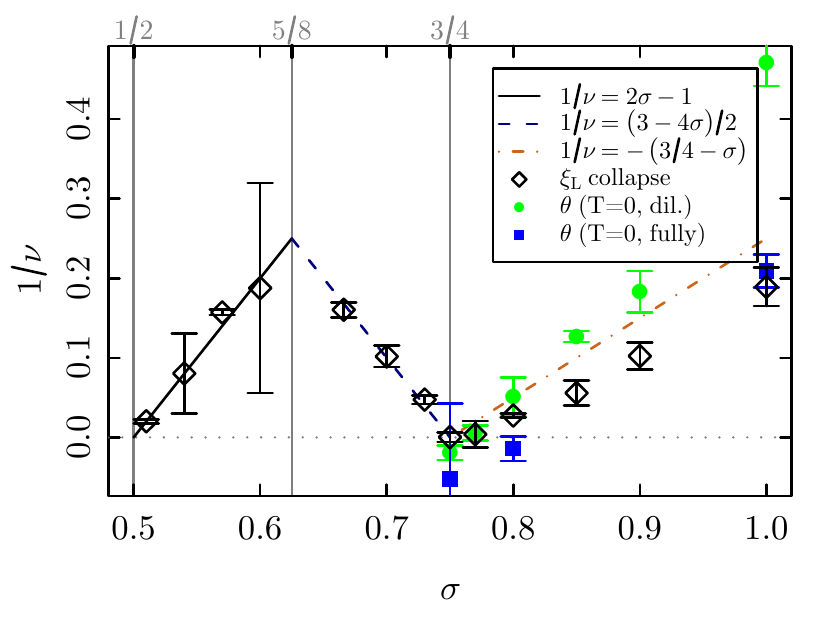}
  \end{center}
  \caption
  { (Color online). Spin-glass transition temperature $T_\mathrm{SG}$ and correlation length exponent
    $\nu$ as estimated from collapses of the correlation length data according
    to the functional form Eq.~\eqref{eq:FSS_corr_len}. Left panel: transition
    temperature $T_\mathrm{SG}$ as compared to the prediction $T_\mathrm{SG} =
    \sqrt{3-4\sigma}$ of Eq.~\eqref{Eq:Tc_prediction_MF_nonMF}. Right panel: $1/\nu$
    and the relations $1/\nu = 2\sigma-1$ valid in the mean-field region
    $1/2<\sigma\le 5/8$, cf.~Eq.~\eqref{eq:mf_nu_ising}, and $1/\nu \approx (3-4\sigma)/2$,
    cf.~Eq.~\eqref{eq:nu_minfty_nonmf}. For $\sigma > 3/4$, where $T_\mathrm{SG} =
    0$, we additionally show $1/\nu = -\theta$ with our estimates of $\theta$
    resulting from the defect-energy calculations, as well as the conjectured $1/\nu
    = -\theta = \sigma-3/4$.
  }
  \label{Fig:xi_data_collapse_results_1}
\end{figure*}

The spin-glass correlation function can be calculated from the spin-spin correlation
function \eqref{Eq:C_as_fct_of_A} via
%\[
\begin{equation}
G_\mathrm{SG}(r) = \frac{1}{L}\sum_{r_{ij}^{\shortmid}=r}\ls C_{ij}^2\rs_\mathrm{av}
= \frac{T^2}{L}\sum_{r_{ij}^{\shortmid}=r}\ls(A^{-1})_{ij}^2\rs_\mathrm{av}.
\label{eq:SG_corrfn}
\end{equation}
%\]
Note that here we use the algebraic graph distance $r_{ij}^{\shortmid} =
\min(|i-j|,L-|i-j|)$ irrespective of whether the ring or chain geometry is
considered. To arrive at the usual second-moment definition of the correlation
length, we use the Fourier decomposition,
%\[
\begin{eqnarray}
  \chi^0_\mathrm{SG}(k) &=&
  \frac{T^2}{L} \sum_{i,j} \ls(A^{-1})_{ij}^2\rs_\mathrm{av}e^{ik[(i-j)\mod L]} \nonumber \\
  &=& \frac{T^2}{L} \sum_{i,j} \ls(A^{-1})_{ij}^2\rs_\mathrm{av} \cos\left(k[(i-j)\mod L]\right) \nonumber \\
  &=& 2\sum_{r=0}^{\lfloor L/2\rfloor} G_\mathrm{SG}(r) \cos(kr),
\end{eqnarray}
%\]
and plug it into Eq.~\eqref{Eq:xi_normal},
\begin{equation}
  \label{eq:xi_normal_est}
  \xi_L =
  \frac{1}{2\sin(k_\mathrm{min}/2)}\left[\frac{\chi^0_\mathrm{SG}(0)}
    {\chi^0_\mathrm{SG}(k_\mathrm{min})}-1\right]^{1/(2\sigma-1)}.
\end{equation}
Here, $k_\mathrm{min} = 2\pi/L$. In practice, we determine $G_\mathrm{SG}(r)$ per
disorder realization from the saddle-point equations. For space efficiency, storing
$G_\mathrm{SG}(r)$ is then preferable over storing $C_{ij}$ directly. Note that since
we are using the disconnected correlation function here, the estimators
\eqref{eq:xi_normal_est} only represent the correlation length above
$T_\mathrm{SG}$. Close to criticality, we expect the scaling form
\begin{equation}
  \label{eq:FSS_corr_len}
  \xi_{L} \sim \left\{
    \begin{array}{rl}
      L^{\nu/4}\mathcal{X}(tL^{1/4}),	& 1/2<\sigma\le5/8,\\
      	L\,\mathcal{\mathcal{X}}(tL^{1/\nu}), & \sigma > 5/8.
    \end{array}
    \right.
\end{equation}
In the ordered phase, on the other hand, $\xi_L$ diverges even more strongly with the
system size \cite{ballesteros:00}. As a consequence, the curves for $\xi_L/L$
($\sigma \ge 5/8$) and $\xi_L/L^{\nu/4}$ ($1/2 < \sigma < 5/8$), respectively, will
cross in the vicinity of the critical temperature.

\begin{figure*}
  \begin{center}
    \includegraphics[scale=1, trim=0 10 5 0]{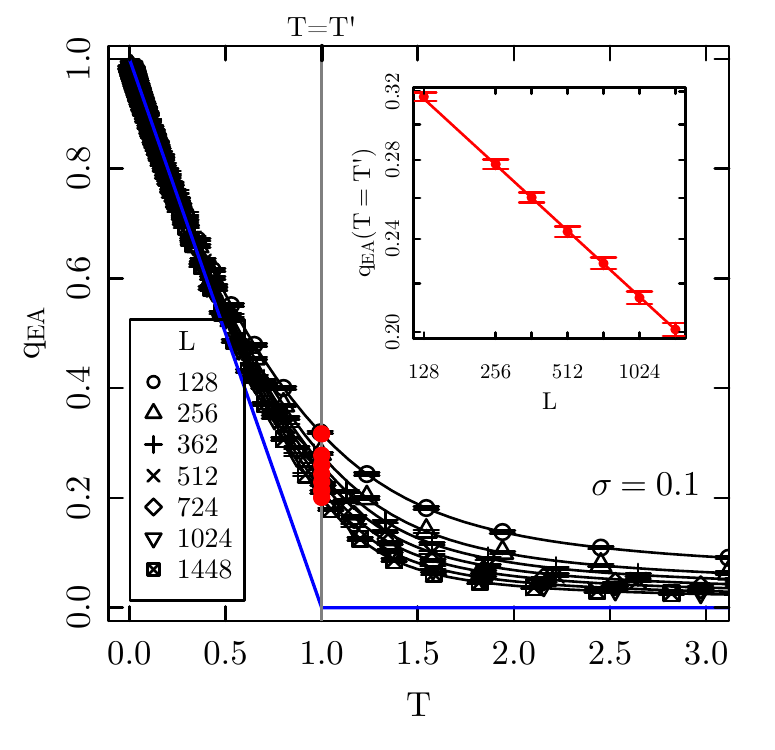}
    \hspace{50pt}
    \includegraphics[scale=1, trim=5 10 0 0]{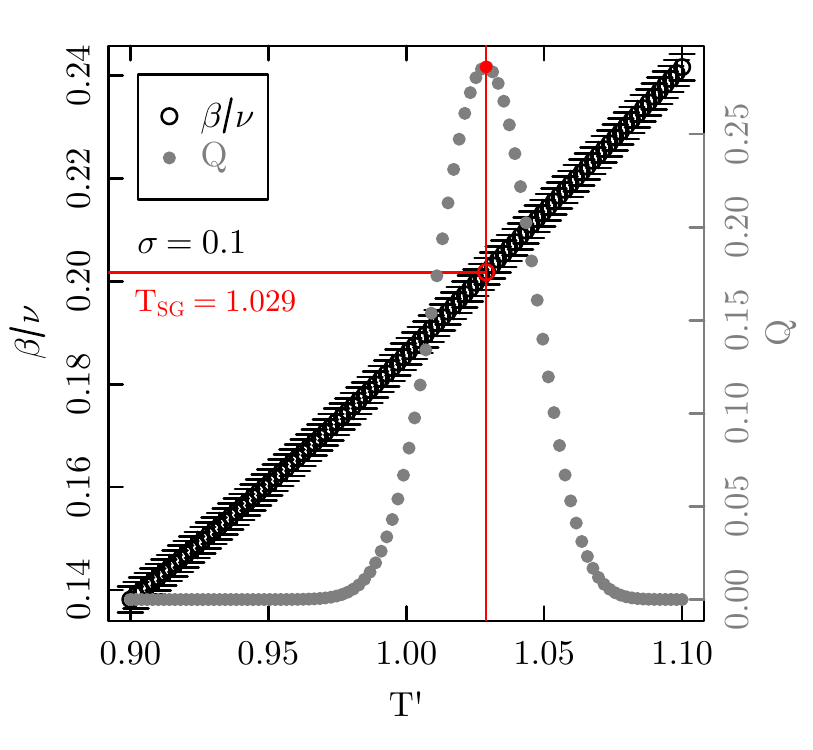}
  \end{center}
  \caption
  { (Color online). Scaling of the Edwards-Anderson order parameter $q_{\mathrm{EA}}$. The left panel
    shows the dependence of $q_{\mathrm{EA}}$ on temperature for $\sigma=0.1$.  For
    $L\to\infty$, we expect the form \eqref{eq:qea_infty} which is indicated by the
    solid blue line. Employing fits of the functional form \eqref{Eq:qEA_FSS_fit} for
    a temperature interval $T_{i}\le T'\le T_{f}$ around the expected critical
    temperature (see inset of left panel), we determine $T_\mathrm{SG}$ and
    $\beta/\nu$ from the point where such fits work best, which is monitored by the
    quality-of-fit parameter $Q$, shown in the right panel.  For a summary of results
    for all $\sigma$ see Fig.~\ref{Fig:results_from_qEA_FSS_1}.
    \label{Fig:qEA_vs_T_showcase_1}
  }
\end{figure*}

Plots of $\xi_L/L$ and $\xi_L/L^{\nu/4}$ resulting from the finite-temperature
calculations with parameters summarized in Tab.~\ref{Tab:corrmat_info} for three
examples of $\sigma$ are shown in Fig.~\ref{Fig:xi_crossing_wo_scaling_1}. Below
$\sigma_u = 3/4$, we find a crossing of the curves as shown for $\sigma = 0.54$. For
$\sigma > 3/4$, on the other hand, the curves only merge in the limit $T\rightarrow
0$. At the critical $\sigma$, we see a merging of the curves with an onset
temperature scaling to $0$ as $L\to\infty$. We note that the scaling of
$\xi_L/L^{\nu/4}$ is hard to observe numerically for $\sigma$ close to $1/2$, where
$\nu = 1/(2\sigma-1) \to \infty$, such that a crossing point of $\xi_L/L^{\nu/4}$ is
not visible for the system sizes considered here for $\sigma = 0.51$. It is possible
to extract estimates for the spin-glass temperature $T_\mathrm{SG}$ and the
correlation length exponent $\nu$ by re-scaling the data for different system sizes
such that they collapse on the scaling function ${\cal X}(x)$. We used two
complementary approaches for performing this collapse: method (a) consists of a joint
fit of all data sets to a third-order polynomial approximating the scaling function
in the chosen regime; method (b) is the collapsing procedure suggested in
Ref.~\onlinecite{houdayer:04} which, in turn, is based on
Ref.~\onlinecite{bhattacharjee:01}. In both cases, we performed the collapse on the
{\em logarithm\/} of the actual data. This turns out to be necessary since, in
particular for small $\sigma$, $\xi_L/L$ spans many orders of magnitude in the range
of temperatures considered here. In some cases, we also employed weights of the data
points involved that decay exponentially away from the adaptively chosen value of the
critical temperature. Statistical errors on the collapse parameters have been
determined by a bootstrap sampling over the whole collapsing procedure
\cite{efron:book}. In the region $\sigma > 5/8$, $\nu$ was determined from the scale
$tL^{1/\nu}$ of the abscissa. On the contrary, for the mean-field region $1/2 <
\sigma \le 5/8$, it was determined from the scaling $\xi_L/L^{\nu/4}$ of the
ordinate. For the latter collapses, we find that the expected scaling of $tL^{1/4}$
of the argument of the scaling function is not very well reproduced, and we allow for
this exponent to fluctuate to accommodate scaling corrections.

As illustrated in Fig.~\ref{Fig:xi_data_collapse_1}, these collapses work rather well
over the whole range of $\sigma$. The resulting estimates of the correlation length
exponent $\nu$ and the critical temperature are summarized in
Fig.~\ref{Fig:xi_data_collapse_results_1}. The transition temperature is consistent
with $T_\mathrm{SG} = 0$ for $\sigma \ge 3/4$ and approaches $T_\mathrm{SG} = 1$ as
$\sigma \to 1/2$. In between, it is compatible with the estimate $T_\mathrm{SG}
\approx \sqrt{3-4\sigma}$ obtained in Sec.~\ref{Sec:1dpl_mInf_SG}. As mentioned
above, in the mean-field regime with $\sigma \to 1/2+$, finite-size corrections become
very pronounced due to the divergent exponent $\nu$. This leads to rather strong
fluctuations of $\nu$ and $T_\mathrm{SG}$ as estimated from the collapsing
procedures, cf.\ Fig.~\ref{Fig:xi_data_collapse_results_1}. In the right panel of
this figure we also compare our result for $1/\nu$
extracted from collapses for $\sigma > 3/4$ with $-\theta$ from the defect-energy
calculations. In general, we find acceptable agreement between zero- and
finite-temperature calculations. The observed systematic deviations give an
indication of the level of unresolved finite-size corrections. As $\sigma \to 1-$,
results for the diluted system start to systematically deviate from those for the fully
connected system due to the observed non-universality discussed in
Sec.~\ref{Sec:Defectenergies}.

Comparing the estimates for $1/\nu$ and $T_\mathrm{SG}$ for the ring, line and
resummed line geometries introduced in Sec.~\ref{sec:geometry}, we find complete
consistency, cf.\ the example results for $\sigma = 3/4$ collected in Table
\ref{Tab:different_geometries}.

\subsubsection{Edwards-Anderson order parameter}

According to the discussion of FSS in our model, we expect
\begin{equation}
  q_{\mathrm{EA}}\sim\left\{
    \begin{array}{rl}
      L^{-1/4}\mathcal{Q}(tL^{1/4}),	& \sigma \le 5/8,\\
      L^{-\beta/\nu}\mathcal{Q}(tL^{1/\nu}), & \sigma > 5/8.
    \end{array}
  \right.
  \label{Eq:qEA_FSS}      
\end{equation}
for temperatures in the scaling window. In the thermodynamic limit, for $\sigma
<1/2$, $\beta =1$, \cite{dealmeida:78,aspelmeier:04a} while for $\sigma >5/8$, $\beta$
is expected to remain close to unity, so that
\begin{equation}
  q_\mathrm{EA} \approx \left\{
    \begin{array}{rl}
      1-T/T_\mathrm{SG}, & T < T_\mathrm{SG},\\
      0, & T\ge T_\mathrm{SG}.
    \end{array}
  \right.
  \label{eq:qea_infty}
\end{equation}
As is illustrated with the unscaled data in the left panel of
Fig.~\ref{Fig:qEA_vs_T_showcase_1}, these expectations are borne out well by our
results. In particular, the thus defined order parameter becomes unity as $T\to 0$,
in contrast to the differently defined $q_\mathrm{EA}^0$ of Eq.~\eqref{eq:EA_wrong}
and Ref.~\onlinecite{lee:05}.

Right at $T_\mathrm{SG}$, $q_\mathrm{EA}$ scales to zero. To extract $T_\mathrm{SG}$
and determine $\beta/\nu$, we again employed scaling collapses.  Due to the observed
instability of the collapsing procedure, we also developed an independent approach
based on the quality of power-law scaling. Since scaling proportional to
$L^{\beta/\nu}$ is only expected at criticality, the critical point might be
determined under the assumption that it coincides with the temperature where
power-law scaling is best observed. We hence performed fits according to the form
\begin{equation}
	q_{\mathrm{EA}}(T=T') = cL^{\beta/\nu}, \label{Eq:qEA_FSS_fit}
\end{equation} 
for an interval of temperatures $T_{i}\le T'\le T_{f}$ around the expected value of
$T_{\mathrm{SG}}$. If power-law scaling only occurs at $T=T_{\mathrm{SG}}$
asymptotically, the quality-of-fit parameter $Q$ should be maximized at this point,
such that the information of both the critical temperature and the exponent
$\beta/\nu$ can be extracted by this procedure. An example for this approach for
$\sigma = 0.1$ is shown in the right panel of Fig.~\ref{Fig:qEA_vs_T_showcase_1}.

\begin{figure*}
  \begin{center}
    \includegraphics[scale=1, trim=0 10 0 0]{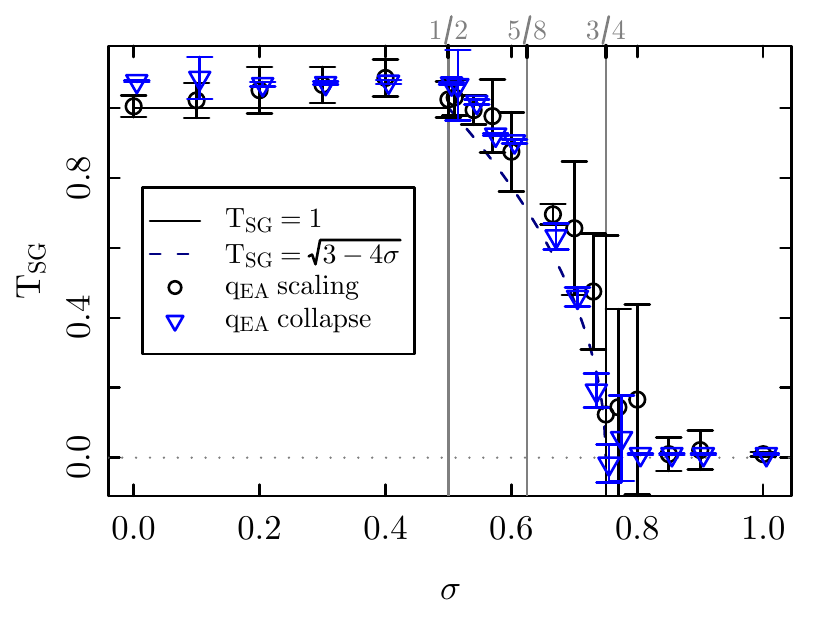} \hspace{0pt}
    \includegraphics[scale=1, trim=0 10 0 0]{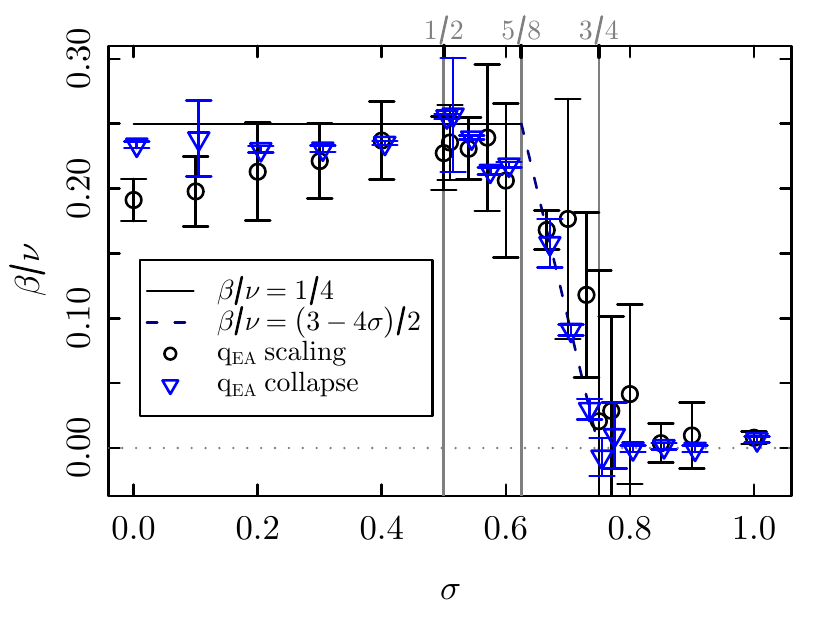}
  \end{center}
  \caption
  { (Color online). Transition temperature $T_\mathrm{SG}$ and critical exponent
    $\beta/\nu$ as extracted from the $Q$-maximization procedure described in
    the main text and in Fig.~\ref{Fig:qEA_vs_T_showcase_1}. As a comparison we
    extracted these quantities from a data collapse as well.}
  \label{Fig:results_from_qEA_FSS_1}
\end{figure*}

The overall results for the transition temperature $T_\mathrm{SG}$ and the critical
exponent $\beta/\nu$ resulting from this analysis are summarized in
Fig.~\ref{Fig:results_from_qEA_FSS_1}, together with the corresponding results of a
collapsing procedure. The estimates of the spin-glass temperature are consistent with
$T_\mathrm{SG} = \sqrt{3-4\sigma}$ in the relevant regime and become constant at
$T_\mathrm{SG} = 1$ for $\sigma\le 1/2$, while they vanish for $\sigma > 3/4$, as
expected. From the present analysis, $T_\mathrm{SG}$ can be resolved with more
precision than from the correlation length particularly in the mean-field regime
$\sigma < 5/8$. The exponent $\beta/\nu$ is consistent with the expectations
summarized in Sec.~\ref{Sec:1dpl_mInf_SG}, i.e., $\beta/\nu = 1/4$ for $\sigma < 5/8$
and $\beta/\nu = (3-4\sigma)/2$ for $5/8< \sigma \le 3/4$. The statistical precision
of our determination, however, is not sufficient to rule out possible different
scenarios and, in particular, to decide whether $\beta/\nu = (3-4\sigma)/2$ might be
exact in the non-mean-field regime. Again, statistical errors are calculated by an
elaborate jackknifing procedure. As shown in Table \ref{Tab:different_geometries}, no
significant deviations between the results for the different models of a 1d geometry
are observed.
\begin{table}[b]
  \caption
  {
    Results for different choices of the geometry of the model, which were introduced 
	in Sec.~\ref{sec:geometry}. There are no significant differences for
	the value $\sigma=3/4$ checked here.
    \label{Tab:different_geometries}
  }
  \begin{ruledtabular}
    \begin{tabular}{ct{7}t{7}t{7}}
      \toprule
      quantity			& \text{ring}	& \text{line}	& \text{summed line}	\\
      \hline
      $\mu$				& 0.372(6)		& 0.374(8)		& 0.371(7)				\\	
      \hline
      $\xi_{L}$			&				&				&						\\	
      $1/\nu$			& 0.000(5)		& -0.01(1)		& -0.03(5)				\\	
      $T_{\mathrm{SG}}$	& 0.00(4)		& 0.04(6)		& 0.01(16)				\\	
      \hline
      $q_{\mathrm{EA}}$	&				& 				&						\\	
      $\beta/\nu$		& 0.01(4)		& -0.02(5)		& 0.00(2)				\\	
      $T_{\mathrm{SG}}$	& 0.06(19)		& -0.09(27)		& 0.00(13)				\\	
    \end{tabular}
  \end{ruledtabular}
\end{table}

\subsubsection{Spin-glass susceptibility}

\begin{figure}
  \begin{center}
    \includegraphics[scale=1, trim=0 10 0 0]{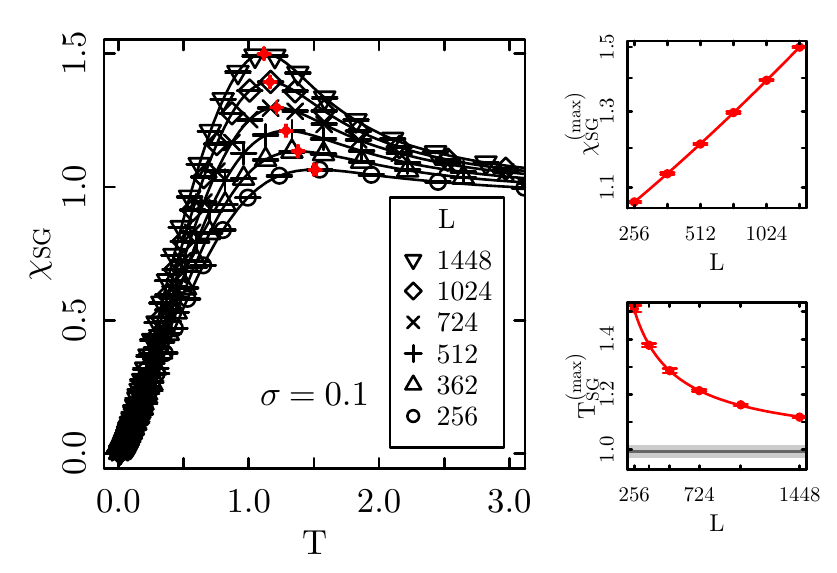} 
  \end{center}
  \caption
  { (Color online). Scaling of maxima of the spin-glass susceptibility
    $\chi_{\mathrm{SG}}$ in a showcase example with $\sigma=0.1$.}
  \label{Fig:chiSG_vs_T_1}
\end{figure}

\begin{figure}
  \begin{center}
    \includegraphics[scale=1]{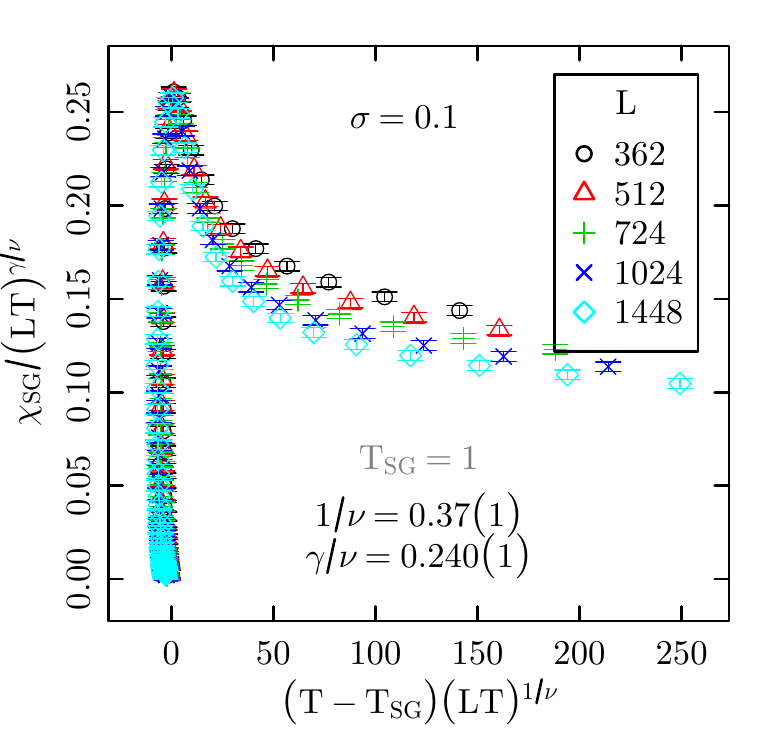}\\
    \includegraphics[scale=0.95,trim=0 20 0 0]{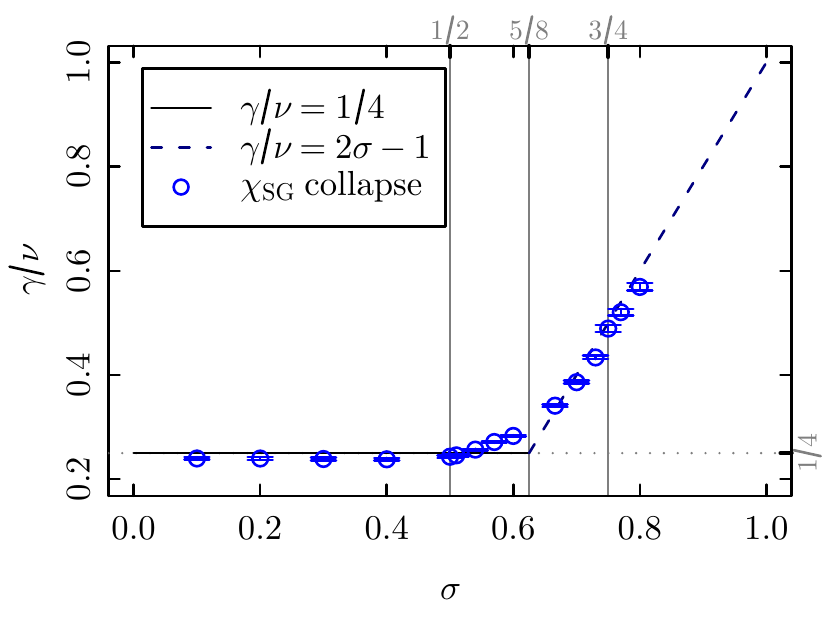}\\[-4ex]
  \end{center}
  \caption
  { (Color online). Top: A collapse of $\chi_\mathrm{SG}$ according to the extended
    scaling form \eqref{eq:extended_scaling} for $\sigma = 0.1$ with adaptively
    determined parameters $\gamma/\nu = 0.24$ and $\nu = 0.37$ ($T_\mathrm{SG} =
    1$). Bottom: estimates of $\gamma/\nu$ resulting from an adaptive collapsing
    routine together with the theoretical expectations $\gamma/\nu = 0.25$ for
    $\sigma \le 5/8$ and $\sigma/\nu = 2\sigma - 1$ for $\sigma > 5/8$, respectively.  }
  \label{Fig:results_from_chiSG_collapse}
\end{figure}

We finally analyzed the scaling behavior of the spin-glass susceptibility as defined
from the connected correlation function in Eq.~\eqref{Eq:chi_SG_cumul_corr_fct}. From
the discussion in Sec.~\ref{Sec:1dpl_mInf_SG} we expect scaling according to
\begin{equation}
  \chi_{\mathrm{SG}} \sim \left\{
    \begin{array}{rl}
      L^{1/4}\mathcal{C}(tL^{1/4}) & \sigma \le 5/8,\\
      L^{\gamma/\nu}\mathcal{C}(tL^{1/\nu}) & \sigma > 5/8.
    \end{array}
  \right.
  \label{Eq:FSS_suscept_minf}
\end{equation}
In contrast to the scaling of $q_\mathrm{EA}$ it is possible here without reference
to numerical derivatives to define a series of pseudo-critical temperatures from the
locations of the maxima of the susceptibility,
\begin{equation}
  T_{\mathrm{SG}}^{(\mathrm{max})} = T_\mathrm{SG} + cL^{-1/\nu}, \label{Eq:chiSG_FSS_fit}
\end{equation}
while the values of $\chi_\mathrm{SG}$ at the maxima should then follow
\begin{equation}
  \chi_{\mathrm{SG}}^{(\mathrm{max})} = cL^{\gamma/\nu}.
\end{equation} 
Fits of the corresponding forms to the data for $\sigma = 0.1$ are shown in
Fig.~\ref{Fig:chiSG_vs_T_1}. We find, however, that the resulting parameter estimates
are afflicted by very strong finite-size corrections. In particular, the resulting
estimates of $1/\nu$ are far off from our theoretical predictions as well as the
results from the analysis of the correlation length.

The presence of strong corrections in the scaling of $\chi_\mathrm{SG}$ is
well-known from studies, e.g., of the Ising spin glass. It has been suggested
\cite{campbell:06} that modified scaling forms incorporating scaling corrections
might contribute towards resolving such corrections and the proposed extended scaling
forms have been successfully applied to the Ising spin glass \cite{katzgraber:06a}. In
particular, one problem of the scaling form \eqref{Eq:FSS_suscept_minf} is that it
cannot reproduce the observed behavior $\chi_\mathrm{SG} \to 1$ as $T\to \infty$,
cf.\ Fig.~\ref{Fig:chiSG_vs_T_1}: assuming that ${\cal C}(x) \sim x^\alpha$ for $x\gg
1$, asymptotic size independence of the data at large $T$ requires that $\alpha =
-\gamma$ and hence $L^{\gamma/\nu}\mathcal{C}(tL^{1/\nu}) \to 0$ as
$T\to\infty$. While this is not in contradiction to scaling theory as the assumed
scaling form should only apply in the critical region, having a scaling form
consistent with the behavior as $T\to 0$ or $T\to \infty$ might allow to extend the
scaling regime or, equivalently, reduce the observed finite-size corrections. A
modified scaling form that serves this purpose is given by \cite{campbell:06}
\begin{equation}
  \label{eq:extended_scaling}
  \chi_\mathrm{SG} = (LT)^{\gamma/\nu}\tilde{\mathcal{C}}[(LT)^{1/\nu}t],
\end{equation}
which is compatible with $\chi_\mathrm{SG}\to 1$ as $T\to \infty$. We used this
extended scaling form to perform collapses of the finite-size data for the spin-glass
susceptibility. Even though some scaling corrections are implicitly included in
Eq.~\eqref{eq:extended_scaling}, these collapses are found to be rather unstable and,
hence, sensitive to the choice of starting values for the parameters and the range of
data points to be included for each lattice size. In view of these uncertainties, we
found it impossible to extract all three parameters, $T_\mathrm{SG}$, $\gamma/\nu$
and $1/\nu$ reliably from a single collapsing procedure. We hence decided to keep
$T_\mathrm{SG}$ fixed at the theoretical prediction $T_\mathrm{SG} =
\sqrt{3-4\sigma}$ which, as is shown in the results of
Fig.~\ref{Fig:qEA_vs_T_showcase_1} for the order parameter, is well compatible with
our numerical results. An example collapse is shown in the left panel of
Fig.~\ref{Fig:results_from_chiSG_collapse}. The right panel shows our resulting
estimates of $\gamma/\nu$ for $\sigma \le 0.8$. These are roughly compatible with our
expectations of $\gamma/\nu = 0.25$ for $\sigma \le 5/8$ and $\sigma/\nu = 2\sigma -
1$ for $\sigma > 5/8$. For $\sigma \gtrsim 3/4$, we do not find stable collapses with
reasonable parameters which we attribute to the fact that, there, $T_\mathrm{SG} =
0$, but our data only reach down to $T_\mathrm{min} = 0.01$. The resulting values of
$1/\nu$ are strongly fluctuating and hence not useful as reliable estimates of this
quantity. An alternative collapsing exercise using a plot of $\chi_\mathrm{SG}$ as a
function of $\xi/L$, which should have the theoretical advantage of involving only a
single adjustable parameter $\gamma/\nu$ did, unfortunately, not lead to more
reliable results.

\section{Conclusions\label{sec:concl}}

We have used extensive numerical simulations together with a number of
phenomenological scaling arguments to give a rather comprehensive account of the
critical behavior of the one-dimensional spin-glass model with power-law interactions
in the limit of an infinite number of spin components $m$. Compared to the more familiar
case with $m < \infty$, we find a number of remarkable differences which are, in
part, related to the shift in the lower and upper critical dimensions of the model.

The lack of metastability in the model allows to perform quasi-exact ground-state
calculations. The resulting defect energies are well described by a long-range
stiffness exponent $\theta = 3/4-\sigma$. This relation can also be deduced from
scaling arguments, but we have not been able to provide a more rigorous
derivation. This relation results in an upper critical $\sigma_u = 3/4$, where
finite-temperature spin-glass transitions first disappear. On lowering $\sigma$,
mean-field behavior sets in at $\sigma_l = 5/8$. These critical interaction ranges
are different from the $\sigma_l = 2/3$ and $\sigma_u = 1$ found for spin glasses
with finite $m$. \cite{kotliar:83} For hypercubic lattices it has been speculated
that $d_l=d_u$ if $m\rightarrow\infty$. \cite{viana:88,lee:05} According to our
analysis of the 1d model theoretical as well as numerical evidence exists showing
that the upper and lower critical $\sigma$ are well separated. We also investigated
the distribution of ground-state energies, and it is found to be Gaussian for the
full range of $\sigma$, again in contrast to the Ising case where non-Gaussian
distributions where found in the mean-field regime
\cite{katzgraber:05b}. Sample-to-sample fluctuations are trivial with $\Theta_f =
1/2$ for $\sigma > 1/2$, but cross over to a value consistent with $\Theta_f = 1/5$
for $\sigma < 1/2$ as conjectured for the SK model in
Ref.~\onlinecite{aspelmeier:10}. 

In Ref.~\onlinecite{leuzzi:08} it has been suggested to study a diluted version of
the 1d long-range spin-glass model to reach even larger system sizes. As we point out
here, however, the two models are not in the same universality class for $\sigma >
1$, where the diluted model becomes equivalent to a short-range 1d system. Right at
$\sigma = 1$, critical exponents depend continuously on the average coordination
number $z$. Also, sample-to-sample fluctuations are trivial for the diluted model
with $\Theta_f = 1/2$ for {\em all\/} $\sigma$, an effect anticipated for the Ising
case \cite{parisi:10}. Additionally, we observe more pronounced scaling corrections
for the diluted model even in the regime $1/2\le \sigma \le 1$ such that we cannot
find an advantage for numerical simulations in the larger system sizes reachable
through the dilution. Additionally, we compared different realizations of the 1d
geometry using rings and chains with and without Ewald summation of interactions.  As
scaling corrections for this class of models are pronounced, one could hope that some
variant of the model leads to a substantial reduction in corrections. For zero
temperature (cf.\ Fig.~\ref{Fig:defE_mu_sig100}) as well as for the critical behavior
at $T> 0$, however, we find no significant differences in the FSS of the different
model variants considered, cf.\ the data collected in Table
\ref{Tab:different_geometries}.

The ground-state calculations are complemented by results from an iterative solution
of the saddle-point equations resulting in the $m\to\infty$ limit, yielding access to
the order parameter, spin-glass susceptibility and correlation length. In contrast to
Ref.~\onlinecite{lee:05}, we argue that using appropriate definitions of the basic
observables the model does show true long-range order in the low-temperature phase,
even for the order of limits $m\to\infty$ before $N\to\infty$ naturally taken in
numerical studies. The critical exponents $\nu$, $\beta$ and $\gamma$ numerically
determined from this approach are consistent with our theoretical arguments in the
full range of $\sigma$. The critical exponents in the non-mean-field regime have been
hard to determine with precision. However, one of the surprises is the utility of the
simple approximate RG scheme first suggested by McMillan \cite{mcmillan:84a} which
seems to work quite well over the entire non-mean field region, $5/8 < \sigma \le
3/4$.

As an aside, we find clear-cut evidence of the recently suggested exactness of
mean-field theory for spin-glass models in the regime $\sigma < 1/2$, \cite{mori:11}
where we see non-universal properties such as the average ground-state energy to be
independent of the interaction range $\sigma$.

The $m \rightarrow \infty$ model studied in this paper is an interesting model in its
own right, partly because it is one of relatively few models known which have a
failure of hyperscaling. The phenomenon of dimensional reduction occurs in its
short-range $ d$-dimensional version which suggests that there might be some elegant
supersymmetry in the model, but this has yet to be discovered. But our chief
motivation in understanding this model was to clear the ground for our future $1/m$
expansion study of spin glasses.

\begin{acknowledgments}
  The authors are indebted to T.~Aspelmeier, H.~Katzgraber and A.~P.~Young for useful
  discussions. We thank T.\ Yavors'kii for useful advice regarding the Ewald
  summation technique. The authors acknowledge computer time provided by NIC J\"ulich
  under grant No.\ hmz18 and funding by the DFG through the Emmy Noether Program
  under contract No. WE4425/1-1.
\end{acknowledgments}

%\bibliography{1dpl_biblio}
\bibliography{citeulike_nourl_noissn}

\end{document}